\documentstyle[seceq,mbf,preprint,epsf]{ptpmod}

\def\aa{x_u}
\def\bb{x_\propto}
\def\cc{x_\infty}
\def\dd{x_\Omega}
\def\ee{x_4}
\def\ff{x_{\rm c}}
\def\xu{x_u}
\def\xa{x_\propto}
\def\xif{x_\infty}
\def\xo{x_\Omega}
\def\xc{x_{\rm c}}
\def\m1{{-1}}
\newcommand{\nn}{\nonumber\\}
\newcommand{\QB}{Q_{\rm B}}
\newcommand{\tQB}{{\tilde Q}_{\rm B}}
\newcommand{\half}{\frac{1}{2}}
\newcommand{\abs}[1]{\left| #1 \right|}
\newcommand{\bra}[1]{\left< #1 \right|}
\newcommand{\ket}[1]{\left| #1 \right>}
\newcommand{\sbra}[1]{\,\langle#1|\,}
\newcommand{\sket}[1]{\,|#1\rangle\,}
\newcommand{\bidx}[1]{{\textstyle {\atop #1}}\!\!}%{}_{#1}\hspace{-0.1em}}
\def\o{{\rm o}}
\def\c{{\rm c}}
\def\LRarrow{\quad \longleftrightarrow \quad }
\def\BRS{``BRS"\ }
\def\B{{\rm B}}
\def\bzm#1{b_0^{{-}(#1^\c)}}
\def\bP#1{{(b_0^-{\cal P})}^{(#1^\c)}}

\newcommand{\Vtho}[1]{\,\langle V_3^{\rm o}(#1)|\,}
\newcommand{\Vfo}[1]{\,\langle V_4^{\rm o}(#1)|\,}
\newcommand{\Vthc}[1]{\,\langle V_3^{\rm c}(#1)|\,}
\newcommand{\UU}[1]{\,\langle U(#1)|\,}
\newcommand{\Valp}[1]{\,\langle V_\propto(#1)|\,}
\newcommand{\Vinf}[1]{\,\langle V_\infty(#1)|\,}
\newcommand{\Uomg}[1]{\,\langle U_\Omega(#1)|\,}
\newcommand{\vfo}[1]{\,\langle v_4^{\rm o}(#1)|\,}
\newcommand{\vtho}[1]{\,\langle v_3^{\rm o}(#1)|\,}
\newcommand{\vthc}[1]{\,\langle v_3^{\rm c}(#1)|\,}
\newcommand{\uv}[1]{\,\langle u(#1)|\,}
\newcommand{\valp}[1]{\,\langle v_\propto(#1)|\,}
\newcommand{\vinf}[1]{\,\langle v_\infty(#1)|\,}
\newcommand{\uomg}[1]{\,\langle u_\Omega(#1)|\,}
\def\op#1{{\cal O}^{(#1)}}
\def\calO{{\cal O}}
\def\VEV#1{\left<#1\right>}
\def\lbl#1{{(#1)}}
\def\of{{\ket{\Psi}}}
\def\cf{{\ket{\Phi}}}
\def\check#1{\stackrel{\vee}{#1}}
\def\sfrac#1#2{\hbox{\large ${#1\over#2}$}}
\def\Idx<#1,#2,#3>{{\mbf #1}^{#2}_{#3}}
\def\Idxc<#1,#2,#3>{{\mbf #1}^{#2}_{#3}}
\def\v#1#2(#3#4#5#6#7){{\xdef\tmpone{#1}\xdef\tmptwo{#2}\xdef\open{o}
\xdef\closed{c}\xdef\oMg{\Omega}\xdef\aLp{\propto}\xdef\iNf{\infty}
\if 3\tmpone
 \ifx \open\tmptwo \vtho{#3,#4,#5} \else
 \ifx \closed\tmptwo \vthc{#3^\c,#4^\c,#5^\c} \else\fi\fi \else
\if 4\tmpone \vfo{#3,#4,#5,#6;#7} \else
\ifx \oMg\tmpone \uomg{#3,#4,#5^\c;#6} \else
\ifx \aLp\tmpone \valp{#3,#4;#5,#6} \else
\ifx \iNf\tmpone \vinf{#3^\c,#4^\c;#5} \else
\if 2\tmpone \uv{#3,#4^\c;#5} \else
\fi\fi\fi\fi\fi\fi}}
\def\V#1#2(#3#4#5#6){{\xdef\tmpone{#1}\xdef\tmptwo{#2}\xdef\open{o}%
\xdef\closed{c}\xdef\oMg{\Omega}\xdef\aLp{\propto}\xdef\iNf{\infty}%
\if3\tmpone%
\ifx\open\tmptwo \Vtho{#3,#4,#5} \else
\ifx\closed\tmptwo \Vthc{#3^\c,#4^\c,#5^\c} \else\fi\fi\else
\if4\tmpone \Vfo{#3,#4,#5,#6} \else
\ifx\oMg\tmpone \Uomg{#3,#4,#5^\c} \else
\ifx\aLp\tmpone \Valp{#3,#4} \else
\ifx\iNf\tmpone \Vinf{#3^\c,#4^\c} \else
\if2\tmpone \UU{#3,#4^\c} \else
\fi\fi\fi\fi\fi\fi}}
\def\VN#1(#2){\left\langle V_{(#1)}(#2)\right|}
\def\Rflo(#1#2){\left\langle R^\o(#1,#2)\right|}
\def\Rflc(#1#2){\left\langle R^\c(#1^\c,#2^\c)\right|}
\def\Gv#1(#2){\bra{v_{\rm #1}(#2)}}
\def\braRo(#1#2){\left\langle R^\o(#1,#2)\right|}
\def\braRc(#1#2){\left\langle R^\c(#1^\c,#2^\c)\right|}
\def\ketRo(#1#2){\left|R^\o(#1,#2)\right\rangle}
\def\ketRc(#1#2){\left|R^\c(#1^\c,#2^\c)\right\rangle}
\def\vfour(#1#2#3#4){{\sbra{v_4^{\o\,(\alpha_{#1},\alpha_{#2},\alpha_{#3},\alpha_{#4})}%
(#1,#2,#3,#4;\sigma_0)}}}
\def\sht#1{{\setbox0=\hbox{\scriptsize$#1$}\ht0=7pt\box0}}
\def\zht#1{{\setbox0=\hbox{\scriptsize$#1$}\ht0=0pt\box0}}
\preprintnumber[3cm]{%<-- [..]: optional width of preprint # column.
KUNS-1508\\ HE(TH)~98/06\\ UT-814\\ hep-th/9807066\\
July}
\title{%        %You can use \\ for explicit line-break
BRS Invariance
of Unoriented\\ Open-Closed String Field Theory
}
\author{%       %Use \sc for the family name
Tsuguhiko {\sc Asakawa},\footnote{E-mail:asakawa@gauge.scphys.kyoto-u.ac.jp}
 Taichiro {\sc Kugo}\footnote{E-mail: kugo@gauge.scphys.kyoto-u.ac.jp}
and Tomohiko {\sc Takahashi}$^{\dagger,}$\footnote{
JSPS Research Fellow. E-mail: tomo@hep-th.phys.s.u-tokyo.ac.jp}
}

\inst{%         %Affiliation, neglected when [addenda] or [errata]
Department of Physics, Kyoto University, Kyoto 606-8502
\\
$^\dagger$Department of Physics,
University of Tokyo, Tokyo 113
}

\recdate{%      %Editorial Office will fill in this.
\today
}

\abst{%       %this abstract is neglected when [addenda] or [errata]
We present the full action for the unoriented open-closed string field 
theory which is based on the $\alpha=p^+$ HIKKO type vertices. 
The BRS invariance of the action is proved up to the terms which are 
expected to cancel the anomalous one-loop contributions.  
This implies that the system is invariant under the gauge transformations 
with open and closed string field parameters up to 
the anomalies.}

\begin{document}

\maketitle

\section{Introduction}

In our previous paper,\cite{rf:KugoTaka} which we refer to as I henceforth, 
we have constructed a consistent
string field theory (SFT) for an unoriented open-closed string 
mixed system to the quadratic order in the string 
fields, and proved the invariance under the gauge transformation with 
closed string field parameter. It was pointed out that the infinity 
cancellation between the disk and projective plane 
amplitudes
\cite{rf:GreenSchwarz,rf:DougGrin,rf:Weinberg,rf:ItoyamaMoxhay,rf:Ohta,rf:DasRey,rf:Tseytlin1} 
plays an essential role for the gauge invariance of the theory. This, in
particular, implies that any {\it oriented} string field theory 
containing open string, where there is no projective plane amplitude 
contribution, cannot be a consistent theory at least on the flat 
background.
\cite{rf:FischSusskind1,rf:FischSusskind2,rf:CLNY,rf:PolCai,rf:FischSussKlebanov,rf:DasRey,rf:Tseytlin2,rf:Pol} 
For the case of light-cone gauge SFT, this means the 
violation of the Lorentz invariance.

In this paper, we continue this task and present the full action 
for this unoriented open-closed string field theory which is an 
$\alpha=p^+$ HIKKO type theory\cite{rf:kugozwie} based on the 
light-cone type vertices. The BRS invariance of the action is thoroughly
proved, up to the terms which are expected to cancel the anomalous 
one-loop contributions. 

\begin{figure}[tb]
   \epsfxsize= .8\textwidth   %or \epsfysize= HEIGHT cm
   \centerline{\epsfbox{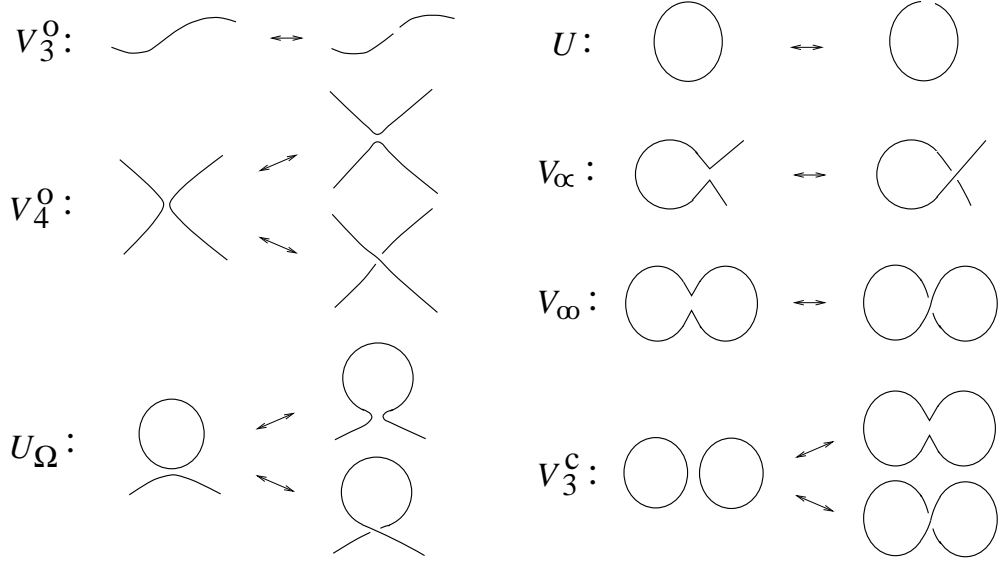}}
 \caption{Seven interaction vertices.}
 \label{fig:vertex}
\end{figure}

The SFT action for such an open-closed string mixed system 
has been known in the case of light-cone gauge and 
{\it oriented} string\cite{rf:KK,rf:ST1,rf:ST2,rf:KikkawaSawada} 
and it had five types of interaction terms, open 3- and 4-string vertices 
$V_3^\o,\ V_4^\o$, closed 3-string vertex $V_3^\c$, open-closed transition 
vertex $U$ and open-open-closed vertex $U_\Omega$.
In the present case of unoriented strings, two additional quadratic 
interaction terms become newly allowed and were studied in detail in I; 
self-intersection interactions $V_\propto$ for open string 
and $V_\infty$ for closed string.\cite{rf:GreenSchwarz2}
Intuitively, the string interactions are 
of only two types if viewed locally on the string world sheet; 
one is the joining-splitting type interaction typically appearing in 
$V_3^\o$ and another is the rearrangement interaction typically 
appearing in $V_3^\c$. If so, these seven vertices already exhaust all the 
possible interaction terms, and are depicted in Fig.~\ref{fig:vertex}. 
%\begin{wrapfigure}[n]{r}{6.6cm}
%\end{wrapfigure}
Taking account of our previous work also, 
the full action of the present system is naturally expected to be given 
by
\begin{eqnarray}
S &=& -{1\over2}\bra{\Psi}\tQB^\o\Pi\ket{\Psi}
-{1\over2}\bra{\Phi}\tQB^\c(b_0^-{\cal P}\Pi)\ket{\Phi} \nn
&&+\frac{g}{3}\V3o(123x)\of_{321} 
+\ee \frac{g^2}{4}\V4(1234)\of_{4321}      
+\bb \hbar\frac{g^2}{2}\V\propto(12xx)\of_{21} \nn
&&+\xc \hbar^{1/2}\frac{g^2}{3!}\V3c(123)\cf_{321}
     +\cc \hbar\frac{g^2}{2}\V\infty(12xx)\cf_{21} \nn
&&+\aa \hbar^{1/2}g \UU{1,2^\c}\cf_{2}\of_{1}
     +\dd \hbar^{1/2}\frac{g^2}{2}\V\Omega(123x)\cf_3\of_{21}\,,
\label{eq:action}
\end{eqnarray}
where $\ee$, $\bb$, $\xc$, $\cc$, $\aa$ and $\dd$ are coupling constants
(relative to the open 3-string coupling constant $g$), and we have 
explicitly shown the power of $\hbar$ (as a loop expansion 
parameter)\cite{rf:ST2,rf:Zwie2}
for each interaction term for clarity although we will suppress them 
henceforth. For notations and conventions, we follow our previous paper 
I. The open and closed string fields are denoted by $\ket{\Psi}$ and 
$\ket{\Phi}$, respectively, both of which are Grassmann {\it odd}. The 
multiple products of string fields are denoted for brevity as 
\begin{equation}
\of_{n\cdots21}\equiv\ket{\Psi}_n\cdots\ket{\Psi}_2\ket{\Psi}_1\,.
\end{equation}
The tilded BRS charges $\tQB$ here, introduced in I, are given by the 
usual BRS charges $\QB$ plus counterterms for the `zero intercept' 
proportional to the squared string length parameter $\alpha^2$:
\begin{equation}
\tQB^\o = \QB^\o + \lambda_\o g^2 \alpha^2 c_0\,, \qquad 
\tQB^\c = \QB^\c + \lambda_\c g^2 \alpha^2 c_0^+\,.
\end{equation}
The ghost zero-modes for the closed string are defined by
$c_0^+\equiv(c_0+\bar c_0)/2,\  c_0^- \equiv c_0-\bar c_0$, and 
$b_0^+\equiv b_0+\bar b_0, \ b_0^-\equiv(b_0-\bar b_0)/2$. 
The string fields are always accompanied by the unoriented projection 
operator $\Pi$, which is given by using twist operator
$\Omega$ in the form $\Pi= (1+\Omega)/2$, where $\Omega$ for open string case means 
also taking transposition of the matrix index. 
The closed string is further accompanied by the projection operator 
${\cal P}$, projecting the $L_0-\bar L_0=0$ modes out,
\begin{equation}
{\cal P}\equiv\int^{2\pi}_0
\frac{d\theta}{2\pi}\exp i\theta(L_0-\bar{L}_0),
\end{equation}
and the corresponding anti-ghost zero-mode factor 
$b_0^-=(b_0-\bar b_0)/2$. 

The main purpose of this paper is to prove the BRS invariance of the 
action (\ref{eq:action}) and to determine the coupling constants $\ee$, 
$\bb$, $\xc$, $\cc$, $\aa$ and $\dd$. As is well known already in 
the light-cone gauge SFT, however, the open-closed mixed system suffers from
the anomaly\cite{rf:ST1,rf:ST2,rf:KK}
and thus the system is not BRS invariant as far as we 
consider the tree action (\ref{eq:action}) alone. Ideally, we should 
also discuss the anomalous loop diagram contributions here. But, since 
the BRS invariance proof is a bit too long already at the `tree level', 
we are obliged to defer the anomaly discussion to the forthcoming paper. 
Therefore, we here content ourselves with doing the following. 
First we classify the terms appearing in the BRS transform 
${\mbf \delta}_\B S$ of the action into groups according to the numbers of the 
external open and closed string fields and the power of coupling constant
$g$. The BRS invariance implies that those terms should cancel
each other separately in each group. The cancellation always occur between 
a pair of the configurations in which the interactions at two 
interaction points take place in an opposite order. 
Then we can see which groups of the terms become the counterterms for the 
anomalous loop diagrams; namely, those `loop' groups contain the terms 
for which the configurations become loop diagrams if the order of the 
two interactions is interchanged. For all the other groups, which 
we call `tree' groups, we prove successively that the cancellations 
between such pairs of configurations indeed occur and the terms in each 
group in ${\mbf \delta}_\B S$ completely cancel out.

The paper is organized as follows. In \S2, we explain in some detail how 
the SFT vertices are constructed, since the signs are very important to 
show the cancellation for proving the BRS invariance. In \S3, we calculate 
the BRS transformation ${\mbf \delta}_\B S$ of the action 
in a systematic way 
and classify the appearing terms into groups mentioned above. 
\S4 is the main part of this paper where we present the BRS invariance 
proof of our action in a manner as explained just above. 
The final section \S5 is devoted to the summary. 
In Appendix A we summarize the general rule for obtaining the BRS and 
gauge transformation laws from the action with a precise treatment of the 
statistics of the open and closed string fields. 
In Appendix B we explain how the ``Generalized Gluing and Resmoothing 
Theorem" (GGRT) proved by LeClair, Peskin and Preitschopf\cite{rf:LPP} 
and the present authors\cite{rf:AKT} for the pure open string system case 
is made applicable to the present open-closed mixed system.

\section{Vertices}

To discuss the BRS invariance of the action (\ref{eq:action}), we must 
show the cancellations between various pairs of terms, as we will do in 
later sections. Therefore it is very important to define the vertices 
very correctly including their signs as well as the weights. 
Fortunately, the definition of the vertices in the manner of 
LeClair, Peskin and Preitschopf (LPP),\cite{rf:LPP} is very powerful and 
convenient also for this purpose. Each vertex of our string field theory
is defined in the form of a product of the LPP vertex corresponding
to a specified way of gluing of strings and the anti-ghost factors 
corresponding to the moduli parameters (interaction points) of the 
vertices. The GGRT,\cite{rf:LPP,rf:AKT} for the LPP vertices makes 
it possible to treat the
weights of the terms without recourse to the detailed expressions for 
the LPP vertices, and the signs can be traced neatly by the anti-ghost 
factors contained in our SFT vertices.

Taking these into account, we give a definition of our SFT vertices 
in this section. For clarity, by taking the open 4-string vertex 
$\V4(1234)$ as a concrete example, we first explain in some details 
%the procedure and convention 
how our SFT vertices are constructed.  

\begin{figure}[tb]
   \epsfxsize= 8cm   %or \epsfysize= HEIGHT cm
   \centerline{\epsfbox{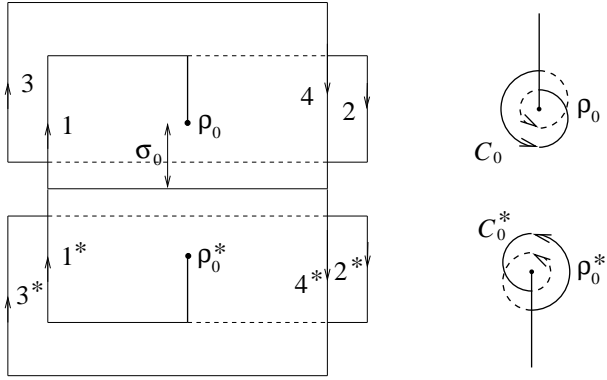}}
 \caption{The $\rho$ plane of the open 4-string vertex $V_4^\o$. 
The integration contours $C_0$ and $C_0^*$ used in 
Eq.~(\protect\ref{eq:QCdeform-b}) for defining 
$b_{\rho_0}$ and $b_{\rho^*_0}$ are also shown. }
 \label{fig:v4}
\end{figure}

The corresponding LPP vertex $\bra{v_4^\o}$ is uniquely given once how the 
participating strings are glued is known. In our case of $\alpha=p^+$ HIKKO 
type theory,\cite{rf:kugozwie}
the gluing is specified by using the string length parameters 
$\alpha_r$ as well as by the moduli parameter $\sigma_0$ specifying the 
interaction point. So, for a given set of the $\alpha$ parameters, 
$(\alpha_1,\alpha_2,\alpha_3,\alpha_4)$, the corresponding LPP vertex is denoted as 
\begin{equation}
\vfour(1234), 
\end{equation}
and is defined by referring to the conformal field theory Green function:
\begin{eqnarray}
&&\vfour(1234)
\op4_4\ket{0}_4\op3_3\ket{0}_3\op2_2\ket{0}_2\op1_1\ket{0}_1\nn
&&\qquad 
= \prod_{r=1}^4 \left({dZ_r\over dw_r}\right)^{d_{\calO_r}}\!\!\cdot\,  
\bigl<\calO_4(Z_4)\calO_3(Z_3)\calO_2(Z_2)\calO_1(Z_1)\bigr>. 
\label{eq:LPPdef}
\end{eqnarray}
We call this type of vertex with string length parameters specified 
`specific LPP vertex', in distinction from `generic one' introduced below. 
This open 4-string vertex exists only for sets of the $\alpha$ parameters 
$(\alpha_1,\alpha_2,\alpha_3,\alpha_4)$ with alternating signs, $(+,-,+,-)$ and 
$(-,+,-,+)$. The string configuration is explicitly depicted in Fig.\ref{fig:v4}
%\begin{wrapfigure}[n]{r}{6.6cm}
%\end{wrapfigure}
for the case of ${\rm sign}(\alpha_1,\alpha_2,\alpha_3,\alpha_4)=(+,-,+,-)$. An important 
property of such specific LPP vertex is 
\begin{eqnarray}
\vfour(1234)&=&\vfour(2341)\nn
&=&\vfour(3412),
\label{eq:LPPprop}
\end{eqnarray}
etc. Namely, once the vertex type is fixed (now, $\bra{v_4^\o}$), 
the specific LPP vertex is uniquely given by specifying which string 
(i.e., Fock space label) $r=1,2,\cdots$ has which length parameter $\alpha_r$. 
The order of the arguments is irrelevant aside from the cyclic ordering 
among open strings (and totally irrelevant for closed strings). This 
property is apparently trivial since those LPP vertices correspond to 
the same mapping of the unit disks $\abs{w_r}\leq1$ of strings $r$ into 
the complex plane $z$. But the equality including the overall sign 
factor is not so trivial in fact, so we demonstrate it from the 
definition (\ref{eq:LPPdef}):
\begin{eqnarray}
&&\vfour(2341)
\op1_1\ket{0}_1\op4_4\ket{0}_4\op3_3\ket{0}_3\op2_2\ket{0}_2 \nn
&&\quad 
= \prod_{r=1}^4 \left({dZ_r\over dw_r}\right)^{d_{\calO_r}}\!\!\cdot\,  
\bigl<\calO_1(Z_1)\calO_4(Z_4)\calO_3(Z_3)\calO_2(Z_2)\bigr> \nn
&&\quad 
= (-1)^{\abs1(\abs2+\abs3+\abs4)}\prod_{r=1}^4 \left({dZ_r\over dw_r}\right)^{d_{\calO_r}}\!\!\cdot\,  
\bigl<\calO_4(Z_4)\calO_3(Z_3)\calO_2(Z_2)\calO_1(Z_1)\bigr> \nn
&&\quad 
= (-1)^{\abs1(\abs2+\abs3+\abs4)}\vfour(1234)
\op4_4\!\ket{0}_4\op3_3\!\ket{0}_3\op2_2\!\ket{0}_2\op1_1\!\ket{0}_1 \nn
&&\quad 
= \vfour(1234)
\op1_1\ket{0}_1\op4_4\ket{0}_4\op3_3\ket{0}_3\op2_2\ket{0}_2,
\end{eqnarray}
where $\abs{r}$ is the statistics index of the operator $\calO_r$ 
which is 0 (1) if $\calO_r$ is bosonic (fermionic). Note that this 
simple property is achieved by the fact that the Fock state 
$\op{r}_r\ket{0}_r$ of string $r$ and the conformal field theory 
operator $\calO_r$ obeys the same statistics thanks to the 
{\it convention} that SL(2;C) vacuum $\ket{0}$ is Grassmann {\it even}. 

We now define `generic LPP vertex' $\v4(1234\sigma_0)$ 
by integrating over the length parameters $\alpha_r$ as follows:
\begin{equation}
\bra{v_4^\o(1,2,3,4;\sigma_0)}
=\int\prod_{r=1}^4 d\alpha_r\, \delta(\sum_r \alpha_r) \vfour(1234) .
\end{equation}
This generic LPP vertex enjoys the cyclic symmetry property because of 
Eq.~(\ref{eq:LPPprop}):
\begin{equation}
\v4(1234\sigma_0) = \v4(2341\sigma_0) = 
\v4(3412\sigma_0) = \v4(4123\sigma_0).
\end{equation}
Henceforth we always mean this generic LPP vertex if we simply call LPP 
vertex. Although the integration is performed over the length parameters 
$\alpha_r$ in this generic LPP vertex, only a single specific LPP vertex is 
picked up if it is contracted with the specific external string states 
$\op{r}_r\ket{0}_r$; usually, the external state operator $\op{r}_r$ 
takes the form $\op{r}_r=\hat\op{r}_r\exp(ip_rX^{(r)})$ so that the 
state
\begin{equation}
\op{r}_r\ket{0}_r=
\hat\op{r}_re^{ip_rX^{(r)}}\ket{0}_r =
\hat\op{r}_r\ket{p}_r
\end{equation}
carries definite momentum $p_r=({\mib p}_r, p^-_r, p^+_r)$ and hence 
definite string length parameter $\alpha_r=2p^+_r$ ($\alpha_r=p^+_r$ for closed 
string). Since the specific LPP vertex $\vfour(1234)$ is constructed on 
the bra state $\prod_r \bidx{r}\bra{\alpha_r}$ and the bra and ket states 
carrying different values of $\alpha_r$ are orthogonal to each other, a 
single specific LPP vertex can survives. 

Finally the vertex $\V4(1234)$ used in the string field theory 
can now be defined by 
\begin{equation}
\label{eq:V4}
\V4(1234) = \int_{\sigma_i}^{\sigma_f}
d\sigma_0 \v4(1234\sigma_0) b_{\sigma_0} \,\prod_r\Pi^{(r)},
\label{eq:SFTvertex}
\end{equation}
where $\sigma_i$ and $\sigma_f$ denote the initial and final points of the 
moduli $\sigma_0$ (interaction point), $\Pi$ is the unoriented
projection operator, and $b_{\sigma_0}$ is the anti-ghost 
factor associated with the quasi-conformal deformation of the Riemann 
surface corresponding to the change of the moduli $\sigma_0$,
\cite{rf:GM,rf:Martinec,rf:DHokerPhong}
which, in this 4-string vertex case, is explicitly given by 
\begin{equation}
b_{\sigma_0}= \left({d\rho_0\over d\sigma_0}\right)b_{\rho_0}+ 
\left({d\rho^*_0\over d\sigma_0}\right)b_{\rho^*_0}\,, \qquad 
b_{\rho_0}= \oint _{C_0}{d\rho\over2\pi i}b(\rho), \quad  
b_{\rho^*_0}= \oint _{C^*_0}{d\rho\over2\pi i}b(\rho) 
\label{eq:QCdeform-b}
\end{equation}
where $C_0$ denotes the closed contour encircling the interaction point 
$\rho_0$ on the $\rho$-plane, and $C_0^*$ and $\rho_0^*$ being their mirrors 
(See Fig.~\ref{fig:v4}).  More generally speaking,
this anti-ghost factor $b_{\sigma_0}$ is characterized by the property that 
its BRS transform,
\begin{equation}
T_{\sigma_0}\equiv\{\,b_{\sigma_0},\, \QB\,\}
=\left({d\rho_0\over d\sigma_0}\right)T_{\rho_0}+
\left({d\rho^*_0\over d\sigma_0}\right)T_{\rho^*_0}\,, \qquad
T_{\rho_0}=\oint _{C_0}{d\rho\over2\pi i}T(\rho),% \ \ {\rm etc.}
\label{eq:QCdeform-T}
\end{equation}
(where $T(\rho)$ is the energy-momentum tensor), be a generator 
of the infinitesimal transformation for the change of the moduli 
$\sigma_0$;\cite{rf:AGMV,rf:KugoSuehiro}
i.e., 
$\v4(1234\sigma_0) T_{\sigma_0} = (d/d\sigma_0)\v4(1234\sigma_0)$. 
We, therefore, have the following important relation 
using the BRS invariance of the LPP vertex,
$\v4(1234\sigma_0) \QB=0$ with $\QB\equiv\sum_r\QB^{(r)}$:
\begin{eqnarray}
&&\v4(1234\sigma_0) b_{\sigma_0} \QB =
\v4(1234\sigma_0) \{\,b_{\sigma_0},\, \QB\,\}\nn
&&\qquad  =
\v4(1234\sigma_0) T_{\sigma_0} = {d\over d\sigma_0}\left\{\v4(1234\sigma_0)\right\}.
\label{eq:QBdsigma}
\end{eqnarray}

We come to another important point here: 
how do we define the moduli $\sigma_0$ explicitly?
We can use as $\sigma_0$ the value of the sigma coordinate $\sigma_0^{(r)}$ of 
any one of the participating string $r$. (For convenience sake, we define 
the coordinate $\sigma_0^{(r)}$ as $\sigma_0^{(r)}\equiv\abs{\alpha_r}{\rm Im}\ln w_r$ 
although the original sigma coordinate of string $r$ is 
$\sigma_r={\rm Im}\ln w_r$, so that the distance measured by $\sigma_0^{(r)}$ is 
equal to that on the $\rho$-plane in magnitude independently of $r$.) 
But the point is that the increasing directions of $\sigma_0^{(r)}$ are 
opposite if the signs $\alpha_r$ are opposite, which causes the sign change 
to the definition (\ref{eq:SFTvertex}). Indeed, if we take two adjacent 
strings 1 and 2 which carry opposite signs of $\alpha_r$, for instance, and 
suppose that the points $\sigma_0^{(1)}$ and $\sigma_0^{(2)}$ correspond to the 
same point on the $\rho$-plane, then the neighboring points $\sigma_0^{(1)}+\varepsilon 
$ and $\sigma_0^{(2)}-\varepsilon$ represent the same point. The contribution of this
infinitesimal region to the integral in the SFT vertex 
(\ref{eq:SFTvertex}) has opposite sign: indeed, since $d\sigma_0^{(1)}=-d\sigma 
_0^{(2)}$ and hence
\begin{equation}
b_{\sigma_0^{(1)}}= \left({d\rho_0\over d\sigma_0^{(1)}}\right)b_{\rho_0} =
-\left({d\rho_0\over d\sigma_0^{(2)}}\right)b_{\rho_0} =
-b_{\sigma_0^{(2)}},
\end{equation}
we have 
\begin{equation}
\int_{\sigma_0^{(1)}}^{\sigma_0^{(1)}+\varepsilon}\!d\sigma_0^{(1)} b_{\sigma_0^{(1)}}
=\int_{\sigma_0^{(2)}}^{\sigma_0^{(2)}-\varepsilon}\!(-d\sigma_0^{(2)}) (-b_{\sigma_0^{(2)}})
=-\int^{\sigma_0^{(2)}}_{\sigma_0^{(2)}-\varepsilon}\!d\sigma_0^{(2)}b_{\sigma_0^{(2)}}.
\end{equation}
Because of this, we generally have the relation
\begin{equation}
\int_{\sigma_i^{(r)}}^{\sigma_f^{(r)}}
\!\!d\sigma_0^{(r)} \v4(1234\sigma_0^{(r)}) b_{\sigma_0^{(r)}}
=
{\rm sign}(\alpha_r\alpha_s)\int_{\sigma_i^{(s)}}^{\sigma_f^{(s)}}
\!\!d\sigma_0^{(s)} \v4(1234\sigma_0^{(s)}) b_{\sigma_0^{(s)}}.
\label{eq:2.12}
\end{equation}
So we must specify which string's $\sigma_0^{(r)}$ coordinate is used 
for the moduli in the definition of the SFT vertex (\ref{eq:SFTvertex}). 
We sometimes use the notation like 
\begin{equation}
\langle{V_4^\o(1,{\stackrel{\downarrow}2},3,4)}| = 
\int_{\sigma_i^{(2)}}^{\sigma_f^{(2)}}
\!d\sigma_0^{(2)} \v4(1234\sigma_0^{(2)}) b_{\sigma_0^{(2)}}
\,\prod_r\Pi^{(r)},
\label{eq:2.13}
\end{equation}
to denote explicitly which string's $\sigma_0^{(r)}$ coordinate is used 
by putting a down arrow on the string label. However, 
we take the {\it convention} that with the SFT vertices 
{\it with the down arrow omitted} we always mean to 
use the $\sigma_0^{(r)}$ coordinate of the open string which appears as the 
{\it first argument} in the SFT vertex. Namely, 
\begin{equation}
\V4(1234) = 
\V4({\stackrel{\downarrow}1}234) = 
\int_{\sigma_i^{(1)}}^{\sigma_f^{(1)}}
\!d\sigma_0^{(1)} \v4(1234\sigma_0^{(1)}) b_{\sigma_0^{(1)}}
\,\prod_r\Pi^{(r)},
\label{eq:SFTvertex2}
\end{equation}
With this convention, the 4-string SFT vertex properly satisfies the 
following {\it anti-cyclic} symmetry
\begin{equation}
\V4x(1234)=-\V4x(2341)=+\V4x(3412)=-\V4x(4123)
\end{equation}
because of the alternating sign property of $(\alpha_1,\alpha_2,\alpha_3,\alpha_4)$ and 
the cyclic symmetry of LPP vertex. Note that this property should 
hold in any case as far as we take the convention that the open 
string field $\ket\Psi$ is Grassmann {\it odd}. This is because 
the SFT vertex appears in the action in the form 
$\V4(1234)\ket{\Psi}_4\ket{\Psi}_3\ket{\Psi}_2\ket{\Psi}_1$, the 
string label $r$ is totally dummy there and so we should have 
\begin{eqnarray}
\V4(1234)\!\ket{\Psi}_{4321}=\V4(2341)\!\ket{\Psi}_{1432}
=-\V4(2341)\!\ket{\Psi}_{4321}.
%\V4(1234)\ket{\Psi}_{4}\ket{\Psi}_3\ket{\Psi}_2\ket{\Psi}_1 
%&=&\V4(2341)\ket{\Psi}_1\ket{\Psi}_4\ket{\Psi}_3\ket{\Psi}_2 \nn
%&=&-\V4(2341)\ket{\Psi}_4\ket{\Psi}_3\ket{\Psi}_2\ket{\Psi}_1.
\hspace{3em}
\end{eqnarray}

In a similar fashion to this $V_4$ example, we can define all
the vertices appearing in the action (\ref{eq:action}).  The quadratic 
vertices $U$, $V_\propto$ and $V_\infty$ have been defined explicitly in
the previous paper I.
For the other cubic interaction vertices also, it have long been known
how the strings are glued (See, e.g., 
Refs.~\citen{rf:HIKKO1,rf:HIKKO2,rf:ShapThorn,rf:Hata-Nojiri}). 
For clarity, we here cite the expressions for all the seven vertices 
following our way of construction and notation.
The cubic interaction vertices $V_3^\o$ for open and $V_3^\c$ for closed 
strings and the open-closed transition vertex $U$ have no moduli parameters:
\begin{eqnarray}
&& \Vtho{1,2,3} = \vtho{1,2,3} \prod_{r=1,2,3} \Pi^{(r)}\,, \nn
&&
 \Vthc{1^\c,2^\c,3^\c} = \vthc{1^\c,2^\c,3^\c} 
 \prod_{r=1^\c,2^\c,3^\c} (b_0^-{\cal P}\Pi)^{(r)}\,, \nn
&&
\bra{U(1,2^\c)}=\bra{u(1,2^\c)}\bP{2}
\prod_{r=1,2^\c}\Pi^{(r)}\,.
\end{eqnarray}
Note that the anti-ghost and projection factors $(b_0^-{\cal P}\Pi)$ for 
the closed strings, each being Grassmann odd, are always multiplied in 
the order as appearing in the argument of the vertex. The 
open-open-closed vertex $U_\Omega$ and closed intersection vertex $V_\infty$, as
well as the open quartic interaction vertex $V_4^\o$ have one moduli 
parameter specifying the interaction point:
\begin{eqnarray}
&& \V4(1234) = \int_{\sigma_i}^{\sigma_f}d\sigma^{(1)}_0
\v4(1234{\sigma_0^{(1)}})b_{\sigma_0^{(1)}} \prod_{r=1}^4 \Pi^{(r)}\,, \nn
&&
 \Uomg{1,2,3^\c} = \int_{\sigma_i}^{\sigma_f}d\sigma^{(1)}_0
  \uomg{1,2,3^\c;\sigma_0^{(1)}}b_{\sigma_0^{(1)}} \Pi^{(1)} \Pi^{(2)}
  (b_0^-{\cal P}\Pi)^{(3^\c)}\,, \nn
&&
\bra{V_\infty(1^\c,2^\c)}=\int_0^{\alpha_{1}\pi/2}\!d\sigma^{(1)}_0
\bra{v_\infty(1^\c,2^\c;\sigma^{(1)}_0)}
b_{\sigma^{(1)}_0}\!\!\prod_{r=1^\c,2^\c}(b_0^-{\cal P}\Pi)^{(r)}.
\end{eqnarray}
Finally the open intersection vertex $V_\propto$ have two moduli parameters 
corresponding to its two interaction points:
\begin{equation}
\bra{V_\propto(1,2)}=\int_{0\leq\sigma^{(1)}_1\leq\sigma^{(1)}_2\leq\pi\alpha_1}
\hspace{-1.5em}d\sigma_1 d\sigma_2\bra{v_\propto(1,2;\sigma^{(1)}_1,\sigma^{(1)}_2)}
b_{\sigma^{(1)}_1}b_{\sigma^{(1)}_2}
\prod_{r=1,2}\Pi^{(r)}\,.
\end{equation}

\section{BRS Transformation}

Once the action is given, there is now a standard procedure for giving 
the BRS and gauge transformations.\cite{rf:Hata1,rf:Hata2,rf:Zwie1,rf:HataZwie}
In this procedure, if the action is 
invariant under the BRS transformation, the invariance under the gauge 
transformation automatically follows. Although this procedure is 
in principle well-known, the details like signs are by no means trivial 
in this case of the mixed system of open and closed strings.  So, 
in Appendix A, we explain the details of this procedure by developing a 
concise notation which can be easily translated into the present bra-ket
notation. Following this procedure, we calculate in this section the BRS
transformation of our action in a systematic way.

Let us write the action (\ref{eq:action}) in the following generic
form:
\begin{eqnarray}
&&S = S^\o_\lbl2 + S^\c_\lbl2 + \sum_i S_\lbl{i} \nn
&&S^\o_\lbl2 = -{1\over2}\bra{\Psi}\tQB^\o\Pi\ket{\Psi}
= -{1\over2}\Rflo(12)\tQB^{\o\,(2)}\Pi^{(2)}\of_{21}\ ,\nn
&&S^\c_\lbl2 = -{1\over2}\bra{\Phi}\tQB^\c(b_0^-{\cal P}\Pi)\ket{\Phi}
= +{1\over2}\Rflc(12)\tQB^{\c\,(2)}(b_0^-{\cal P}\Pi)^{(2)}\cf_{2^\c1^\c}\ , \nn
&&S_\lbl{i}= {g_\lbl{i}\over c(i)!\, o(i)} 
\VN{i}(J_1,\cdots,J_{o(i)};I^\c_1,\cdots,I^\c_{c(i)})
\cf_{I_{c(i)}^\c\cdots I_1^\c} \of_{J_{o(i)}\cdots J_1} \nn
&&\quad \ \ = {g_\lbl{i}\over c(i)!\, o(i)} 
\VN{i}(\Idx<J,1,{o(i)}>;\Idxc<I,1,{c(i)}>)
\cf_{\Idxc<I,c(i),1>} \of_{\Idx<J,o(i),1>}\ ,
\end{eqnarray}
where $c(i)$ and $o(i)$ are the numbers of the closed and open string fields,
respectively, appearing in the $i$-th type vertex $\sbra{V_\lbl{i}}$, 
and the vectors like $\Idx<J,1,{o(i)}>$ and $\Idxc<I,{c(i)},1>$ are 
abbreviations for the ordered sets of indices $(J_1,\cdots,J_{o(i)})$ and 
$(I_{c(i)}^\c,\cdots,I_1^\c)$.  
According to Eq.~(\ref{eq:BRStrf}) in Appendix A, the BRS transformation
$\mib\delta_\B$ of (ket) string field is given by the differentiation of the
action $S$ with respect to the bra string field. Using the rule of 
differentiation explained also in Appendix A and, in particular, noting 
that $\delta/\delta\bra{\Psi}$ is Grassmann even and $\delta/\delta\bra{\Phi}$ is Grassmann odd,
and using $(\delta/\delta\bidx{a}\bra{\Psi})\ket{\Psi}_b=\ketRo(ab)$ and 
$(\delta/\delta\bidx{a^\c}\bra{\Phi})\ket{\Phi}_{b^\c}=\ketRc(ab)$,
we find the following BRS transformation law for open and closed string 
fields, respectively:
\begin{eqnarray}
&&\mib\delta_\B^{\rm open} \of_a 
= {\delta\over\delta\bidx{a}\bra{\Psi}} S 
= -\tQB^\o \Pi\of_a + \sum_j \mib\delta_{\B\,(j)}^{\rm open} \of_a \nn
&&\ \mib\delta_{\B\,(j)}^{\rm open} \of_a 
\equiv{\delta\over\delta\bidx{a}\bra{\Psi}} S_\lbl{j}
= {g_\lbl{j}\over c(j)!} 
\VN{j}(\Idx<L,1,{o(j){-}1}>,b\,;\Idxc<K,1,{c(j)}>)
\cf_{\!\Idxc<K,c(j),1>}\! \ketRo(ab)\of_{\Idx<L,{o(j){-}1},1>}, \nn
&&\mib\delta_\B^{\rm closed} b_0^-\cf_{a^\c} 
= {\delta\over\delta\bidx{a^\c}\bra{\Phi}} S 
= -\tQB^\c (b_0^-{\cal P}\Pi)\cf_{a^\c} 
+ \sum_j \mib\delta_{\B\,(j)}^{\rm closed} b_0^-\cf_{a^\c} \nn
&&\ \mib\delta_{\B\,(j)}^{\rm closed} b_0^-\cf_{a^\c} 
\equiv{\delta\over\delta\bidx{a^\c}\bra{\Phi}} S_\lbl{j}
=(b_0^-c_0^-)^{(a^\c)}{\delta\over\delta\bidx{a^\c}\bra{\Phi}} S_\lbl{j} \nn
&&\ = {-b_0^-}^{(a^\c)}\!\!\!\!{g_\lbl{j}\over(c(j){-}1)!o(j)} 
\bigl\langle 
V_{(j)}(\Idx<L,1,{o(j)}>;\Idxc<K,1,{c(j){-}1}>,\check{b^\c})\bigr|
\left|R^\c(a^\c\!,b^\c)\right\rangle%\ketRc(ab)
\cf_{\!\Idxc<K,{c(j){-}1},1>} \of_{\Idx<L,{o(j)},1>} .
\end{eqnarray} 
Here we have used the fact that $(\delta/\delta\bidx{a^\c}\bra{\Phi}) S_\lbl{j}$ 
always contains the anti-ghost factor ${b_0^-}^{(a^\c)}$ from the structure 
of our vertices so that the factor of $(b_0^-c_0^-)^{(a^\c)}$ multiplied 
to it equals effectively 1 since $b_0^-c_0^-b_0^-=b_0^-$. Moreover the 
${b_0^-}^{(b^\c)}$ factor contained in the vertex
\begin{equation}
\VN{j}(\Idx<L,1,{o(j)}>;\Idxc<K,1,{c(j){-}1}>,{b^\c}) \equiv 
\bigl\langle V_{(j)}(\Idx<L,1,{o(j)}>;\Idxc<K,1,{c(j){-}1}>,
\check{b^\c})\bigr| \ {b_0^-}^{(b^\c)} 
\end{equation}
has been eliminated together with ${c_0^-}^{(a^\c)}$ by using an equality
\begin{equation}
(b_0^-c_0^-)^{(a^\c)}{b_0^-}^{(b^\c)}\ketRc(ab)
=(b_0^-c_0^-)^{(a^\c)}{b_0^-}^{(a^\c)}\ketRc(ab)
={b_0^-}^{(a^\c)}\ketRc(ab).
\end{equation}
$\mib\delta_{\B\,(j)}^{\rm open}$ and $\mib\delta_{\B\,(j)}^{\rm closed}$
are the parts of the BRS transformation coming from the $S_\lbl{j}$ 
term of the action. 
The free part BRS transformation 
$\mib\delta_{\B\,(2)}^{\rm open}\of_a\equiv-\tQB^\o \Pi\of_a$ and 
$\mib\delta_{\B\,(2)}^{\rm closed}b_0^- \cf_{a^\c}
\equiv-\tQB^\c (b_0^-{\cal P}\Pi)\cf_{a^\c}$  on the action $S$ 
can be easily calculated to yield
\begin{eqnarray}
&&(\mib\delta_{\B\,(2)}^{\rm open}+\mib\delta_{\B\,(2)}^{\rm closed}) S \nn
&&\quad\  = +\Rflo(12)\tQB^{\o\lbl2}\tQB^{\o\lbl2}\Pi\ket{\Psi}_{21} +
\Rflc(12)\tQB^{\c\lbl{2^\c}}\tQB^{\c\lbl{2^\c}}
(b_0^-{\cal P}\Pi)^{\lbl{2^\c}}\ket{\Phi}_{2^\c1^\c} \nn
&&\qquad \ +\sum_i (-)^{o(i)+c(i)+1}
\!{g_\lbl{i}\over c(i)!o(i)} 
\VN{i}(\Idx<J,1,{o(i)}>;\Idxc<I,1,{c(i)}>) \!\sum_{a=I,J} \QB^{(a)}
\cf_{\Idxc<I,{c(i)},1>} \of_{\Idx<J,{o(i)},1>}\,,
\hspace{2em}
\label{eq:freeBRS}
\end{eqnarray}
where use has been made of the commutativity of $\tQB$ with the 
projection operators $\Pi$ and ${\cal P}$. 
Note that $\sum \tQB$ acting on $\sbra{V_\lbl{i}}$ has become $\sum \QB$. 
This holds since the difference between them 
$\sum \lambda\alpha^2 c_0$ vanishes generally on the vertex $\sbra{V_\lbl{i}}$:
\begin{eqnarray}
&&\VN{i}(\Idx<J,1,{o(i)}>;\Idxc<I,1,{c(i)}>)
\bigl(\sum_{k=1}^{c(i)} \lambda_\c\alpha_{I_k^\c}^2 c_0^{+\,\lbl{I_k^\c}}+
\sum_{k=1}^{o(i)} \lambda_\o\alpha_{J_k}^2 c_0^{(J_k)}\bigr) \nn
&&\quad = 
\VN{i}(\Idx<J,1,{o(i)}>;\Idxc<I,1,{c(i)}>)
\lambda_\o\oint_{C_{\rho_0}} {d\rho\over2\pi i} c(\rho)
=0.
\end{eqnarray}
Here $\lambda_\c=2\lambda_\o$ has been used and $C_{\rho_0}$ is a closed contour 
encircling all the interaction points on the $\rho$ plane. The presence of
the anti-ghost factors $b_{\rho_0}$ sitting at the interaction points 
$\rho_0$ is potentially dangerous since they yield poles 
$\VEV{b(z_0)c(z)}=1/(z_0-z)$ on the $z$ plane. But when going to the $z$ 
plane, $\oint_{C_{\rho_0}} (d\rho/2\pi i) c(\rho)$ becomes $\oint_{C_{z_0}} 
(dz/2\pi i) (d\rho/dz)^2 c(z)$ and the $(d\rho/dz)^2$ contains double zeros 
$(z_0-z)^2$ there. Therefore the integrand is regular even at 
interaction points and hence vanishes.
Note that the squares of the tilded BRS operators in Eq.~(\ref{eq:freeBRS}) 
become 
$\tQB^{\o}\tQB^{\o} = \lambda_\o\alpha^2g^2\{\QB^{\o},\,c_0\}$ and 
$\tQB^{\c}\tQB^{\c}=\lambda_\c\alpha^2g^2\{\QB^{\c},\,c_0^+\}$ for 
the open and closed string cases, respectively, by using the
nilpotency of the usual $\QB$ as well as of $c_0$.
 
Now calculate the $j$-th open BRS transformation part 
$\mib\delta_{\B\,(j)}^{\rm open}$ of the $i$-th action term $S_\lbl{i}$:
noting that $\mib\delta_{\B\,(j)}$ and $\ketRo(ab)$ are Grassmann odd 
and that the Grassmann even-oddness of $\sbra{V_{(j)}}$ is 
$(-1)^{o(j)+c(j)}$, we find
\begin{eqnarray}
\mib\delta_{\B\,(j)}^{\rm open}S_{(i)}
&=&
{g_\lbl{i}\over c(i)!} 
\VN{i}(a,\Idx<J,2,{o(i)}>;\Idxc<I,1,{c(i)}>)
\cf_{\Idxc<I,{c(i)},1>} \of_{\Idx<J,{o(i)},2>}
\left(-\mib\delta_{\B\,(j)}^{\rm open}\of_a\right) \nn
&=&
-{g_\lbl{i}\over c(i)!} 
\VN{i}(a,\Idx<J,2,{o(i)}>;\Idxc<I,1,{c(i)}>)
\cf_{\Idxc<I,{c(i)},1>} \of_{\Idx<J,{o(i)},2>} \nn
&&\qquad\qquad \quad  \times{g_\lbl{j}\over c(j)!} 
\VN{j}(\Idx<L,1,{o(j){-}1}>,b;\Idxc<K,1,{c(j)}>)
\cf_{\Idxc<K,{c(j)},1>} \ketRo(ab)\of_{\Idx<L,{o(j){-}1},1>} \nn
&=&(-)^{1+o(j)+c(j)}{g_\lbl{i}g_\lbl{j}\over c(i)!c(j)!} 
\VN{j}(\Idx<L,1,{o(j){-}1}>,b;\Idxc<K,1,{c(j)}>)
\VN{i}(a,\Idx<J,2,{o(i)}>;\Idxc<I,1,{c(i)}>) \nn
&&\qquad\qquad \quad  \times 
\cf_{\Idxc<I,{c(i)},1>} \of_{\Idx<J,{o(i)},2>} 
\cf_{\Idxc<K,{c(j)},1>} \ketRo(ab)\of_{\Idx<L,{o(j){-}1},1>} \nn
&=&(-)^{1+o(j)+c(j)}{g_\lbl{i}g_\lbl{j}\over c(i)!c(j)!} 
\VN{j}(\Idx<L,1,{o(j){-}1}>,b;\Idxc<K,1,{c(j)}>)
\VN{i}(a,\Idx<J,2,{o(i)}>;\Idxc<I,1,{c(i)}>) \nn
&&\!\!\!\times(-)^{c(i)+o(i)-1+c(j)}\ketRo(ab)
\cf_{\Idxc<I,{c(i)},1>}
(-)^{c(j)(o(i)-1)}
\cf_{\Idxc<K,{c(j)},1>} \of_{\Idx<J,{o(i)},2>} 
 \of_{\Idx<L,{o(j){-}1},1>} \nn
&=& C^{\rm open}_{ji}
\VN{j}(\Idx<L,1,{o(j){-}1}>,a;\Idxc<K,1,{c(j)}>)
\VN{i}(b,\Idx<J,2,{o(i)}>;\Idxc<I,1,{c(i)}>)\ketRo(ab)  \nn
&&\hspace{15em}\times 
\cf_{\Idxc<I,{c(i)},1>\!\!,\,\Idxc<K,{c(j)},1>}
\of_{\Idx<J,{o(i)},2>\!\!,\,\Idx<L,{o(j){-}1},1>}
\end{eqnarray}
with the final coefficient given by 
\begin{equation}
C^{\rm open}_{ji}=(-)^{(c(j)+1)(o(i)+1)+o(j)+c(i)}
{g_\lbl{i}g_\lbl{j}\over c(i)!c(j)!}\ . 
\end{equation}
Note that, in going to the last line, we have exchanged the arguments 
$a$ and $b$ of $\ketRo(ba)$ using the anti-symmetry property, 
$\ketRo(ba)=-\ketRo(ab)$. This was done for the later convenience in 
applying the GGRT. In the same way we can find 
the $j$-th closed BRS transformation part 
$\mib\delta_{\B\,(j)}^{\rm closed}$ of the $i$-th action term $S_\lbl{i}$:
\begin{eqnarray}
\mib\delta_{\B\,(j)}^{\rm closed}S_{(i)}
&=&C^{\rm closed}_{ji}
\VN{j}(\Idx<L,1,{o(j)}>;\Idxc<K,1,{c(j){-}1}>,a^\c)
\VN{i}(\Idx<J,1,{o(i)}>;b^\c,\Idxc<I,2,{c(i)}>) \ketRc(ab) \nn
&&\hspace{14em}\times\cf_{\Idxc<I,{c(i)},2>\!\!,\,\Idxc<K,{c(j){-}1},1>}
\of_{\Idx<J,{o(i)},1>\!\!,\,\Idx<L,{o(j)},1>}
\end{eqnarray}
with the coefficient
\begin{equation}
C^{\rm closed}_{ji}
=(-)^{c(j)(o(i)+1)+o(j)}
{g_\lbl{i}g_\lbl{j}\over(c(i)-1)!(c(j)-1)!o(i)o(j)} \ .
\end{equation}
The resultant coefficients $C^{\rm open}_{ji}$ and 
$C^{\rm closed}_{ji}$ for 
$\mib\delta_{\B\,(j)}^{\rm open}S_{(i)}$ and 
$\mib\delta_{\B\,(j)}^{\rm closed}S_{(i)}$
are summarized in Tables 1 and 2, respectively.
\begin{table}[tb]
%\begin{wraptable}{l}{\halftext}
\caption{Coefficients $C^{\rm open}_{ji}$ of 
$\mib\delta_{\B\,(j)}^{\rm open}S_{(i)}$.}
\label{table:1}
\begin{center}
\begin{tabular}{ccc|c|c|c|c|c} \hline\hline
&&$\lbl{i}$ & $V_3^\o(g)$ & $U(\xu g)$ & $V_4^\o(x_4g^2)$ & $V_\propto(\xa g^2)$ & $U_\Omega(\xo g^2)$\\
&&$o(i)+1$ & $4\equiv0$ & $2\equiv0$ & $5\equiv1$ & $3\equiv1$ & $3\equiv1$ \\
&&$c(i)$   & $0$    & $1$    & $0$    & $0$    & $1$    \\
\cline{1-3}
$\lbl{j}$          &$c(j)+1$ &$o(j)$  & & &  & & \\ \hline
$V_3^\o(g)$      & 1       & $3\equiv1$ &
  $-g^2$ & $+\xu g^2$ & $+x_4g^3$ & $+\xa g^3$ & $-\xo g^3$ \\  \hline
$U(\xu g)$       & $2\equiv0$  & 1      &
  $-\xu g^2$ & $+\xu^2g^2$ & $-\xu x_4g^3$ & $-\xu\xa g^3$ & $+\xu\xo g^3$ \\  \hline
$V_4^\o(x_4g^2)$ & 1       & $4\equiv0$ &
  $+x_4g^3$ & $-x_4\xu g^3$ & $-x_4^2g^4$ & $-x_4\xa g^4$ & $+x_4\xo g^4$ \\  \hline
$V_\propto(\xa g^2)$  & 1       & $2\equiv0$ &
  $+\xa g^3$ & $-\xa\xu g^3$ & $-\xa x_4g^4$ & $-\xa^2g^4$ & $+\xa\xo g^4$ \\  \hline
$U_\Omega(\xo g^2)$  & $2\equiv0$  & $2\equiv0$ &
  $+\xo g^3$ & $-\xo\xu g^3$ & $+\xo x_4g^4$ & $+\xo\xa g^4$ & $-\xo^2 g^4$ \\  \hline
\end{tabular}
\end{center}
%\end{wraptable}
\end{table} 

\begin{table}[tb]
%\begin{wraptable}{l}{\halftext}
\caption{Coefficients $C^{\rm closed}_{ji}$ of 
$\mib\delta_{\B\,(j)}^{\rm closed}S_{(i)}$.}
\label{table:2}
\begin{center}
\begin{tabular}{ccc|c|c|c|c} \hline\hline
&&$\lbl{i}$ & $U(\xu g)$ & $V_\infty(\xif g^2)$ & $V_3^\c(\sfrac1{2!}\xc g^2)$ & 
$U_\Omega(\sfrac12\xo g^2)$ \\
&&$o(i)+1$ & $2\equiv0$ & $1$    & $1$    & $3\equiv1$ \\
\cline{1-3}
$\lbl{j}$ &$c(j)$ &$o(j)$  & & &  & \\ \hline
\rule[-1.5ex]{0pt}{3.5ex}$\check U(\xu g)$      & 1       & $1$    &
  $-\xu^2g^2$ & $+\xu\xif g^2$ & $+\sfrac1{2!}\xu\xc g^3$ & $+\sfrac12\xu\xo g^3$  \\  \hline
\rule[-1.5ex]{0pt}{3.5ex}$\check V_\infty(\xif g^2)$ & $2\equiv0$  & 0      &
  $+\xif\xu g^3$ & $+\xif^2g^4$ & $+\sfrac1{2!}\xif\xc g^4$ & $+\sfrac12\xif\xo g^4$  \\  \hline
\rule[-1.5ex]{0pt}{3.5ex}$\check {V^\c_3}(\sfrac1{2!}\xc g^2)$  & $3\equiv1$ & $0$ &
  $+\sfrac1{2!}\xc\xu g^3$ & $-\sfrac1{2!}\xc\xif g^4$ & $-\sfrac1{2!2!}\xc^2 g^4$ & $-\sfrac1{2!2}\xc\xo g^4$  \\  \hline
\rule[-1.5ex]{0pt}{3.5ex}$\check U_\Omega(\sfrac12\xo g^2)$  & $1$  & $2\equiv0$ &
  $+\sfrac12\xo\xu g^3$ & $-\sfrac12\xo\xif g^4$ & $-\sfrac1{2!2}\xo\xc g^4$ & $-\sfrac14\xo^2 g^4$  \\  \hline
\end{tabular}
\end{center}
%\end{wraptable}
\end{table} 

It can be easily shown that 
$\mib\delta_{\B\,(j)}S_{(i)}=\mib\delta_{\B\,(i)}S_{(j)}$
both for $\mib\delta_{\B\,(j)}^{\rm open}$ and $\mib\delta_{\B\,(j)}^{\rm closed}$,
so that we have to retain only half number of the terms 
$\mib\delta_{\B\,(j)}S_{(i)}$ for $i\not=j$ by using the coefficients 
multiplied by 2. We thus can write the explicit form for the full BRS 
transformation $\mib \delta_\B S$ of the action by arranging the terms 
with the same number of open and closed external states and the same 
powers of $g$ as given below: there are 6 vertices containing open string
fields and 5 vertices containing closed string fields (including the
2-point kinetic terms), so that $6\times7/2 + 5\times6/2=36$ terms appear in
total. ($\bra{U}\sum\QB$ and $\bra{U_\Omega}\sum\QB$ terms below should be
counted as two terms each since they contain both $\QB^\o$ and
$\QB^\c$.) 

O($g$)
\begin{eqnarray} 
\label{eq:1-1}
   ({\rm T1})&&\qquad \sfrac{2}{3}\Vtho{1,2,3}
       \big(\QB^{(1)}+\QB^{(2)}+\QB^{(3)}\big)
                 \ket{\Psi}_{321} \\
\label{eq:1-2}
   ({\rm T2})&&\qquad -2\aa\UU{1,2^\c}\big(\QB^{(1)}+\QB^{(2^\c)}\big)
                 \ket{\Phi}_{2^\c}\ket{\Psi}_1
\end{eqnarray}

O($g^2$)
\begin{eqnarray}
\label{eq:2-2}
   ({\rm T3})&&\quad \sfrac{2}{3}\ff\V3c(123)
         \big(\QB^{(1)}+\QB^{(2)}+\QB^{(3)}\big)
            \ket{\Phi}_{3^\c2^\c1^\c} \\
\noalign{\vskip .7ex}
\label{eq:2-1}
   ({\rm T4})&&\quad \big[- \sfrac12\ee\V4(1234)
                 \big(\QB^{(1)}+\QB^{(2)}+\QB^{(3)}+\QB^{(4)}\big) \nn
      && \qquad -\V3o(12a)\V3o(b34)\ketRo(ab)
             \big] \ket{\Psi}_{4321} \\
\noalign{\vskip .7ex}
\label{eq:2-3}
   ({\rm T5})&&\quad \big[ 
            \dd\Uomg{1,2,3^\c}\big(\QB^{(1)}+\QB^{(2)}+\QB^{(3^\c)}\big) \nn
      && \qquad -2\aa\UU{a,3^\c}\V3o(b12)\ketRo(ab) \big]
            \ket{\Phi}_{3^\c}\ket{\Psi}_{21} \\
\noalign{\vskip .7ex}
\label{eq:2-4}
   ({\rm L1})&&\quad \big[-\bb\V\propto(12xx) \big(\QB^{(1)}+\QB^{(2)}\big) \nn
      && \qquad -\aa^2\UU{1,\check{a^\c}}\UU{2,{b^\c}}\ketRc(ab) \nn
      && \qquad +\lambda_{\o}\alpha_2^2\Rflo(12) \{\QB^{(2)},{c_0}^{(2)}\}
             \hspace{5em}\big] \ket{\Psi}_{21}  \\ 
\noalign{\vskip .7ex}
\label{eq:2-5}
   ({\rm T6})&&\quad \big[
            -\cc\V\infty(12xx) \big(\QB^{(1)}+\QB^{(2)}\big) \nn
      && \qquad +{\aa}^2\UU{a,1^\c}\UU{b,2^\c}\ketRo(ab) \nn
      && \qquad +\lambda_{\c}\alpha_{2^\c}^2\Rflc(12) \{\QB^{(2^\c)},{c_0}^{+(2^\c)}\}
               \bP{2} \big] \ket{\Phi}_{2^\c1^\c} 
\end{eqnarray}

O($g^3$)
\begin{eqnarray}
\label{eq:3-1}
   ({\rm T7})&&\quad +2\ee\V3o(12a)\V4(b345)\ketRo(ab) \ket{\Psi}_{54321} \\
\noalign{\vskip .7ex}
\label{eq:3-2}
   ({\rm T8})&&\quad \big[
            2\dd\Uomg{1,a,4^\c}\V3o(b23)\ketRo(ab) \nn
      && \qquad -2\aa\ee\UU{a,4^\c}\V4(b123)\ketRo(ab)
             \big] \ket{\Phi}_{4^\c}\ket{\Psi}_{321} \hspace{1em}\\
\noalign{\vskip .7ex}
\label{eq:3-3}
   ({\rm T9})&&\quad \big[
            \aa\ff\UU{1,\check{a^\c}}\V3c(b23)\ketRc(ab) \nn
      && \qquad -2\dd\aa\Uomg{1,a,2^\c}\UU{b,3^\c}\ketRo(ab)
            \big] \ket{\Phi}_{3^\c2^\c}\ket{\Psi}_1 \\
\noalign{\vskip .7ex}
\label{eq:3-4}
   ({\rm L2})&&\quad \big[ 2\bb \V\propto(1axx) \V3o(b23) \ketRo(ab) \nn
      && \qquad +\dd\aa\Uomg{1,2,\check{a^\c}}\UU{3,{b^\c}}\ketRc(ab)
            \big]\ket{\Psi}_{321} \\
\noalign{\vskip .7ex}
\label{eq:3-5}
   ({\rm T10})&&\quad \big[
            -2\aa\bb\UU{a,1^\c}\V\propto(b2xx)\ketRo(ab) \nn
      && \qquad +2\aa\cc\UU{2,\check{a^\c}}\V\infty(b1xx)\ketRc(ab)
            \big] \ket{\Phi}_{1^\c}\ket{\Psi}_2
\end{eqnarray}

O($g^4$)
\begin{eqnarray}
\label{eq:4-1}
   ({\rm T11})&&\quad -{\ee}^2\V4(123a)\V4(b456)\ketRo(ab)
            \ket{\Psi}_{654321} \\
\noalign{\vskip .7ex}
\label{eq:4-2}
   ({\rm T12})&&\quad 2\dd\ee\Uomg{1,a,5^\c}\V4(b234)\ketRo(ab)
            \ket{\Phi}_{5^\c}\ket{\Psi}_{4321} \\
\noalign{\vskip .7ex}
\label{eq:4-3}
   ({\rm T13})&&\quad -\sfrac14{\ff}^2
         \Vthc{1^\c,2^\c,\check{a^\c}}\V3c(b34)\ketRc(ab)
            \ket{\Phi}_{4^\c3^\c2^\c1^\c} \\
\noalign{\vskip .7ex}
\label{eq:4-4}
   ({\rm T14})&&\quad \big[
          -{\dd}^2\Uomg{1,a,3^\c}\Uomg{b,2,4^\c}\ketRo(ab) \nn
      && \quad -\sfrac12\dd\ff\Uomg{1,2,\check{a^\c}}%\V3c(b34)
        \bra{V_3^\c(b^\c\!,3^\c\!,4^\c)}\ket{R^\c(a^\c\!,b^\c)}
          \big] \ket{\Phi}_{4^\c3^\c}\ket{\Psi}_{21} \hspace{2em}\\
\noalign{\vskip .7ex}
\label{eq:4-5}
   ({\rm L3})&&\quad \big[
            -2\bb\ee\V\propto(1axx)\V4(b234)\ketRo(ab) \nn
      && \qquad -\sfrac14{\dd}^2\Uomg{1,2,\check{a^\c}}\Uomg{3,4,{b^\c}}
            \ketRc(ab) \big] \ket{\Psi}_{4321} \\
\noalign{\vskip .7ex}
\label{eq:4-6}
   ({\rm T15})&&\quad -\ff\cc
         \Vthc{1^\c,2^\c,\check{a^\c}}\V\infty(b34x)\ketRc(ab)
            \ket{\Phi}_{3^\c2^\c1^\c} \\
\noalign{\vskip .7ex}
\label{eq:4-7}
   ({\rm T16})&&\quad \big[
            2\dd\bb\Uomg{1,a,3^\c}\V\propto(b2xx)\ketRo(ab) \nn
      && \qquad -\dd\cc\Uomg{1,2,\check{a^\c}}\V\infty(b3xx)\ketRc(ab)
            \big] \ket{\Phi}_{3^\c}\ket{\Psi}_{21} \\
\noalign{\vskip .7ex}
\label{eq:4-8}
   ({\rm L4})&&\quad  -{\bb}^2\Valp{1,a}\Valp{b,2}\ketRo(ab)
            \ket{\Psi}_{21} \\
\noalign{\vskip .7ex}
\label{eq:4-9}
   ({\rm L5})&&\quad {\cc}^2\Vinf{1^\c,\check{a^\c}}\V\infty(b2xx)\ketRc(ab)
            \ket{\Phi}_{2^\c1^\c}
\end{eqnarray}

\section{BRS invariance}

The light-cone gauge string field theory for open-closed mixed system 
has long been known to have an anomaly which breaks the Lorentz invariance 
at the one-loop level.\cite{rf:ST1,rf:ST2,rf:KikkawaSawada}
This anomaly was present even in the oriented string system and required 
the existence of the open-closed transition interaction $U$ to cancel it.
In our framework of $\alpha=p^+$ HIKKO\cite{rf:kugozwie}
type unoriented open-closed string 
theory, this is reflected in the fact that the BRS (and gauge) 
invariance suffers from the anomaly. 

\begin{figure}[tb]
   \epsfxsize= 13.5cm   %or \epsfysize= HEIGHT cm
   \centerline{\epsfbox{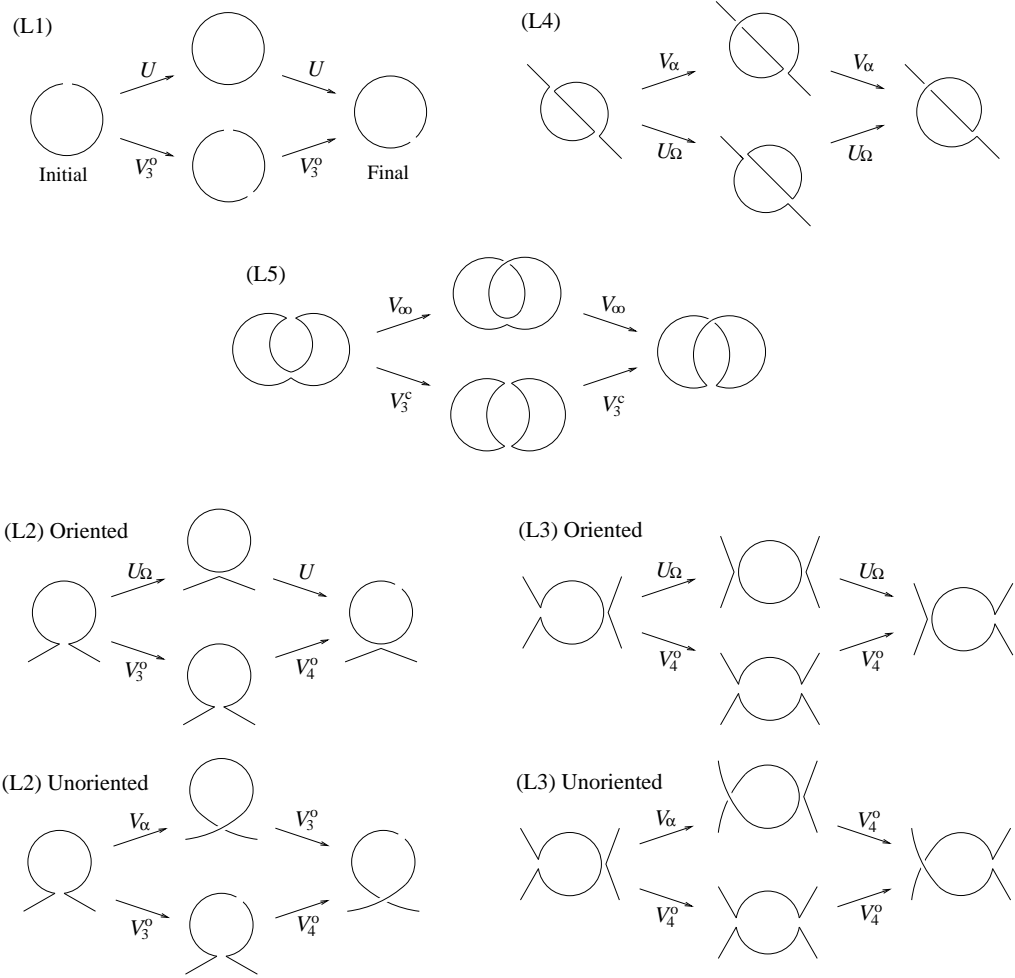}}
\vspace{1mm}
 \caption{Configurations requiring the loop counterterms. The
upper paths correspond to the `tree' terms appearing in (L1) -- (L5), 
respectively, and the lower paths to the loop diagrams.}
\vspace{-2mm}
\label{fig:loop}
\end{figure}

The five terms labeled as (L1) -- (L5) in Eqs.~(\ref{eq:2-4}), 
(\ref{eq:3-4}), (\ref{eq:4-5}), (\ref{eq:4-8}) and (\ref{eq:4-9}), 
do not vanish by 
themselves and will be cancelled by the anomalous contributions of 
one-loop diagrams.\footnote{In the previous paper I, we have 
erroneously claimed that the (L5) term, 
$\bra{V_\infty}\bra{V_\infty}\ket{R^\c}$, cancels out totally by itself. 
But actually the configuration shown in (L5) in Fig.~\ref{fig:loop}, which 
corresponds to the case where the two crosscap cuts overlap, does not 
cancel and needs the loop counterterm.}
More naturally, we should say this oppositely; if we
started with the theory possessing only the $V_3^\o, V_4^\o$ and 
$V_3^\c$ interactions, then the theory is safely BRS invariant at tree 
level. At quantum level, however, the BRS invariance is violated by some
anomalous one-loop diagrams, and the other interaction vertices $U, U_\Omega 
$ and $V_\propto, V_\infty$ are required to be introduced to cancel those 
anomalies. Their coupling strengths are found to be of the order 
\begin{equation}
U,\ U_\Omega\sim O(\hbar^{1/2}),\qquad 
V_\propto,\ V_\infty\sim O(\hbar^{1}),
%x_3^\c \sim O(\hbar^{1/2}),
\end{equation}
in $\hbar$ as a loop expansion parameter, as already shown in the action 
(\ref{eq:action}).\cite{rf:Zwie2}
Therefore, the non-vanishingness of the (L1) -- (L5) 
terms is by no means unwelcome, but rather gives the very raison 
d'\'etre of the interaction vertices $U, U_\Omega$ and $V_\propto, V_\infty$.

In this paper we confine ourselves to proving the BRS invariance only at
the tree level and defer the proof of cancellations between the anomalous 
one-loop contributions and the (L1) -- (L5) terms to the forthcoming paper. 
We, therefore, do not discuss the five terms (L1) -- (L5) in any detail. 
But, here, let us just see what types of one-loop diagrams the five terms 
(L1) -- (L5) will cancel. This is shown in Fig.~\ref{fig:loop}. 
%\begin{wrapfigure}[n]{r}{6.6cm}
%\end{wrapfigure}
There the examples of string configurations are given which need loop 
counterterms. In each diagram, there are two interaction points and, 
depending on which interaction of the two takes place earlier than the
other, two different intermediate states appear for a given set of
initial and final states; the upper paths correspond to the `tree'
terms appearing in (L1) -- (L5), respectively and the lower paths to
the loop diagrams. (Whether the path corresponds to loop or
tree diagram can be judged by considering whether the momenta of the
intermediate states are uniquely determined or not when those of the
initial and final states are given.) Look at the first configuration
(L1), for example. The upper path represents a possible configuration
for the term $\UU{1,\check{a^\c}}\UU{2,{b^\c}}\ketRc(ab)$ in
Eq.~(\ref{eq:2-4}), and the same configuration of the initial and
final strings can be realized by choosing the other intermediate state
shown in the lower path of the figure, which corresponds to the
one-loop (non-planar but orientable) diagram constructed by using open
3-string vertices $\bra{V_3^\o}$ twice. This (L1) example is just the
same one as the Lorentz anomaly in the case of light-cone gauge string
field theory mentioned above.

In this section, we prove that the theory has the BRS symmetry at the tree
level if the parameters $\lambda_\c$, $\lambda_\o$, $\aa$, $\bb$, $\cc$, $\dd$,
$\ee$ and $\ff$ in the action satisfy
\begin{eqnarray}
 && \lambda_\c=2\lambda_\o= \lim_{\epsilon\rightarrow0}
      \frac{32n\aa^2}{\epsilon^2} \label{eq:rel-1}\\
 && \cc=-n\aa^2=4\pi i\bb \label{eq:rel-2}\\
 && \ee=1 \label{eq:rel-3},\\
 && \aa=\dd,\label{eq:rel-4}\\
 && \ff=8\pi i\dd\,,\label{eq:rel-5}
\end{eqnarray}
where Eqs.~(\ref{eq:rel-1}) and (\ref{eq:rel-2}) are the relations 
derived already in the previous paper I.

The order $g$ terms (T1) and (T2) in Eqs.~(\ref{eq:1-1}), (\ref{eq:1-2})
and the order $g^2$ term (T3) in Eq.~(\ref{eq:2-2})
vanish by the BRS invariance of the vertices $V_3^\o, U$ 
and $V_3^\c$, respectively; each of these has no moduli 
parameters and hence is essentially identical with the corresponding 
LPP vertex which is manifestly BRS invariant by construction. 
The terms (T6) and (T10) in Eqs.~(\ref{eq:2-5}) and (\ref{eq:3-5}), 
containing only the quadratic interaction vertices 
$U$, $V_\propto$ and $V_\infty$, were already proved to vanish in our previous 
paper I if the coupling relations (\ref{eq:rel-1})  
and (\ref{eq:rel-2}) are satisfied. 
So, we now discuss the remaining eleven terms (T4), (T5), (T7--9) and 
(T11--16) successively and show that they indeed vanish in the following.

In this section we will often use the GGRT, 
which was originally proved in Refs.~\citen{rf:LPP}and 
\citen{rf:AKT} for the simplest cases of purely open string system. 
But here we need more general formulas. Indeed we have various types of
vertices containing closed strings also and must treat the contractions 
of such vertices by closed string reflector $\ketRc(ab)$ as well as by 
open one $\ketRo(ab)$. We, however, show in the Appendix B that almost 
the same form of GGRT formulas actually hold for all the cases we need.
So we shall use such formulas freely in the following.

\subsection{O($g^2$) invariance}
\noindent
\subsubsection{T4 terms}%Eq.~(\ref{eq:2-1})}

The cancellation of the two terms in (T4) in Eq.~(\ref{eq:2-1}) have long 
been known since the first proof by HIKKO in Ref.\citen{rf:HIKKO1}. 
However, we here prove it again to demonstrate how much the proof is 
simplified by the use of our present machinery of LPP vertex. This 
will determine $\ee$ to be $1$ in the present definition of the vertex.

\begin{figure}[tb]
   \epsfxsize= 10 cm   %or \epsfysize= HEIGHT cm
   \centerline{\epsfbox{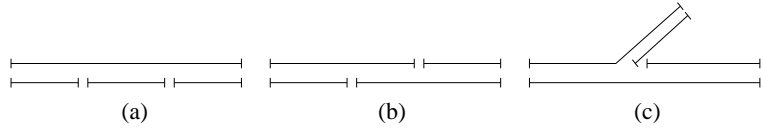}}
 \caption{The possible three types of configurations obtained by gluing two 
open 3-string vertices.}
 \label{fig:v3v3}
\end{figure}

The second term of (T4) corresponds to the gluing of two 3-string 
LPP vertices $\bra{v_3^\o}$. For this simplest gluing, we have the GGRT 
formula
\begin{equation}
\v3o(12ax)\v3o(b34x)\ketRo(ab) 
= \bra{\tilde v^{\o}_4(1,2,3,4)}.
\label{eq:GGRTv3}
\end{equation}
The $\sbra{\tilde v^{\o}_4(1,2,3,4)}$ is a generic LPP vertex for four 
open-strings, given by an integration of specific 
LPP vertex $\sbra{\tilde v^{\o\,(\alpha_1,\alpha_2,\alpha_3,\alpha_4)}_4(1,2,3,4)}$ 
over the string length parameters $\alpha_1,\alpha_2,\alpha_3$ and $\alpha_4$. 
The specific vertex $\sbra{\tilde v^{\o\,(\alpha_1,\alpha_2,\alpha_3,\alpha_4)}_4}$ 
represents the LPP vertices which correspond to various ways of gluing 
of the four open strings depending on the set of the values of 
the parameters $\alpha_1, \alpha_2,\alpha_3$ and $\alpha_4$. 
The possible 4-string configurations which can be realized by gluing
two 3-string vertices for all possible choices of string length parameters, 
fall into three types (a), (b) and (c) drawn in
Fig.~\ref{fig:v3v3}.

\begin{figure}[tb]
   \epsfxsize= 9 cm   %or \epsfysize= HEIGHT cm
   \centerline{\epsfbox{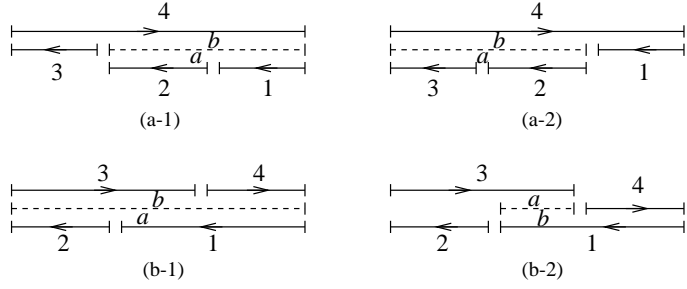}}
 \caption{Two ways of gluing realizing the type (a) and (b) 
configurations, respectively. The dashed lines denote the intermediate 
strings.}
 \label{fig:v3v3-2}
\end{figure}

Consider the type (a) configuration first, and  
name the four strings in the (a) configuration 1, 2, 3, and 4 as drawn
in (a-1) in Fig.~\ref{fig:v3v3-2}. 
But this configuration can be
realized in two ways by using two 3-string vertices as drawn in (a-1)
and (a-2) in Fig.~\ref{fig:v3v3-2}, where the dashed lines denote the
intermediate strings $a$ and $b$ which are glued together by
$\ketRo(ab)$. So this vertex $\sbra{\tilde v^{\o(\alpha_1,\alpha_2,\alpha_3,\alpha_
4)}_4(1,2,3,4)}$ appears twice in the second term in (T4); 
%
%\begin{wrapfigure}[n]{r}{6.6cm}
%\end{wrapfigure}
%\begin{wrapfigure}[n]{r}{6.6cm}
%\end{wrapfigure}
%
Namely, one term corresponding to the configuration (a-1) is contained 
in the second term of (T4) in the form
\begin{eqnarray}
&& -\langle{v_3^\o}^{(\alpha_1,\alpha_2,\alpha_a)}(1,2,a)|\,
 \langle{v_3^\o}^{(\alpha_b,\alpha_3,\alpha_4)}(b,3,4)|\,
 \ketRo(ab)\ket{\Psi}_{4321} \nn
&&\qquad \quad = -\langle{\tilde v_4}^{\o\,(\alpha_1,\alpha_2,\alpha_3,\alpha_4)}
 (1,2,3,4)|\,\ket{\Psi}_{4321} 
\label{eq:4.8}
\end{eqnarray}
and the other term corresponding to (a-2) in the form
\begin{eqnarray}
&& -\langle{v_3^\o}^{(\alpha_2,\alpha_3,\alpha_a)}(2,3,a)|\,
 \langle{v_3^\o}^{(\alpha_b,\alpha_4,\alpha_1)}(b,4,1)|\,
 \ketRo(ab)\ket{\Psi}_{1432} \nn
&&\ \ = -\langle{\tilde v_4}^{\o\,(\alpha_2,\alpha_3,\alpha_4,\alpha_1)}
 (2,3,4,1)|\,\ket{\Psi}_{1432} %\nn &&\qquad \quad 
= +\langle{\tilde v_4}^{\o(\alpha_1,\alpha_2,\alpha_3,\alpha_4)}
 (1,2,3,4)|\ket{\Psi}_{4321},
\hspace{3em}
\label{eq:4.9}
\end{eqnarray}
where we have used the GGRT (\ref{eq:GGRTv3}), 
the cyclic symmetry of the LPP 4-string 
vertex $\bra{\tilde v_4^\o}$ similar to Eq.~(\ref{eq:LPPprop}), 
and $\ket{\Psi}_{1432}=-\ket{\Psi}_{4321}$ because of the Grassmann odd 
property of the open string fields $\ket{\Psi}$. We see that these two 
terms have {\it opposite signs} and cancel each other. Consequently, we 
have proven that the second term in (T4) actually contains no terms $\propto 
\langle{\tilde v_4}^{\o(\alpha_1,\alpha_2,\alpha_3,\alpha_4)}(1,2,3,4)|$ corresponding to 
the type (a) configuration. Similarly, the terms corresponding to the 
type (b) configuration, realized in two ways, (b-1) and (b-2) in 
Fig.~\ref{fig:v3v3-2}, can be seen to cancel out in the second term in 
(T4). (Actually the same Eqs.~(\ref{eq:4.8}) and (\ref{eq:4.9}) apply to
the (b-1) and (b-2) terms, respectively, if we name the strings as 
drawn in Fig.~\ref{fig:v3v3-2}.)

Thus, now only remaining are the terms corresponding to the type (c) 
configuration in Fig.~\ref{fig:v3v3}, called 
`horn diagram' by HIKKO.\cite{rf:HIKKO1}
Contrary to the previous types (a) and (b),
this configuration is realized by using two 3-string vertices 
in a {\it unique} 
%\begin{wrapfigure}[6]{r}{6.6cm}
\begin{figure}[tb]
   \epsfxsize= 3.5cm   %or \epsfysize= HEIGHT cm
   \centerline{\epsfbox{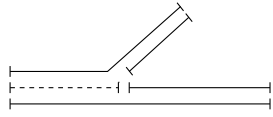}}
 \caption{The unique way for drawing an intermediate string in (c).}
 \label{fig:horn-2}
\end{figure}
%\end{wrapfigure}
way as drawn in Fig.~\ref{fig:horn-2}. 
Therefore the terms of this configuration must be cancelled by 
other new contribution than the second term in (T4). This is just 
given by the first term in (T4), as we now see. 

\begin{figure}[tb]
   \epsfxsize= 10 cm   %or \epsfysize= HEIGHT cm
   \centerline{\epsfbox{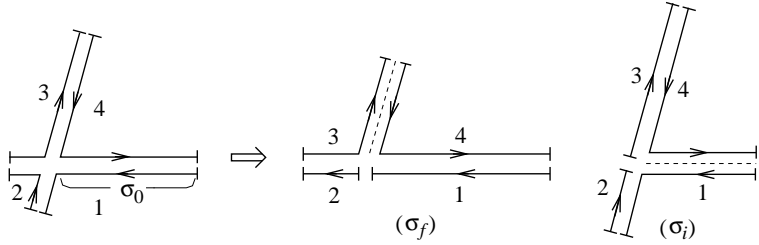}}
 \caption{Two configurations of the 4-string vertex $\vfo{\sigma_0}$
realized at the end-points $\sigma_0=\sigma_f$ and $\sigma_i$. The dotted lines
indicate the intermediate string in the case they are realized by
gluing two 3-string vertices. }
 \label{fig:v4-if}
\end{figure}

The first term in (T4) can be rewritten as
\begin{eqnarray}
&& -\frac{\ee}{2}
\Vfo{1,2,3,4}\sum_{r}\QB^{(r)}\ket{\Psi}_{4321}
=-\frac{\ee}{2}
\int_{\sigma_i}^{\sigma_f}\!\!d\sigma_0\, {d\over d\sigma_0}\bigl(\v4(1234\sigma_0)\bigr)
\ket{\Psi}_{4321}
\nn
  &&\hspace{2em}
  =-\frac{\ee}{2}\bigl(\vfo{1,2,3,4;\sigma_f}
  -\vfo{1,2,3,4;\sigma_i}\bigr)\prod_r\Pi^{(r)}\ket{\Psi}_{4321},
\label{eq:surface}
\end{eqnarray}
using the property (\ref{eq:QBdsigma}) that the BRS charge $\QB$ acts on 
the SFT vertex as a differential operator with respect to the moduli 
parameter. Here we have omitted the unoriented 
projection operators $\prod_{r=1}^4\!\Pi^{(r)}$ for brevity, which we shall 
do also henceforth without notice unless they become important. We 
immediately recognize that the appearing surface terms, 
%$\vfo{1,\kern -1pt 2,\kern -1pt 3,\kern -1pt 4;\sigma_0}$ at the end-points 
$\vfo{\sigma_0}$ at the end-points 
$\sigma_0=\sigma_f$ and $\sigma_i$, just realize the same 
string-configurations as the horn diagram, as depicted in 
Fig.~\ref{fig:v4-if}. The Fig.~\ref{fig:v4-if} is drawn assuming that 
the string 4 carries the maximum string length $\abs{\alpha}$ among the 
four. 
%, without loss of generality because of the cyclic
%property of $\bra{v_4^\o}$.
Note that these specific configurations with $\abs{\alpha_4}$ being maximum are 
contained four times for each with $\sigma_0=\sigma_f$ and $\sigma_i$ 
in Eq.~(\ref{eq:surface}), since the labels 1 --- 4 are dummy there and 
the vertex $\bra{v_4^\o}$ is cyclic symmetric. 
%\begin{wrapfigure}[n]{r}{6.6cm}
%\end{wrapfigure}
On the other hand, in the second term of (T4) using two 3-string 
vertices, the (c) terms corresponding to these horn diagram configurations 
appear twice for each of $\sigma_f$ and $\sigma_i$; indeed, in view of 
Fig.~\ref{fig:v4-if}, we have the following two terms for the 
configuration $(\sigma_f)$, (omitting the superfices like $(\alpha_2,\alpha_3,\alpha_a)$ 
of the vertices for brevity, here and henceforth),
\begin{eqnarray}
-\langle{v_3^\o}(2,3,a)|\, \langle{v_3^\o}(b,4,1)|\,
 \ketRo(ab)\ket{\Psi}_{1432}
&=& +\langle{v_4^\o} (1,2,3,4;\sigma_f)|\ket{\Psi}_{4321} \nn
-\langle{v_3^\o}(4,1,a)|\,\langle{v_3^\o}(b,2,3)|\,
 \ketRo(ab)\ket{\Psi}_{3214}
&=& +\langle{v_4^\o} (1,2,3,4;\sigma_f)|\ket{\Psi}_{4321}
\hspace{2em}
\label{eq:4.12}
\end{eqnarray}
by using the GGRT (\ref{eq:GGRTv3}) and 
$\ket{\Psi}_{1432}=-\ket{\Psi}_{4321}$ etc, 
and, similarly, two terms of 
$-\langle{v_4^\o} (1,2,3,4;\sigma_i)|\ket{\Psi}_{4321}$ for 
the configuration $(\sigma_i)$.\footnote{Note that, although the first equation 
here in (\ref{eq:4.12}) and the previous Eq.~(\ref{eq:4.9}) look 
the same, they actually represent different quantities corresponding
to the different regions of string length parameters; here the
string fields $\ket{\Psi}_i$ ($i=1,2,3,4$) carry an alternating signs of 
string length parameters, i.e., $\{\alpha_1, \alpha_3{>}0,\ \alpha_2,\alpha_4{<}0\}$ 
or $\{\alpha_1, \alpha_3{<}0,\ \alpha_2,\alpha_4{>}0\}$, while, in Eq.~(\ref{eq:4.9}), 
they carry those in the region $\{\alpha_1, \alpha_2,\alpha_3{>}0, \ \alpha_4{<}0\}$ or 
$\{\alpha_1, \alpha_2,\alpha_3{<}0, \ \alpha_4{>}0\}$.}
Therefore, the terms corresponding to 
these horn diagram configurations cancels between the first and second 
terms in (T4), Eq.~(\ref{eq:2-1}), if the 4-string coupling constant 
$\ee$ satisfies
\begin{eqnarray}
 \bigl(-{\ee\over2}\bigr)\times4 + (+1)\times2=0\quad  \Rightarrow\quad \ee=1.
\label{eq:x4}
\end{eqnarray}

\subsubsection{T5 terms}%\underline{Eq.~(\ref{eq:2-3})}

The vanishingness of the (T5) terms in Eq.~(\ref{eq:2-3}) can be
proved in a very similar manner as in the previous case.

\begin{figure}[tb]
   \epsfxsize= 7.5 cm   %or \epsfysize= HEIGHT cm
   \centerline{\epsfbox{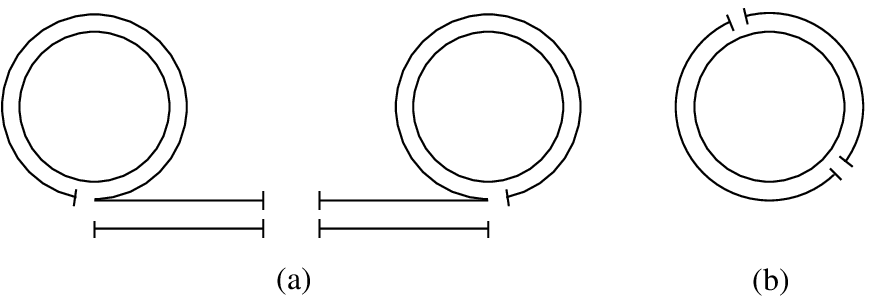}}
 \caption{Two types of configurations for $\bra{U}\bra{V_3^\o}\ket{R^\o}$.}
 \label{fig:uv3-1}
\end{figure}

The second term of (T5) has three possible configurations
as depicted in Fig.~\ref{fig:uv3-1}.
%\begin{wrapfigure}[n]{r}{6.6cm}
%\end{wrapfigure}
%\begin{wrapfigure}[n]{r}{6.6cm}
\begin{figure}[tb]
   \epsfxsize= 5 cm   %or \epsfysize= HEIGHT cm
   \centerline{\epsfbox{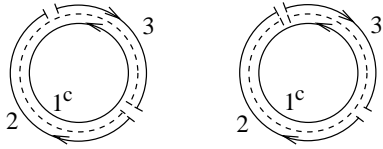}}
 \caption{Two ways of gluing yielding the configuration (b) in 
Fig.~\protect\ref{fig:uv3-1}.}
 \label{fig:uv3-2}
\end{figure}
%\end{wrapfigure}
The type (b) configuration can be realized by gluing the two
vertices $\bra{U}$ and $\bra{V_3^\o}$ again in two ways as drawn in 
Fig.~\ref{fig:uv3-2}, and appear in the second term in (T5) in the forms
\begin{eqnarray}
&& \bra{U(a,1^c)}\bra{V_3^\o(b,2,3)}\ketRo(ab)
 \ket{\Phi}_{1^\c}\ket{\Psi}_{32} =
\bra{\tilde v(2,3,1^c)} \ket{\Phi}_{1^\c}\ket{\Psi}_{32}, \nn
&& \bra{U(a,1^c)}\bra{V_3^\o(b,3,2)}\ketRo(ab)
 \ket{\Phi}_{1^\c}\ket{\Psi}_{23} =
\bra{\tilde v(3,2,1^c)} \ket{\Phi}_{1^\c}\ket{\Psi}_{23},
\hspace{2em}
\label{eq:4.13}
\end{eqnarray}
respectively. Here $\bra{\tilde v(2,3,1^c)}$ denotes the LPP
vertex for one closed and two open strings resultant from this
gluing. This vertex is cyclic symmetric with respect to the two 
open string arguments, 
$\bra{\tilde v(2,3,1^c)}=\bra{\tilde v(3,2,1^c)}$, since the matrix 
indices of the two open strings are contracted between the two.  
Since $\ket{\Psi}_{23}=-\ket{\Psi}_{32}$, the two terms in (\ref{eq:4.13}) 
clearly cancel each other.

\begin{figure}[tb]
   \epsfxsize= 10cm   %or \epsfysize= HEIGHT cm
   \centerline{\epsfbox{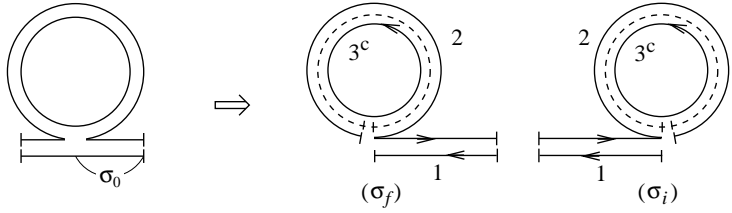}}
 \caption{Two configurations of $\uomg{\sigma_0}$ realized at 
the end-points $\sigma_0=\sigma_f$ and $\sigma_i$. 
%The dotted lines indicate the intermediate string in the case they are 
%realized by gluing two 3-string vertices. 
} 
\label{fig:uomg-if}
\end{figure}

Remaining are the terms of type (a) configurations, which are again 
to be cancelled by the first term in (T5). The first term of (T5)
is rewritten as follows in the same way as in Eq.~(\ref{eq:surface}):
\begin{eqnarray}
&& \dd \Uomg{1,2,3^\c}\QB
 \ket{\Phi}_{3^\c}\ket{\Psi}_{21}\nn
&& \quad =  -\dd \int_{\sigma_i}^{\sigma_f}  \uomg{1,2,3^\c;\sigma_0} \{\,b_{\sigma_0},\ \QB\,\}
  (b_0^-{\cal P}\Pi)^{(3^\c)}\ket{\Phi}_{3^\c}\ket{\Psi}_{21}\nn
&& \quad =  -\dd \bigl\{\uomg{1,2,3^\c;\sigma_f}
      -\uomg{1,2,3^\c;\sigma_i} \bigr\}(b_0^-{\cal P}\Pi)^{(3^\c)}
      \ket{\Phi}_{3^\c}\ket{\Psi}_{21}.
\hspace{2em}
\label{eq:4.14}
\end{eqnarray}
These surface terms $\uomg{1,2,3^\c;\sigma_0}$ with $\sigma_0=\sigma_f$ and $\sigma_i$ 
have the same string-configurations as the type (a) as 
drawn in Fig.~\ref{fig:uomg-if}. 
%\begin{wrapfigure}[n]{r}{6.6cm}
%\end{wrapfigure}
Each of these terms with the string
lengths specified and satisfying $\abs{\alpha_2}>\abs{\alpha_1}$ appears 
twice in this Eq.~(\ref{eq:4.14}). And the corresponding (a) terms in the 
second term in (T5) are given, by the help of GGRT, as
\begin{eqnarray}
&& -2x_u\bra{U(a,3^c)}\bra{V_3^\o(b,2,1)}\ketRo(ab)
 \ket{\Phi}_{3^\c}\ket{\Psi}_{12}  
\nn &&\qquad 
=
-2x_u\uomg{2,1,3^\c;\sigma_f}(b_0^-{\cal P}\Pi)^{(3^\c)} 
\ket{\Phi}_{3^\c}\ket{\Psi}_{12}, \nn
&& -2x_u\bra{U(a,3^c)}\bra{V_3^\o(b,1,2)}\ketRo(ab)
 \ket{\Phi}_{3^\c}\ket{\Psi}_{21} 
\nn &&\qquad 
=
-2x_u\uomg{1,2,3^\c;\sigma_i} (b_0^-{\cal P}\Pi)^{(3^\c)}
  \ket{\Phi}_{3^\c}\ket{\Psi}_{21} \,,
\label{eq:4.15}
\end{eqnarray}
where we have used the fact that both $\bra{V_3^\o}$ and $\ket{R^\o}$
are Grassmann odd. 
Noting that $\uomg{2,1,3^\c;\sigma_0}=\uomg{1,2,3^\c;\sigma_0}$ 
and $\ket{\Psi}_{21}=-\ket{\Psi}_{12}$, we see that these terms in 
(\ref{eq:4.14}) and (\ref{eq:4.15}) exactly cancel each other if
\begin{eqnarray}
\label{eq:xu-xomg}
 -\dd\times2 + 2\aa=0 \quad \Rightarrow\quad \aa=\dd.
\end{eqnarray}

\subsection{$O(g^3)$ invariance}

\subsubsection{T7 term}%{Eq.~(\ref{eq:3-1})}

\begin{figure}[tb]
   \epsfxsize= 12cm   %or \epsfysize= HEIGHT cm
   \centerline{\epsfbox{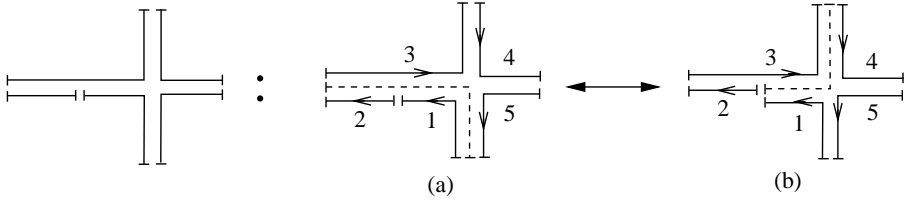}}
 \caption{The unique configuration for $\bra{V_3^\o}\bra{V_4^\o}\ket{R^\o}$ 
and two ways of gluing realizing it.}
 \label{fig:v4v3-2}
\end{figure}

In this case the generic configuration is unique, of the type drawn in
Fig.~\ref{fig:v4v3-2}. 
Name the five strings 1, 2, ---, 5 in a cyclic order as shown there. 
Then, for any single configuration with 
a definite set of string length parameters $\alpha_1$ --- $\alpha_5$,  
there are always two ways to realize it by gluing the vertices 
$\bra{V_4^\o}$ and $\bra{V_3^\o}$, as shown in Fig.~\ref{fig:v4v3-2}. 
The term corresponding to the (a) diagram is contained in (T7), 
Eq.~(\ref{eq:3-1}), in the form
\begin{eqnarray}
&&\V3o(12a)\V4(b345)\ketRo(ab) \ket{\Psi}_{54321} \nn
&&\qquad  =+\V3o(12a)\V4(b3{\stackrel{\downarrow}4}5)\ketRo(ab) \ket{\Psi}_{54321}\nn
&&\qquad  =+\int_{\sigma_i^{(4)}}^{\sigma_f^{(4)}}d\sigma_0^{(4)}
\v3o(12ax)\v4(b345{\sigma_0^{(4)}})
b_{\sigma_0^{(4)}}\ketRo(ab) \ket{\Psi}_{54321}\nn
&&\qquad =-\int_{\sigma_i^{(4)}}^{\sigma_f^{(4)}}d\sigma_0^{(4)}
\sbra{\tilde v(12345;\sigma_0^{(4)})}b_{\sigma_0^{(4)}} \ket{\Psi}_{54321},
\label{eq:4.17}
\end{eqnarray}
where ${\stackrel{\zht{\downarrow}}4}$ means that the anti-ghost factor 
$b_{\sht{\sigma_0^{(4)}}}$ with string-4 moduli $\sigma_0^{(4)}$ is used as 
explained in 
Eq.~(\ref{eq:2.13}), and the identity (\ref{eq:2.12}) has been used. 
$\sbra{\tilde v(12345;\sigma_0^{(4)})}$ is the LPP vertex for the five 
strings resultant from this gluing. Note that the sign change has 
occured to the last expression since we have changed the order of 
$b_{\sht{\sigma_0^{(4)}}}$ and $\ket{R^\o}$ before applying the GGRT. The 
term corresponding to the (b) diagram is, on the other hand, contained 
in (T7) in the form:
\begin{eqnarray}
&&\V3o(23a)\V4(b451)\ketRo(ab) \ket{\Psi}_{15432} \nn
&&\qquad  =-\V3o(23a)\V4(b{\stackrel{\downarrow}4}51)\ketRo(ab) \ket{\Psi}_{15432}\nn
&&\qquad =-\int_{\sigma_i^{(4)}}^{\sigma_f^{(4)}}d\sigma_0^{(4)}
\v3o(23ax)\v4(b451{\sigma_0^{(4)}})b_{\sigma_0^{(4)}}\ketRo(ab) \ket{\Psi}_{54321} \nn
&&\qquad =+\int_{\sigma_i^{(4)}}^{\sigma_f^{(4)}}d\sigma_0^{(4)}
\sbra{\tilde v(12345;\sigma_0^{(4)})}b_{\sigma_0^{(4)}} \ket{\Psi}_{54321},
\hspace{3em}
\label{eq:4.18}
\end{eqnarray}
where the cyclic symmetry property of the LPP vertex 
$\sbra{\tilde v(12345;\sigma_0^{(4)})}$ and 
$\ket{\Psi}_{15432}\linebreak =+\ket{\Psi}_{54321}$ have been used. 
The negative sign 
here has appeared since the identity (\ref{eq:2.12}) says
\begin{equation}
\V4({\stackrel{\zht{\downarrow}}b}451)=-\V4(b{\stackrel{\zht{\downarrow}}4}51).
\end{equation}
Thus the two contributions (\ref{eq:4.17}) and (\ref{eq:4.18}) 
cancel each other, proving that (T7) vanishes.

\subsubsection{T8 terms}%{Eq.~(\ref{eq:3-2})}

\begin{figure}[tb]
   \epsfxsize= 11cm   %or \epsfysize= HEIGHT cm
   \centerline{\epsfbox{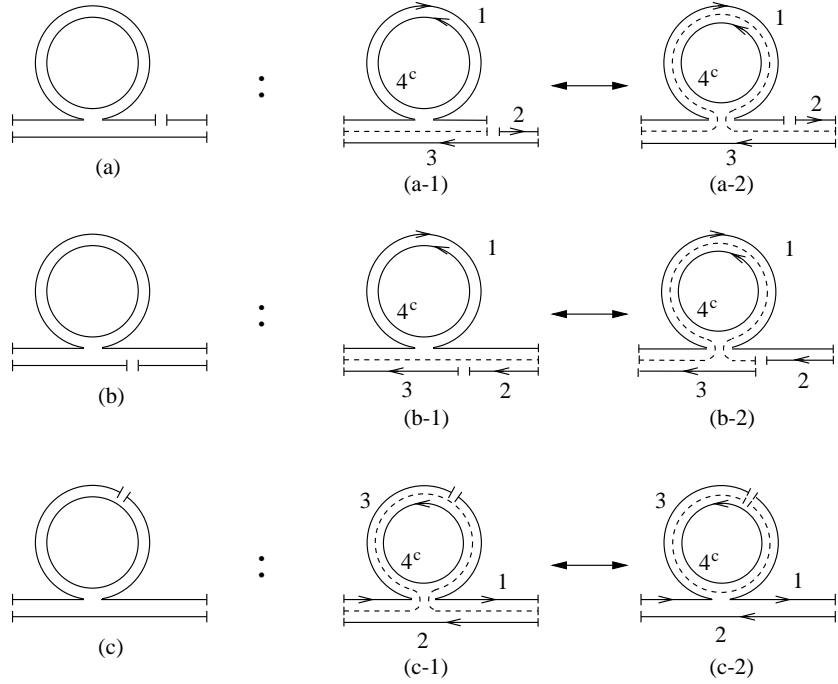}}
 \caption{Three configurations for (T8) terms.}
 \label{fig:v3uomg-2}
\end{figure}

Generic configurations resultant from the contraction of 
two vertices $\bra{U_\Omega}$ and $\bra{V_3^\o}$, or 
$\bra{U}$ and $\bra{V_4^\o}$, fall into 
three types, (a), (b) and (c), depicted in Fig.~\ref{fig:v3uomg-2},
%\begin{wrapfigure}[n]{r}{6.6cm}
%\end{wrapfigure}
each of which is realized in two ways 
as also shown in Fig.~\ref{fig:v3uomg-2};
only (c-2) diagram is given by gluing $\bra{U}$ and $\bra{V_4^\o}$ and 
all the others are by gluing $\bra{U_\Omega}$ and $\bra{V_3^\o}$.
As in the previous cases, cancellations occur between the two 
ways of gluing in each pair. Denoting the LPP vertex resultant from this
gluing by $\bra{\tilde v(123,4^\c)}$ generically, the pair of 
(a-1) and (a-2) is contained in (T8) in the following form:
\begin{eqnarray}
&&\hbox{(a-1)}:\nn
&&\ 
  2\dd\Uomg{1,a,4^\c}\V3o(b23)\ketRo(ab)\ket{\Phi}_{4^\c}\ket{\Psi}_{321}\nn
  &&\ \ 
  = 2\dd\int d\sigma_0^{(1)}\uomg{1,a,4^\c;\sigma_0^{(1)}}b_{\sigma_0^{(1)}}
      \bP{4}\!\v3o(b23x)\!\ketRo(ab)\ket{\Phi}_{4^\c}\ket{\Psi}_{321}\nn
  &&\ \ 
  = 2\dd \int d\sigma_0^{(1)}\bra{\tilde v(123,4^\c)}b_{\sigma_0^{(1)}}
       \bP{4}\ket{\Phi}_{4^\c}\ket{\Psi}_{321}\nn
&&\hbox{(a-2)}:\nn
&&\ 
  2\dd\Uomg{3,a,4^\c}\V3o(b12)\ketRo(ab)\ket{\Phi}_{4^\c}\ket{\Psi}_{213}\nn
  &&\ \ 
  = 2\dd\int d\sigma_0^{(3)}\uomg{3,a,4^\c;\sigma_0^{(3)}}b_{\sigma_0^{(3)}}
     \bP{4}\!\v3o(b12x)\!\ketRo(ab)\ket{\Phi}_{4^\c}\ket{\Psi}_{213}\nn
  &&\ \ 
  = 2\dd \int d\sigma_0^{(3)}\bra{\tilde v(123,4^\c)}b_{\sigma_0^{(3)}}
         \bP{4}\ket{\Phi}_{4^\c}\ket{\Psi}_{213}.
\label{eq:4.20}
\end{eqnarray}
Although the states have the same sign $\ket{\Psi}_{213}=+\ket{\Psi}_{321}$, 
the anti-ghost factors have opposite signs, 
$\int d\sigma_0^{(3)}b_{\sht{\sigma_0^{(3)}}}=-\int d\sigma_0^{(1)}b_{\sht{\sigma_0^{(1)}}}$ 
since the increasing directions of $\sigma_0^{(3)}$ and $\sigma_0^{(1)}$ are 
opposite in order to keep the common configuration, similarly to 
Eq.~(\ref{eq:2.12}) in the open 4-string vertex case. Thus (a-1) and 
(a-2) cancel each other. The same equations (\ref{eq:4.20}) also apply 
to the (b-1) and (b-2) diagrams, respectively, if we name the strings as
shown in Fig.~\ref{fig:v3uomg-2}, so that the (b) configuration also 
cancels out. Cancellation between (c-1) and (c-2), on the other hand, 
occurs if the condition
\begin{eqnarray}
 \dd = \xu x_4\,,
\label{eq:xu-xomega-x4}
\end{eqnarray}
holds. Indeed, (c-1) diagram is contained in (T8) in the form
\begin{eqnarray}
&&  2\dd\Uomg{2,a,4^\c}\V3o(b31)\ketRo(ab)\ket{\Phi}_{4^\c}\ket{\Psi}_{132}\nn
  &&\quad    
 = 2\dd\int d\sigma_0^{(2)}\uomg{2,a,4^\c;\sigma_0^{(2)}}
    b_{\sigma_0^{(2)}}\bP{4}\!\v3o(b31x)\!\ketRo(ab)
          \ket{\Phi}_{4^\c}\ket{\Psi}_{132}\nn
  &&\quad    
  = 2\dd \int d\sigma_0^{(2)}\bra{\tilde v(123,4^\c)}b_{\sigma_0^{(2)}}
    \bP{4}\ket{\Phi}_{4^\c}\ket{\Psi}_{321}
\end{eqnarray}
while (c-2) is contained in (T8) in the form
\begin{eqnarray}
&&  -2\aa\ee\UU{a,4^\c}\V4(b123)\ketRo(ab)\ket{\Phi}_{4^\c}\ket{\Psi}_{321}\nn
  &&\quad    
  =-2\aa\ee\UU{a,4^\c}\V4(b1{\stackrel{\downarrow}2}3)\ketRo(ab)
  \ket{\Phi}_{4^\c}\ket{\Psi}_{321}\nn
  &&\quad    
  =-2\aa\ee\int d\sigma_0^{(2)}\uv{a,4^\c}
   \bP{4}\!\v4(b123{\sigma_0^{(2)}})b_{\sigma_0^{(2)}}\!\ketRo(ab)
  \ket{\Phi}_{4^\c}\ket{\Psi}_{321}\nn
  &&\quad    
  =-2\aa\ee \int d\sigma_0^{(2)}\bra{\tilde v(123,4^\c)}b_{\sigma_0^{(2)}}\bP{4}
    \ket{\Phi}_{4^\c}\ket{\Psi}_{321}.
\end{eqnarray}
Here use has been made of the Grassmann oddness of $\bra{v_4^\o}$ and 
$\ket{R^\o}$. 
The required condition (\ref{eq:xu-xomega-x4}) is actually satisfied 
by the relations $x_4=1$ and $x_u=x_\Omega$ already determined 
in Eqs.~(\ref{eq:x4}) and (\ref{eq:xu-xomg}).

\subsubsection{T9 terms}%{Eq.~(\ref{eq:3-3})}

\begin{figure}[tb]
   \epsfxsize= 11cm   %or \epsfysize= HEIGHT cm
   \centerline{\epsfbox{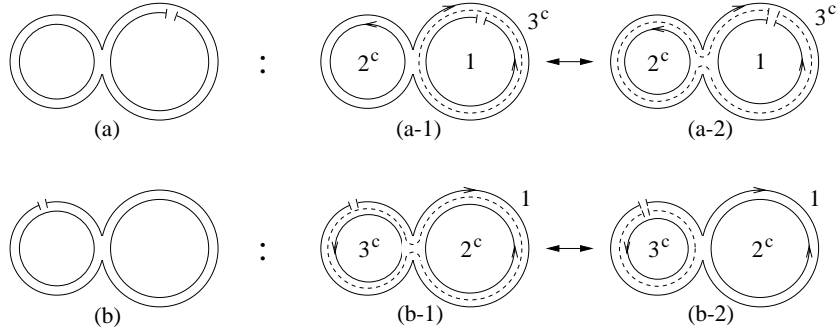}}
 \caption{Two configurations exist for (T9) terms.}
 \label{fig:uv3c-2}
\end{figure}

\begin{figure}[tb]
   \epsfxsize= 13.5cm %\textwidth   %or \epsfysize= HEIGHT cm
   \centerline{\epsfbox{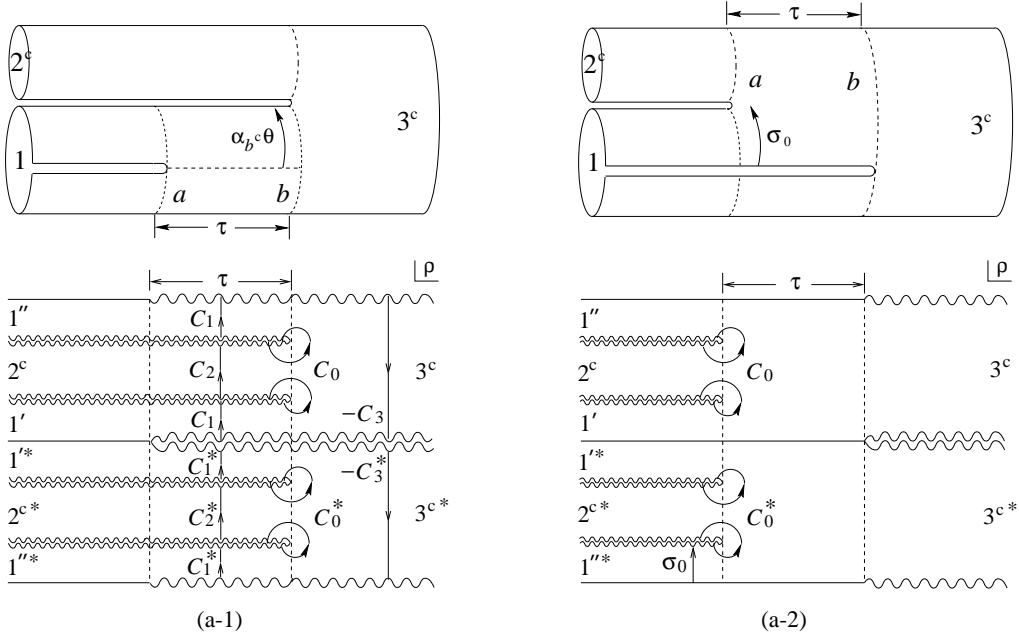}}
 \caption{$\rho$ planes for the diagrams which reduce to (a-1) and (a-2)
       in Fig.~\protect\ref{fig:uv3c-2} at $\tau=0$.}
 \label{fig:t9}
\end{figure}

The generic configurations obtained by contracting
the vertex $\bra{U}$ with $\bra{V_3^\c}$ or $\bra{U_\Omega}$ 
fall into two types, (a) and (b), depicted in Fig.~\ref{fig:uv3c-2}.
Again each of them are realized in two ways 
as also shown in Fig.~\ref{fig:uv3c-2}.
%\begin{wrapfigure}[n]{r}{6.6cm}
%\end{wrapfigure}
As in the previous cases, cancellations occur between 
(a-1) and (a-2), and (b-1) and (b-2). 
Denoting the LPP vertex corresponding to the glued configuration (a) in 
Fig.~\ref{fig:uv3c-2} by $\bra{\tilde v(1,2^\c3^\c;\sigma_0)}$, the pair of 
diagrams (a-1) and (a-2) are
contained in (T9) in the following form:
\begin{eqnarray}
&&\hbox{(a-1)}:\nn
&&\ 
  \aa\ff\UU{1,\check{a^\c}}\V3c(b23)\ketRc(ab) 
             \ket{\Phi}_{3^\c2^\c}\ket{\Psi}_1 \nn
  &&\ =
  \aa\ff \bra{u(1a^\c)}\!\bra{v_3^\c(b^\c2^\c3^\c)}\bzm{b}
\!\!\int\!{d\theta\over2\pi}e^{i\theta(L-\bar L)^{(b^\c)}}
\!\!\!\prod_{r=2,3}\!\bP{r}\ket{R^\c(a^\c b^\c)}
\ket{\Phi}_{3^\c2^\c}\!\ket{\Psi}_1 \nn
  &&\ =
  \aa\ff \!\int\!{d\theta\over2\pi}
\bra{u(1a^\c)}\!\bra{v_3^\c(b^\c2^\c3^\c)}e^{i\theta(L-\bar L)^{(b^\c)}}
\!\ket{R^\c(a^\c b^\c)}
 \bzm{b}\!\!\!\prod_{r=2,3}\!\bP{r}\ket{\Phi}_{3^\c2^\c}\!\ket{\Psi}_1 \nn
  &&\ =
  \aa\ff \int{d\theta\over2\pi}
     \bra{\tilde v(1,2^\c3^\c;\theta)}\bzm{b} 
          \prod_{r=2,3}\bP{r}\ket{\Phi}_{3^\c2^\c}\ket{\Psi}_1 
\label{eq:a1}
\\
&&\hbox{(a-2)}:\nn
&&\ 
  -2\dd\aa\Uomg{1,a,2^\c}\UU{b,3^\c}\ketRo(ab)\ket{\Phi}_{3^\c2^\c}\ket{\Psi}_1 \nn
  &&\     
  = -2\dd\aa \int\!d\sigma_0\bra{u_\Omega(1a2^\c;\sigma_0)}b_{\sigma_0}
     \bP{2}\!\bra{u(b3^\c)}\bP{3}\!\ket{R^\o(ab)}
\ket{\Phi}_{3^\c2^\c}\ket{\Psi}_1 \nn
  &&\     
  = +2\dd\aa \int\!d\sigma_0\bra{u_\Omega(1a2^\c;\sigma_0)}\!\bra{u(b3^\c)}
      \!\ket{R^\o(ab)}
            b_{\sigma_0}\bP{2}\bP{3}\!\ket{\Phi}_{3^\c2^\c}\ket{\Psi}_1 \nn
  &&\    
  = +2\dd\aa \int d\sigma_0
     \bra{\tilde v(1,2^\c3^\c;\sigma_0)}b_{\sigma_0}
            \prod_{r=2,3}\bP{r}\ket{\Phi}_{3^\c2^\c}\ket{\Psi}_1 \ ,
\label{eq:a2}
\end{eqnarray}
where some of the commas in the string arguments of the vertices are 
omitted for brevity, and we have used the Grassmann oddness of 
$\ket{R^\o}$ and the GGRT. The resultant LPP vertices for the glued 
configurations (a-1) and (a-2) are clearly the same 
(See Fig.~\ref{fig:t9} at $\tau=0$): 
\begin{equation}
     \bra{\tilde v(1,2^\c3^\c;\theta)}=
     \bra{\tilde v(1,2^\c3^\c;\sigma_0)} 
\qquad {\rm for}\quad \alpha_{b^\c}\theta=\sigma_0\,.
\label{eq:thetasigma}
\end{equation}
We, therefore, have only to compare the anti-ghost factors 
$\bzm{b}$ and $b_{\sigma_0}$ appearing in Eqs.~(\ref{eq:a1}) and 
(\ref{eq:a2}), respectively. 
This comparison is actually very similar to that performed already in the 
previous paper I for the cancellation of (T10) between 
$\bra{U}\bra{V_\propto}\ket{R^\o}$ and $\bra{U}\bra{V_\infty}\ket{R^\o}$. 
For this purpose, look at the 
$\rho$-plane diagrams drawn in Fig.~\ref{fig:t9}, where the 
figures represent the configurations which reduce to the present ones 
(a-1) and (a-2), respectively,  as the time interval $\tau$ goes to zero.
%\begin{wrapfigure}[n]{r}{6.6cm}
%\end{wrapfigure}
First as performed in the previous paper, the anti-ghost factor 
$\bzm{b}$ is replaced by 
$\widetilde\bzm{b}\equiv 
\bzm{b}+(\alpha_{b^\c}/\alpha_{2^\c})\bzm{2}+(\alpha_{b^\c}/\alpha_{3^\c})\bzm{3}$. 
This is possible 
since $\bzm{2}$ and $\bzm{3}$ are zero in front of the factors 
$\prod_{r=2,3}\bP{r}$ present in Eq.~(\ref{eq:a1}). Then 
note that
\begin{eqnarray}
\bzm{r}&\equiv&\half (b_0^{(r)}-\bar b_0^{(r)})=
\half \left(\oint {d\rho_r\over2\pi i} b^{(r)}(\rho_r)-{\rm a.h.}\right) \nn
&=&\half \left(\oint {d\rho\over2\pi i}{d\rho\over d\rho_r} b(\rho)-{\rm a.h.}\right)
=\half \alpha_r \left(\oint {d\rho\over2\pi i} b(\rho)-{\rm a.h.}\right)\,,
\end{eqnarray}
where we have used the fact that the $\rho$ coordinate is identified with 
$\rho=\alpha_r\rho_r+$const in the region of string $r$. Hence 
$\widetilde\bzm{b}$ can be seen to reduce to the following expression by
making a deformation of the integration contour:
\begin{eqnarray}
\widetilde\bzm{b}&=& 
\half\alpha_{b^\c}\left(\oint_{C_1+C_2-C_3}-\oint_{C_1^*+C_2^*-C_3^*}\right) 
{d\rho\over2\pi i}b(\rho)\nn
&=&-\half\alpha_{b^\c}\left(\oint_{C_0}-\oint_{C_0^*}\right) 
{d\rho\over2\pi i}b(\rho)
\equiv-\half\alpha_{b^\c}\left(b_{\rho_0}-b_{\rho_0^*}\right) \,,
\label{eq:bzerominus}
\end{eqnarray}
where the contours $C_i$ and $C_i^*$ are shown in Fig.~\ref{fig:t9}. 

On the other hand, the anti-ghost factor $b_{\sigma_0}$ appearing 
in Eq.~(\ref{eq:a2}) is written in the form 
\begin{eqnarray}
b_{\sigma_0}
=\left(d\rho_0\over d\sigma_0\right)\oint_{C_0}{d\rho\over2\pi i}b(\rho)
+\left(d\rho_0^*\over d\sigma_0\right)\oint_{C_0^*}{d\rho\over2\pi i}b(\rho)
=i\left(b_{\rho_0}-b_{\rho_0^*}\right)\,,
\label{eq:bsigmazero}
\end{eqnarray}
using $d\rho_0/d\sigma_0=i$ and $d\rho_0^*/d\sigma_0=-i$.
Therefore, (\ref{eq:a1}) and (\ref{eq:a2}) cancel each other if 
\begin{equation}
  \aa\ff \int_0^{2\pi}{d\theta\over2\pi}
     \left(-\half\alpha_{b^\c}\right)\, \cdots%\left(b_{\rho_0}-b_{\rho_0^*}\right)
  = -2i\dd\aa \int_0^{\alpha_1\pi}d\sigma_0\, \cdots 
\label{eq:4.31}
\end{equation}
since the integrands are the same for $\alpha_{b^\c}\theta=\sigma_0$ 
by Eq.~(\ref{eq:thetasigma}). 
If we note the relation  $\alpha_{b^\c}=\alpha_1/2$ (since the strings 
$b^\c$ and $1$ 
are closed and open strings, respectively, in the (a-1) case), we see that
the integration regions on both sides in Eq.~(\ref{eq:4.31}) coincide, 
$\alpha_{b^\c}\int_0^{2\pi}d\theta= \int_0^{\alpha_1\pi}d(\alpha_{b^\c}\theta)
=\int_0^{\alpha_1\pi}d\sigma_0$, and thus the cancellation is complete if 
\begin{equation}
  -{1\over4\pi}\ff  = -2i\dd  \quad \Rightarrow\quad \ff=8\pi i\dd\,.
\label{eq:xcxomega}
\end{equation}
The cancellation between (b-1) and (b-2) is also seen in quite the same way 
and thus (T9) has been proved to vanish.

\subsection{$O(g^4)$ invariance}

\subsubsection{T11 term}%{Eq.~(\ref{eq:4-1})}

\begin{figure}[tb]
   \epsfxsize= 12.6cm   %or \epsfysize= HEIGHT cm
   \centerline{\epsfbox{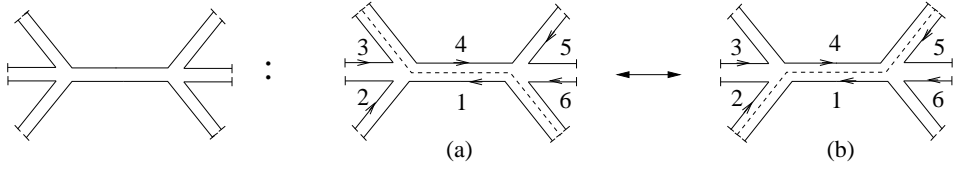}}
 \caption{The unique configuration for (T11) term.}
 \label{fig:v4v4-2}
\end{figure}

This (T11) term has already been proved to vanish by HIKKO.\cite{rf:HIKKO1}
\ The configuration obtained by contracting two $\bra{V_4^\o}$ vertices 
is unique, of the type drawn in Fig.~\ref{fig:v4v4-2}.
Name the six open strings 1, 2, ---, 6 in a cyclic order as 
shown there.  For any single configuration with 
a definite set of string length parameters $\alpha_1$ --- $\alpha_6$,  
there are always two ways to realize it by gluing the vertices 
$\bra{V_4^\o}$ and $\bra{V_4^\o}$, as also shown in Fig.~\ref{fig:v4v4-2}. 
%\end{wrapfigure}
The terms corresponding to the two configurations (a) and (b) are 
contained in (T11) in the following forms, respectively: 
\begin{eqnarray}
&&\V4(123a)\V4(b456)\ketRo(ab) \ket{\Psi}_{654321} \nn
&&\  =(-)
\V4(1{\stackrel{\downarrow}2}3a)(-)\V4(b45{\stackrel{\downarrow}6})\ketRo(ab) \ket{\Psi}_{654321}\nn
&&\ = +\!\int\!d\sigma_0^{(2)} d\sigma_0^{(6)}\!
\sbra{v_4^\o(123a;\sigma_0^{(2)})}b_{\sigma_0^{(2)}}\!
\sbra{v_4^\o(b456;\sigma_0^{(6)})}b_{\sigma_0^{(6)}}\!\sket{R^\o(ab)}
 \ket{\Psi}_{654321}\nn
&&\V4(234a)\V4(b561)\ketRo(ab) \ket{\Psi}_{165432} \nn
&&\ =
\V4({\stackrel{\downarrow}2}34a)(+)\V4(b5{\stackrel{\downarrow}6}1)\ketRo(ab) (-)\ket{\Psi}_{654321}\nn
&&\ = -\!\int\!d\sigma_0^{(2)} d\sigma_0^{(6)}\!
\sbra{v_4^\o(234a;\sigma_0^{(2)})}b_{\sigma_0^{(2)}}\!
\sbra{v_4^\o(b561;\sigma_0^{(6)})}b_{\sigma_0^{(6)}}\!
\ket{R^\o(ab)} \ket{\Psi}_{654321}.
\hspace{4em}
\end{eqnarray}
Note that the minus sign in the last expression has come from 
$\ket{\Psi}_{165432}=-\ket{\Psi}_{654321}$, giving the relatively opposite 
signs between the two terms. Apply the GGRT to both terms there.
Then, clearly, they yield the same LPP vertex 
$\sbra{\tilde v(123456;\sigma_0^{(2)},\sigma_0^{(6)})}$ keeping the relatively 
opposite overall signs, so that they turn out to cancel each other.

\subsubsection{T12 term}%{Eq.~(\ref{eq:4-2})}

\begin{figure}[tbh]
   \epsfxsize= 12.6cm   %or \epsfysize= HEIGHT cm
   \centerline{\epsfbox{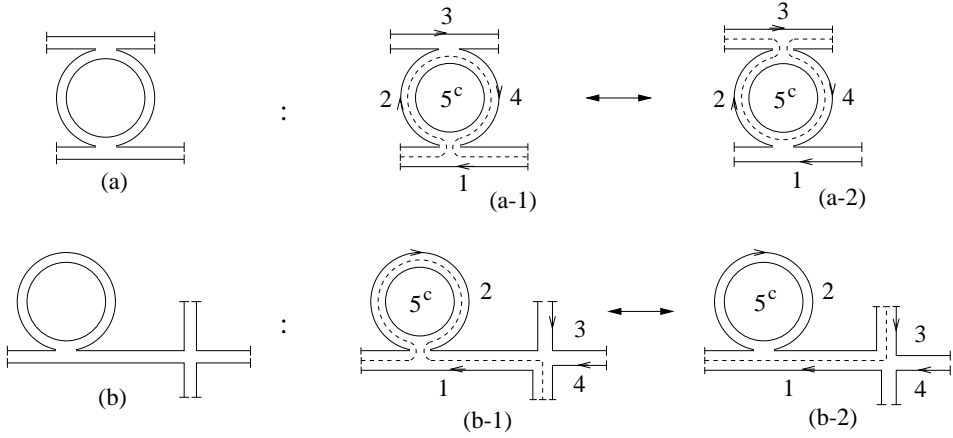}}
 \caption{Two configurations for (T12) term.}
\label{fig:v4uomg-2}
\end{figure}

There appear two distinct configurations when contracting the vertices 
$\bra{U_\Omega}$ and $\bra{V_4^\o}$, which are the types (a) and (b) 
given in Fig.~\ref{fig:v4uomg-2}.
Again they are each realized in two ways as also drawn in 
Fig.~\ref{fig:v4uomg-2}. 
%\begin{wrapfigure}[n]{r}{6.6cm}
%\end{wrapfigure}
The terms of the diagrams (a-1) and (a-2) are 
contained in (T12) in the following forms, respectively: 
\begin{eqnarray}
&&\Uomg{1,a,5^\c}\V4(b234)\ketRo(ab)\ket{\Phi}_{5^\c}\ket{\Psi}_{4321} \nn
&&\quad  =
\sbra{u_\Omega(1a5^\c;\sigma_0^{(1)})}b_{\sigma_0^{(1)}}
\sbra{v_4^\o(b234;\sigma_0^{(3)})}b_{\sigma_0^{(3)}}
\ket{R^\o(ab)}\ket{\Phi}_{5^\c}\ket{\Psi}_{4321} 
\label{eq:4.34}\\
&&\Uomg{3,a,5^\c}\V4(b412)\ketRo(ab)\ket{\Phi}_{5^\c}\ket{\Psi}_{2143} \nn
&&\quad  =
\sbra{u_\Omega(3a5^\c;\sigma_0^{(3)})}b_{\sigma_0^{(3)}}
\sbra{v_4^\o(b412;\sigma_0^{(1)})}b_{\sigma_0^{(1)}}
\ket{R^\o(ab)}\ket{\Phi}_{5^\c}\ket{\Psi}_{2143}, 
\hspace{2em}
\label{eq:4.35}
\end{eqnarray}
where the common integration symbols $\int d\sigma_0^{(1)}d\sigma_0^{(3)}$ have been 
suppressed and use has been made of 
$\V4(b234)=+\V4(b2{\stackrel{\downarrow}3}4)$. 
In this case, the states are the same, $\ket{\Psi}_{4321}=\ket{\Psi}_{2143}$, 
and the GGRT gives a common LPP vertex for the glued configuration (a), 
but the orders of the anti-ghost factor 
$b_{\sht{\sigma_0^{(1)}}}b_{\sht{\sigma_0^{(3)}}}$ 
are opposite between the two. They thus cancel each other. 
For the case of (b) configuration,  (b-1) term is given by the same 
Eq.~(\ref{eq:4.34}) while (b-2) by
\begin{eqnarray}
&&\Uomg{2,a,5^\c}\V4(b341)\ketRo(ab)\ket{\Phi}_{5^\c}\ket{\Psi}_{1432} \nn
&&\quad  =
\sbra{u_\Omega(2a5^\c;\sigma_0^{\prime(2)})}b_{\sigma_0^{\prime(2)}}
\sbra{v_4^\o(b341;\sigma_0^{(3)})}(-b_{\sigma_0^{(3)}})
\ket{R^\o(ab)}\ket{\Phi}_{5^\c}\ket{\Psi}_{1432},
\hspace{3em}
\end{eqnarray}
with the symbol $\int d\sigma_0^{\prime(2)}d\sigma_0^{(3)}$ suppressed again.
Comparing the diagrams (b-1) and (b-2), we see that the interaction 
point $\sigma_0^{\prime(2)}$ here of $\uomg{2,a,5^\c;\sigma_0^{\prime(2)}}$ is 
the same as that of $\uomg{1,a,5^\c;\sigma_0^{(1)}}$ in Eq.~(\ref{eq:4.34}) 
so that $b_{\sht{\sigma_0^{\prime(2)}}}=-b_{\sht{\sigma_0^{(1)}}}$ 
(directions are opposite). 
Therefore the anti-ghost factors are the same between them, 
$b_{\sht{\sigma_0^{\prime(2)}}}(-b_{\sht{\sigma_0^{(3)}}})
=b_{\sht{\sigma_0^{(1)}}}b_{\sht{\sigma_0^{(3)}}}$, but the 
states have opposite signs, $\ket{\Psi}_{1432}=-\ket{\Psi}_{4321}$. Thus 
(T12) is also proved to vanish.

\vspace{3pt}

\subsubsection{T13 term}%{Eq.~(\ref{eq:4-3})}

This term was already analyzed intensively and shown to vanish by 
HIIKKO.\cite{rf:HIKKO2} \ 
So let us show this fact in our terminology briefly.

%\begin{wrapfigure}[n]{r}{6.6cm}
\begin{figure}[tb]
   \epsfxsize= 13cm   %or \epsfysize= HEIGHT cm
   \centerline{\epsfbox{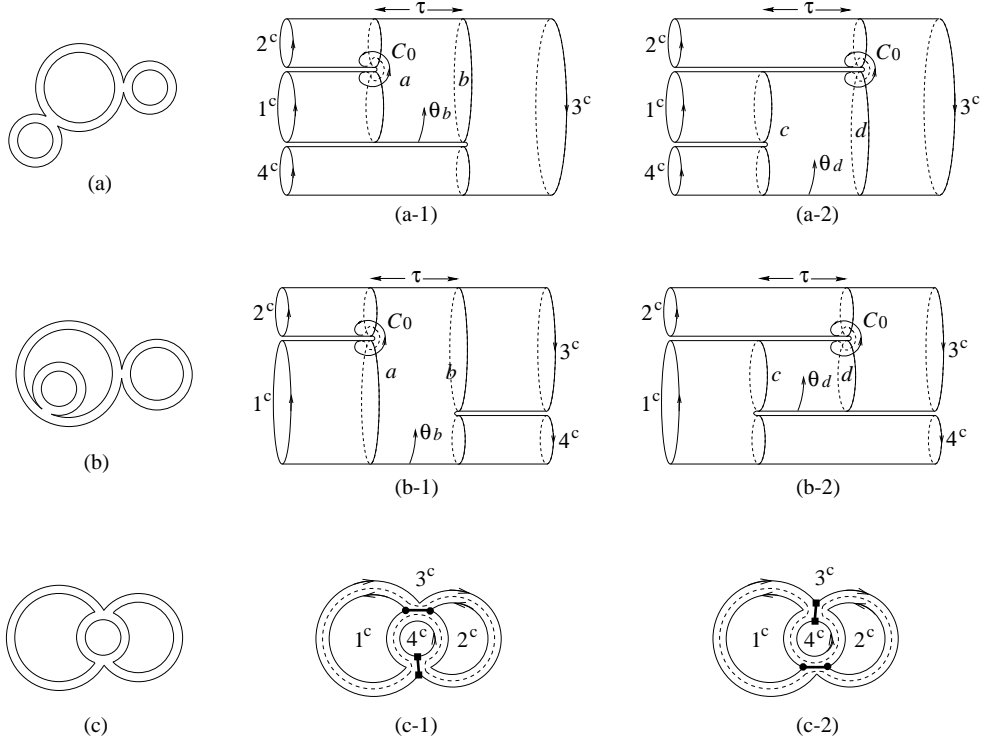}}
 \caption{Three configurations for (T13) term. In (c-1) and (c-2), the 
bonds connecting the solid dots and solid squares denote 
the $1^\c$-$2^\c$ and $3^\c$-$4^\c$ interaction points, respectively.}
 \label{fig:t13}
\end{figure}
%\end{wrapfigure}
The generic configurations obtained by gluing two closed 3-string 
vertices $\bra{V_3^\c}$ fall into three types, (a), (b) and (c), in 
Fig.~\ref{fig:t13}, each of which is realized in two ways of gluing, as 
also shown there. The (a-1) and (a-2) terms for (a), (and (b-1) and 
(b-2) for (b) as well, if the strings are named as shown in 
Fig.~\ref{fig:t13},) are contained in the (T13) term, (\ref{eq:4-3}), in
the following forms, respectively:
\begin{eqnarray}
%&&\hbox{(a-1), (b-1):} \nn
&&\quad  
\Vthc{1^\c,2^\c,\check{a^\c}}\V3c(b34)\ketRc(ab)
            \ket{\Phi}_{4^\c3^\c2^\c1^\c} \nn
&&\qquad = 
\v3c(12axx)\v3c(b34xx)\bzm{b}
\int{d\theta_b\over2\pi}e^{i\theta_b(L-\bar L)^{(b^\c)}}\ketRc(ab) \nn
&&\qquad \qquad \qquad \times\prod_{r=1^\c,2^\c,3^\c,4^\c}\bP{r}
\ket{\Phi}_{4^\c3^\c2^\c1^\c} 
\label{eq:t13a-1}
\\
%&&\hbox{(a-2), (b-2):} \nn
&&\quad \ 
\Vthc{4^\c,1^\c,\check{c^\c}}\V3c(d23)\ketRc(cd)
            \ket{\Phi}_{3^\c2^\c1^\c4^\c} \nn
&&\qquad = 
\v3c(41cxx)\v3c(d23xx)\bzm{d}
\int{d\theta_d\over2\pi}e^{i\theta_d(L-\bar L)^{(d^\c)}}\ketRc(cd) \nn
&&\qquad \qquad \qquad \times\prod_{r=4^\c,1^\c,2^\c,3^\c}\bP{r}
\ket{\Phi}_{3^\c2^\c1^\c4^\c}
\hspace{2em}
\label{eq:t13a-2}
\end{eqnarray}
The external states and the $\bzm{r}$ factors associated with them 
are common as a whole to these two terms; 
$\prod_{r=1^\c,2^\c,3^\c,4^\c}\bP{r}
=-\prod_{r=4^\c,1^\c,2^\c,3^\c}\bP{r}$ and 
$\ket{\Phi}_{4^\c3^\c2^\c1^\c}=-\ket{\Phi}_{3^\c2^\c1^\c4^\c}$. 
So the sign difference should come from $\bzm{b}$ and $\bzm{d}$.
By a similar reasoning to the (T9) (a-1) case, these anti-ghost factors 
can be converted into
\begin{eqnarray}
%\bzm{b} &\Rightarrow& +\half\alpha_{b^\c}(b_{\rho_0}-\bar b_{\rho_0^*})\,, \nn
%\bzm{d} &\Rightarrow& -\half\alpha_{d^\c}(b_{\rho_0}-\bar b_{\rho_0^*})\,,
\bzm{b} \Rightarrow+\half\alpha_{b^\c}(b_{\rho_0}-\bar b_{\rho_0^*})\,, 
\qquad 
\bzm{d} \Rightarrow-\half\alpha_{d^\c}(b_{\rho_0}-\bar b_{\rho_0^*})\,,
\label{eq:bfactor}
\end{eqnarray}
in the presence of $\bzm{1}\bzm{2}$ and of $\bzm{2}\bzm{3}$ 
respectively, where 
$b_{\rho_0}=\oint_{C_0}(d\rho/2\pi)\linebreak b(\rho)$ is the anti-ghost factor 
corresponding to the shift of the $1^\c$-$2^\c$ string interaction points
drawn in Fig.~\ref{fig:t13} (and $\bar b_{\rho^*_0}$ is its anti-holomorphic 
counterpart). 
On the other hand, 
comparing the diagrams (a-1) and (a-2), and (b-1) and (b-2), we easily 
see that these pairs of diagrams realize the same glued configurations 
when the twisting angles $\theta_b$ and $\theta_d$ of the intermediate closed 
strings satisfy the relation
\begin{equation}
\alpha_b\theta_b = -\alpha_d\theta_d
\end{equation}
if the origins of $\theta$'s are chosen suitably. Therefore, to keep the 
same glued configurations, the increasing directions of $\theta_b$ and 
$\theta_d$ are opposite, and the opposite signs (as well as their weights) 
between $\bzm{b}$ and $\bzm{d}$ in Eq.~(\ref{eq:bfactor}) reflect this 
fact. Note that the (a) configuration, by definition, corresponds 
to the twisting angle regions $-\alpha_1\pi\leq\alpha_b\theta_b\leq\alpha_1\pi$ and 
$-\alpha_1\pi\leq\alpha_d\theta_d\leq\alpha_1\pi$ for (a-1) and (a-2) diagrams, respectively,
and that the (b) configuration corresponds 
to the twisting angle regions 
$-(\alpha_1-\abs{\alpha_4})\pi\leq\alpha_b\theta_b\leq(\alpha_1-\abs{\alpha_4})\pi$ and 
$-(\alpha_1-\abs{\alpha_4})\pi\leq\alpha_d\theta_d\leq(\alpha_1-\abs{\alpha_4})\pi$ 
for (b-1) and (b-2) diagrams, respectively.
We thus have shown that (a-1) and (a-2) terms, and (b-1) and (b-2) as well, 
cancel each other between these integration regions. 

If the twisting angle $\theta_b$ in the (b-1) diagram exceeds the above limit 
and falls into the region 
$(\alpha_1-\abs{\alpha_4})\pi\leq\alpha_b\abs{\theta_b}\leq(\alpha_2+\abs{\alpha_3})\pi$  
(note that $(\alpha_1-\abs{\alpha_4})+2\alpha_2=\alpha_2+\abs{\alpha_3}$), 
then (b-1) diagram turns into the (c) configuration. The (c-1) and 
(c-2) diagrams in Fig.~\ref{fig:t13} correspond to the positive and 
negative $\theta_b$, respectively, in this region. Then it is clear from 
the figure that the $1^\c$-$2^\c$ string interaction point
of (c-1) corresponds to the $3^\c$-$4^\c$ string interaction point 
of (c-2) and vice versa. But, if the anti-ghost factor $\bzm{b}$ is 
expressed by $b_{\rho_0}$ of the $3^\c$-$4^\c$ interaction point,
it has an extra minus sign relative to the above 
$1^\c$-$2^\c$ interaction point case (see the (b-1) diagram),
which again reflects the fact that the increasing directions of 
$\theta_b$ and $-\theta_b$ (realizing the same configuration) are opposite. 
Thus (c-1) and (c-2) are also seen to cancel each other. 

\subsubsection{T14 terms}%{Eq.~(\ref{eq:4-4})}

\begin{figure}[tb]
   \epsfxsize= 12.6cm   %or \epsfysize= HEIGHT cm
   \centerline{\epsfbox{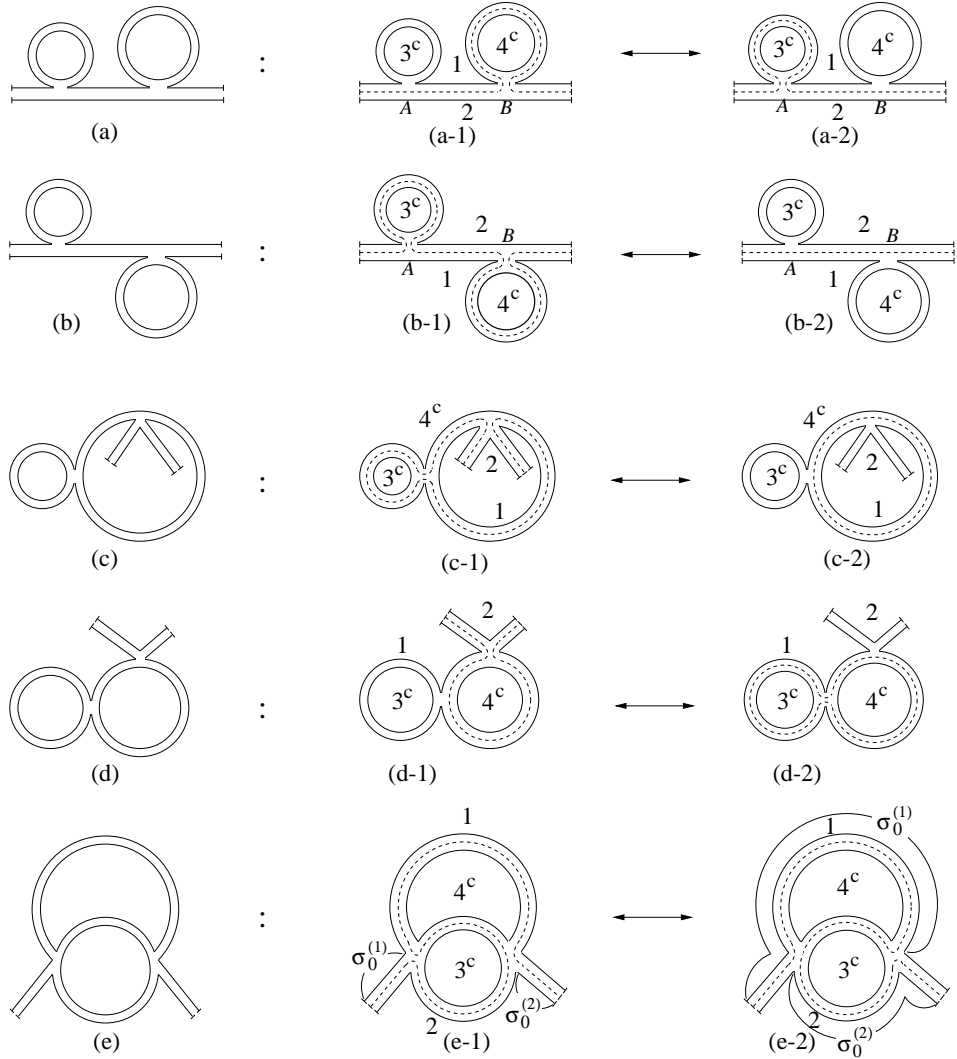}}
 \caption{Five configurations for (T14) terms.}
 \label{fig:uomguomg-2}
\end{figure}

There are 5 relevant configurations in this case, (a) --- (e), shown 
in Fig.~\ref{fig:uomguomg-2}, each of which is 
realized in two ways as also drawn there.
%\begin{wrapfigure}[n]{r}{6.6cm}
%\end{wrapfigure}
The terms (a-1) and (a-2) for (a), and (b-1) and 
(b-2) for (b) as well by naming the strings as shown in 
Fig.~\ref{fig:uomguomg-2}, are 
contained in the first term in (T14) in the forms
\begin{eqnarray}
&&\Uomg{1,a,3^\c}\Uomg{b,2,4^\c}\ketRo(ab)\ket{\Phi}_{4^\c3^\c}
\ket{\Psi}_{21} \nn
&&\quad =\sbra{u_\Omega(1a3^\c;\sigma_{0A}^{(1)})}b^A_{\sigma_0^{(1)}}
(-)\sbra{u_\Omega(b24^\c;\sigma_{0B}^{(2)})}b^B_{\sigma_0^{(2)}}\ket{R^\o(ab)}
\ket{\Phi}_{4^\c3^\c}\ket{\Psi}_{21} \nn
&&
\Uomg{2,a,3^\c}\Uomg{b,1,4^\c}\ketRo(ab)
           \ket{\Phi}_{4^\c3^\c}\ket{\Psi}_{12} \nn
&&\quad =\sbra{u_\Omega(2a3^\c;\sigma_{0A}^{(2)})}b^A_{\sigma_0^{(2)}}
(-)\sbra{u_\Omega(b14^\c;\sigma_{0B}^{(1)})}b^B_{\sigma_0^{(1)}}\ket{R^\o(ab)}
           \ket{\Phi}_{4^\c3^\c}\ket{\Psi}_{12},
\hspace{3em}
\end{eqnarray}
respectively, where the integration symbols 
$\int d\sigma_0^{(1)}d\sigma_0^{(2)}$ are omitted,
$\Uomg{b,r,4^\c}=(-)\Uomg{b,{\stackrel{\downarrow}r},4^\c}$ 
with $r=1$ and $2$ have been used, and the labels $A$ and $B$ attached to 
the anti-ghost factors to distinguish the two interaction points appearing 
in Fig.~\ref{fig:uomguomg-2}. So despite the appearance, the 
anti-ghost factors have the same signs between the two terms, 
$b^A_{\sht{\sigma_0^{(1)}}}b^B_{\sht{\sigma_0^{(2)}}}
=b^A_{\sht{\sigma_0^{(2)}}}b^B_{\sht{\sigma_0^{(1)}}}$, 
since we have $b^A_{\sht{\sigma_0^{(1)}}}=-b^A_{\sht{\sigma_0^{(2)}}}$ and 
$b^B_{\sht{\sigma_0^{(2)}}}=-b^B_{\sht{\sigma_0^{(1)}}}$. 
The states, on the other hand, have 
opposite signs, $\ket{\Psi}_{21}=-\ket{\Psi}_{12}$, and hence the (a) and (b)
terms vanish in (T14). 

The cancellations in other two cases of (c) and (d), are those between 
contractions 
$\bra{U_\Omega}\bra{U_\Omega}\ket{R^\o}$ and $\bra{U_\Omega}\bra{V_3^\c}\ket{R^\c}$.
The terms (c-1) and (c-2),  (and (d-1) and (d-2) as well, if 
the strings are named as shown in Fig.~\ref{fig:uomguomg-2},) are 
contained in the first and second terms in (T14) in the forms
\begin{eqnarray}
&&
-\dd^2\Uomg{1,a,3^\c}\Uomg{b,2,4^\c}\ketRo(ab)
             \ket{\Phi}_{4^\c3^\c}\ket{\Psi}_{21} \nn
&&\ =
-\dd^2\int d\sigma_0^{(1)}d\sigma_0^{(2)}
\uomg{1,a,3^\c;\sigma_0^{(1)}}b_{\sigma_0^{(1)}}\bP{3} \nn
&&\hspace{5em}\times 
      (-)\uomg{b,2,4^\c;\sigma_0^{(2)}}b_{\sigma_0^{(2)}}\bP{4}\ketRo(ab)
             \ket{\Phi}_{4^\c3^\c}\ket{\Psi}_{21} \nn
&&\ =
+\dd^2\!\int\!d\sigma_0^{(2)}d\sigma_0^{(1)}
\sbra{\tilde v(123^\c4^\c;\sigma_0^{(2)},\sigma_0^{(1)})}
          b_{\sigma_0^{(2)}}b_{\sigma_0^{(1)}}
\!\!\prod_{r=3,4}\!\!\bP{r}\!\ket{\Phi}_{4^\c3^\c}\!\ket{\Psi}_{21}  
\label{eq:4.38} \\
&&
-\sfrac12\dd\ff\Uomg{1,2,\check{a^\c}}\V3c(b34)\ketRc(ab)
             \ket{\Phi}_{4^\c3^\c}\ket{\Psi}_{21} \nn
&&\ =
-\sfrac12\dd\ff(-)\int d\sigma_0^{(2)}\uomg{1,2,a;\sigma_0^{(2)}}b_{\sigma_0^{(2)}} \nn
&&\qquad
      \times\v3c(b34x)\bzm{b}\int{d\theta\over2\pi}e^{i\theta(L-\bar L)^{(b^\c)}} 
        \prod_{r=3,4}\bP{r}\ketRc(ab)
           \ket{\Phi}_{4^\c3^\c}\ket{\Psi}_{21} \nn
&&\ =
+\sfrac12\dd\ff\!\int\!d\sigma_0^{(2)}{d\theta\over2\pi}
\sbra{\tilde v(123^\c4^\c;\sigma_0^{(2)},\theta)}
        b_{\sigma_0^{(2)}}\bzm{b}%\ket{{\rm ext}}_{4^\c3^\c21}
\!\!\prod_{r=3,4}\!\!\bP{r}\! \ket{\Phi}_{4^\c3^\c}\!\ket{\Psi}_{21},
\hspace{3em}
\label{eq:4.39}
\end{eqnarray}
respectively, where 
a care has been taken of the signs 
and the identities similar to (\ref{eq:2.12}) have been used for 
$\bra{U_\Omega}$. Here the LPP glued vertices denote 
\begin{eqnarray}
\sbra{\tilde v(123^\c4^\c;\sigma_0^{(2)},\sigma_0^{(1)})}
&\equiv&\sbra{u_\Omega(1a3^\c;\sigma_0^{(1)})}\sbra{u_\Omega(b24^\c;\sigma_0^{(2)})}
\ket{R^\o(ab)} \nn
\sbra{\tilde v(123^\c4^\c;\sigma_0^{(2)},\theta)}
&\equiv&\sbra{u_\Omega(12a^\c;\sigma_0^{(2)})}\sbra{v_3^\c(b^\c3^\c4^\c)}
e^{i\theta(L-\bar L)^{(b^\c)}}\ket{R^\c(a^\c b^\c)}\,.
\hspace{4em}
\end{eqnarray}
By drawing the $\rho$ plane diagram corresponding to the 
present configurations (c-1) and (c-2) as shown in Fig.~\ref{fig:t14}, 
we see that the glued configurations, 
and hence these LPP vertices, coincide with each other when $\sigma_0^{(1)}$ 
and $\theta$ satisfy a relation:
%\begin{wrapfigure}[n]{r}{6.6cm}
\begin{figure}[tb]
   \epsfxsize= 13cm%\textwidth   %or \epsfysize= HEIGHT cm
   \centerline{\epsfbox{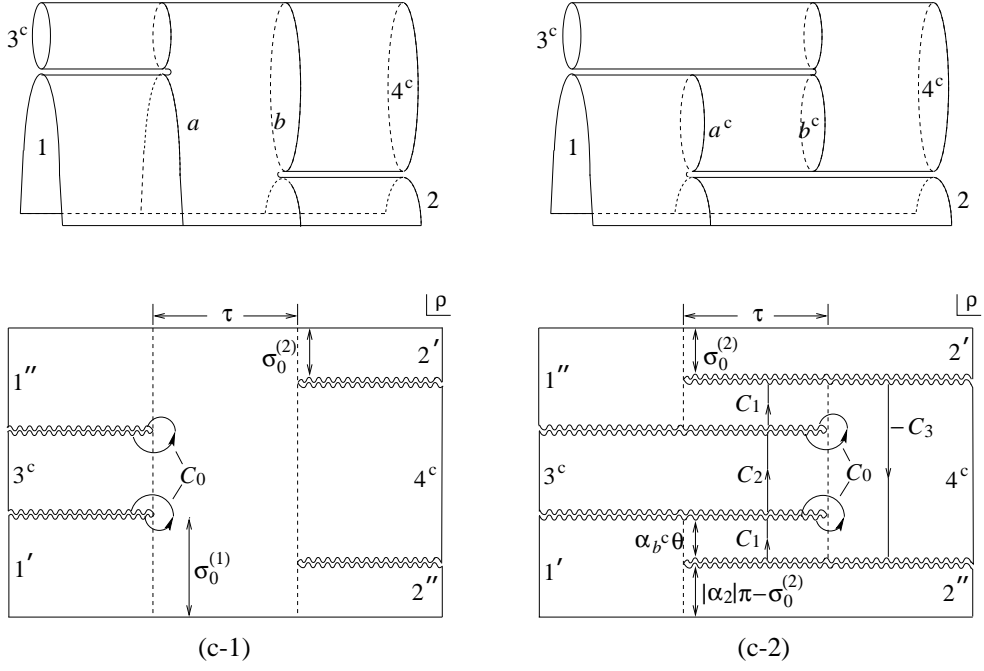}}
 \caption{$\rho$ planes for the (c-1) and (c-2) diagrams in 
Fig.~\protect\ref{fig:uomguomg-2}.}
 \label{fig:t14}
\end{figure}
%\end{wrapfigure}
\begin{equation}
\sbra{\tilde v(123^\c4^\c;\sigma_0^{(2)},\theta)}
=\sbra{\tilde v(123^\c4^\c;\sigma_0^{(2)},\sigma_0^{(1)})}
\qquad {\rm for}\quad \alpha_{b^\c}\theta+\abs{\alpha_2}\pi-\sigma_0^{(2)}=\sigma_0^{(1)}\,.
\end{equation}
So, in view of Eqs.~(\ref{eq:4.38}) and (\ref{eq:4.39}), 
we must again compare the anti-ghost factors 
$b_{\sht{\sigma_0^{(1)}}}$ and $\bzm{b}$ appearing there.
This is actually quite the same situation as encountered in (T9) case 
above. Indeed, if we compare the $\rho$ plane diagrams Fig.~\ref{fig:t14}
for the present case and Fig.~\ref{fig:t9} for the (T9) case, 
we can see an exact parallelism. Therefore, from 
Eqs.~(\ref{eq:bzerominus}) and (\ref{eq:bsigmazero}), we have the 
equality 
\begin{equation}
\bzm{b}= 
-\half\alpha_{b^\c}\left(b_{\rho_0}-b_{\rho_0^*}\right) = 
+\half i\alpha_{b^\c} b_{\sigma_0^{(1)}}\,.
\end{equation}
and we also see that the full region of (c-2)
with $0\leq\theta<2\pi$ 
corresponds to a part of the region of (c-1)
with $\sigma_0^{(1)}$, 
\begin{equation}
\abs{\alpha_2}\pi-\sigma_0^{(2)}\leq\sigma_0^{(1)}\leq\abs{\alpha_1}\pi-\sigma_0^{(2)}.
\label{eq:c1region}
\end{equation}
(The same is true also for the configuration (d): 
the full region of (d-2) with $0\leq\theta<2\pi$ 
corresponds to a part of (d-1) in 
$\abs{\alpha_2}\pi-\sigma_0^{(2)}\leq\sigma_0^{(1)}\leq\abs{\alpha_1}\pi-\sigma_0^{(2)}$.)\ \ 
Thus the two terms 
in these regions in Eq.~(\ref{eq:4.38}) cancel each other if
\begin{equation}
+\dd^2 \int^{\abs{\alpha_1}\pi-\sigma_0^{(2)}}_{\abs{\alpha_2}\pi-\sigma_0^{(2)}}d\sigma_0^{(1)}
=
-\frac12\dd\ff\int_0^{2\pi}{d\theta\over2\pi}\half i\alpha_{b^\c}
\end{equation}
holds. That is, since $\alpha_{b^\c}d\theta=d\sigma_0^{(1)}$, we find 
\begin{equation}
\dd = -i\ff{1\over8\pi}
\quad \Rightarrow\quad \ff = 8\pi i\dd,
\end{equation}
the same condition as Eq.~(\ref{eq:xcxomega}) obtained above.

What happens, then, to the configuration (c-1), or (d-1), 
if $\sigma_0^{(1)}$ goes outside 
the region of Eq.~(\ref{eq:c1region})? Consider the 
case (c-1) first. A little inspection 
of the diagram (c-1) in Fig.~\ref{fig:uomguomg-2} 
(or in Fig.~\ref{fig:t14}) shows 
that it yields the configurations (e-1) and (e-2) for the regions
\begin{eqnarray}
&&\hbox{(e-1):}
\qquad \sigma_0^{(1)}\leq\abs{\alpha_1}\pi-\sigma_0^{(2)}
\leq\sigma_0^{(1)}+ \abs{\alpha_3^\c}2\pi\nn
&&\quad \qquad \qquad \Rightarrow\quad 
(\abs{\alpha_1}-2\abs{\alpha_3^\c})\pi 
\leq\sigma_0^{(1)}+\sigma_0^{(2)}\leq\abs{\alpha_1}\pi\,,\nn
&&\hbox{(e-2):}\qquad 
\abs{\alpha_1}\pi-\sigma_0^{(1)}\leq\sigma_0^{(2)}
\leq\abs{\alpha_1}\pi-\sigma_0^{(1)}+ \abs{\alpha_3^\c}2\pi\nn
&&\quad \qquad \qquad \Rightarrow\quad 
\abs{\alpha_1}\pi\leq\sigma_0^{(1)}+\sigma_0^{(2)}
\leq(\abs{\alpha_1}+2\abs{\alpha_3^\c})\pi\,.
\label{eq:e1e2region}
\end{eqnarray}
and the configurations (b-1) and (b-2) for the rest regions 
$\sigma_0^{(1)}\leq(\abs{\alpha_1}-2\abs{\alpha_3^\c})\pi-\sigma_0^{(2)}$ and 
$\sigma_0^{(1)}\geq(\abs{\alpha_1}+2\abs{\alpha_3^\c})\pi-\sigma_0^{(2)}$.
Consideration of the configuration (d-1) shows that the whole region 
outside (\ref{eq:c1region}) yields (a-1) and (a-2). 
The configurations (a-1) and (a-2), as well as (b-1) and (b-2), 
have been shown to cancel with each other already in the above.

Let us now show that 
the two configurations (e-1) and (e-2) also cancel each other.
We already know from Eq.~(\ref{eq:e1e2region}) that the two 
configurations (e-1) and (e-2) come from a single term 
(\ref{eq:4.38}) in different regions of the two parameters 
$(\sigma_0^{(1)}, \ \sigma_0^{(2)})$. Moreover, from the diagrams (e-1) and 
(e-2) in Fig.~\ref{fig:uomguomg-2}, they clearly give the same glued 
configuration (e) when 
\begin{equation}
(\sigma_0^{(1)},\ \sigma_0^{(2)}) \ \ \hbox{in (e-1)}
\qquad \leftrightarrow \qquad 
(\abs{\alpha_2}\pi-\sigma_0^{(2)},\  \abs{\alpha_1}\pi-\sigma_0^{(1)}) \ \ \hbox{in (e-2)},
\end{equation} 
which indeed gives a one-to-one mapping between the two regions 
(\ref{eq:e1e2region}) for (e-1) and (e-2).
And thus the anti-ghost factors also have the correspondence 
\begin{equation}
(b_{\sigma_0^{(1)}},\ b_{\sigma_0^{(2)}}) \ \ \hbox{in (e-1)}
\qquad \leftrightarrow \qquad 
(-b_{\sigma_0^{(2)}},\  -b_{\sigma_0^{(1)}}) \ \ \hbox{in (e-2)},
\end{equation} 
where the minus signs come from the fact that the increasing directions of 
$\sigma_0^{(r)}$ are opposite between the two. Therefore, the 
anti-ghost factor $b_{\sht{\sigma_0^{(2)}}}b_{\sht{\sigma_0^{(1)}}}$ 
in Eq.~(\ref{eq:4.38}) is the same but has opposite order for the 
two configurations (e-1) and (e-2), so that they exactly cancel each other.

\subsubsection{T15 term}%{Eq.~(\ref{eq:4-7})}

\begin{figure}[tb]
   \epsfxsize= 13cm   %or \epsfysize= HEIGHT cm
   \centerline{\epsfbox{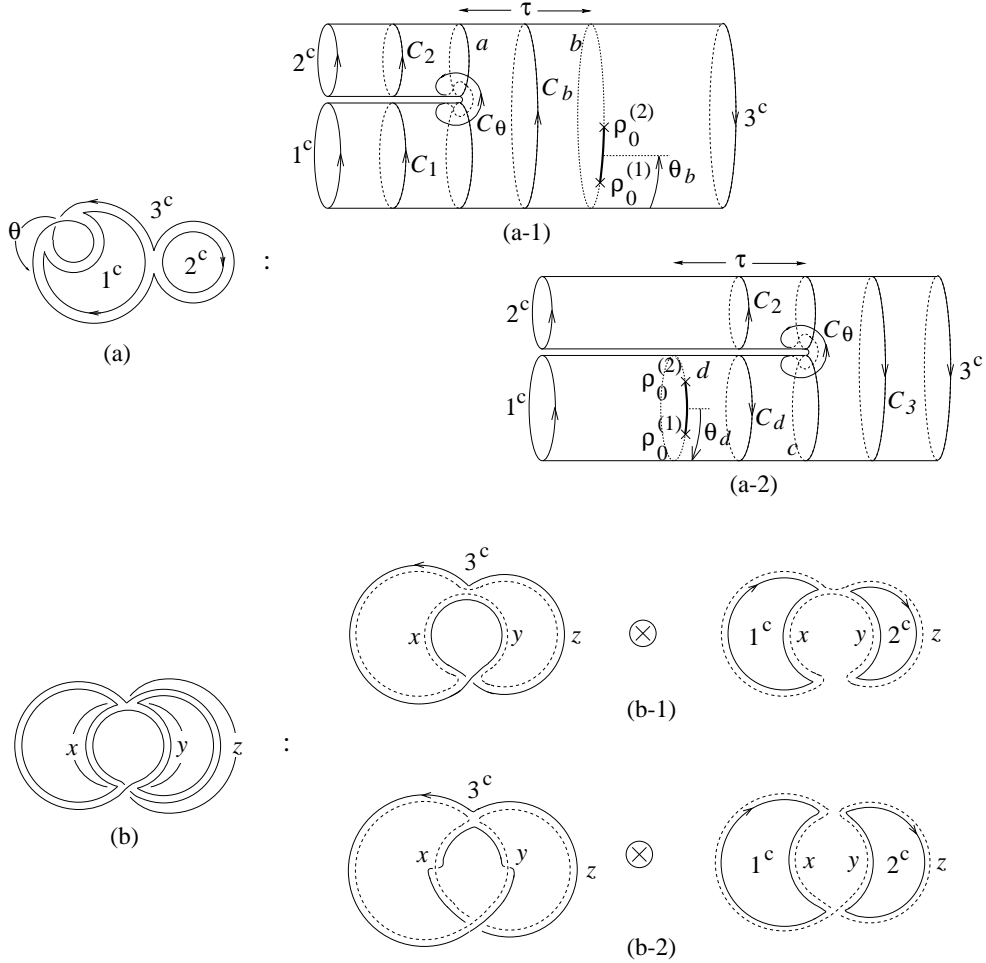}}
 \caption{Two configurations for (T15) term. The diagrams (b-1) and (b-2) 
are drawn by decomposing the process into the initial-to-intermediate and 
intermediate-to-final transition parts, for clarity.}
 \label{fig:vinfv3c-2}
\vspace{0.5cm}
\end{figure}

When contracting $\bra{V_3^\c}$ and $\bra{V_\infty}$, there appear two 
glued configurations, (a) and (b), as drawn in Fig.~\ref{fig:vinfv3c-2}. 
%\begin{wrapfigure}[n]{r}{6.6cm}
%\end{wrapfigure}
For the former configuration (a),
it is easy to see the cancellation between the two way gluing (a-1) 
and (a-2) giving a common configuration: they appear in the (T15) term 
(\ref{eq:4-6}) in the following forms, respectively:
\begin{eqnarray}
&&\Vthc{1^\c,2^\c,\check{a^\c}}\Vinf{b^\c,3^\c}\ketRc(ab)
            \ket{\Phi}_{3^\c2^\c1^\c} \nn
&&\qquad =\int{d\theta_b\over2\pi}\v3c(12ax)\vinf{b^\c,3^\c;\sigma_0}b_{\sigma_0}
       \bzm{b}e^{i\theta_b(L-\bar L)^{(b^\c)}}\bP{3} \nn
 &&\qquad \qquad \times\ketRc(ab)
            \prod_{r=1,2}\bP{r}\ket{\Phi}_{3^\c2^\c1^\c} \nn
&&\qquad =\int{d\theta_b\over2\pi}\bra{\tilde v(1^\c2^\c3^\c;\sigma_0,\theta_b)}
b_{\sigma_0}\bzm{b}\prod_{r=1,2,3}\bP{r}\ket{\Phi}_{3^\c2^\c1^\c} 
\label{eq:4.49}
\\
&&\Vthc{2^\c,3^\c,\check{c^\c}}\Vinf{d^\c,1^\c}\ketRc(cd)
            \ket{\Phi}_{1^\c3^\c2^\c} \nn
&&\qquad =\int{d\theta_d\over2\pi}\v3c(23cx)\vinf{d^\c,1^\c;\sigma_0}b_{\sigma_0}
\bzm{d}e^{i\theta_d(L-\bar L)^{(d^\c)}}\bP{1} \nn
&&\qquad \qquad \times\ketRc(cd)
            \prod_{r=2,3}\bP{r}\ket{\Phi}_{1^\c3^\c2^\c} \nn
&&\qquad =\int{d\theta_d\over2\pi}\bra{\tilde v(1^\c2^\c3^\c;\sigma_0,\theta_d)}
b_{\sigma_0}\bzm{d}
            \prod_{r=1,2,3}\bP{r}\ket{\Phi}_{3^\c2^\c1^\c}\,,
\label{eq:4.50}
\end{eqnarray}
where the LPP vertices for the glued configurations are defined by
\begin{equation}
\bra{\tilde v(1^\c2^\c3^\c;\sigma_0,\theta_b)}
\equiv\v3c(12ax)\vinf{b^\c,3^\c;\sigma_0}e^{i\theta_b(L-\bar L)^{(b^\c)}}\ketRc(ab)
\end{equation}
and similar one for $\bra{\tilde v(1^\c2^\c3^\c;\sigma_0,\theta_d)}$.
For the common configuration (a), the anti-ghost factor $b_{\sigma_0}$ 
coming from the $\bra{V_\infty}$ vertex is common between the two terms 
(\ref{eq:4.49}) and (\ref{eq:4.50}). So we have only to compare 
the anti-ghost factors $\bzm{b}$ and $\bzm{d}$. By the same method as used 
in the (T9) term around Eq.~(\ref{eq:bzerominus}), they can be replaced by 
\begin{eqnarray}
\bzm{b}\ &\Rightarrow&\  
\half{\alpha_{b^\c}}\left(\oint_{C_b+C_1+C_2}-{\rm a.h.}\right) 
{d\rho\over2\pi i}b(\rho)=+\half\abs{\alpha_{3^\c}}
\left(b_{\rho_0}-b_{\rho_0^*}\right) \,,\nn
\bzm{d}\ &\Rightarrow&\  
\half\alpha_{d^\c}\left(\oint_{C_d+C_2+C_3}-{\rm a.h.}\right) 
{d\rho\over2\pi i}b(\rho)=-\half\alpha_{1^\c}
\left(b_{\rho_0}-b_{\rho_0^*}\right) \,,
\end{eqnarray}
where the contours $C_b,\ C_d$ and $C_i$ ($i=1,2,3$) are drawn in 
Fig.\ref{fig:vinfv3c-2} and $b_{\rho_0}=\oint_{C_\theta}(d\rho/2\pi i)b(\rho)$ with 
the contour $C_\theta$ encircling the 3-closed-string interaction point
$\rho_0=i\alpha_1\pi$.
We have used the fact that $\alpha_{b^\c}=\abs{\alpha_3}>0$ and 
$\alpha_{d^\c}=-\alpha_1<0$ 
(or $\alpha_{b^\c}\cdot \alpha_{d^\c}<0$, more generally).  Because of this 
the anti-ghost factors $\bzm{b}$ and $\bzm{d}$, with integration measures 
$d(\abs{\alpha_3}\theta_b)$ and $d(\alpha_1\theta_d)$, respectively, are equal but have 
opposite signs. This reflects the fact that the intermediate closed 
strings in the two configurations (a-1) and (a-2) must be twisted 
to the opposite directions (by amounts $\abs{\alpha_3\theta_b}=\abs{\alpha_1\theta_d}$),
in order to keep the common glued configuration as seen in 
Fig.\ref{fig:vinfv3c-2}. Thus the two terms (a-1) and (a-2),
(\ref{eq:4.49}) and (\ref{eq:4.50}), cancel each other. 

Note however that this cancellation occurs between 
(a-1) in the restricted region 
\begin{equation}
-\alpha_1\pi+\sigma_0\leq\abs{\alpha_3}{\theta_b}\leq\alpha_1\pi-\sigma_0
\end{equation}
and (a-2) in the full region $-\pi\leq\theta_d<\pi$. 
If the twisting angle $\theta_b$ comes into the regions
\begin{eqnarray}
R_+&:& \qquad \alpha_1\pi-\sigma_0\leq\abs{\alpha_3}{\theta_b}\leq\alpha_1\pi+\sigma_0 \nn
R_-&:& \qquad -(\alpha_1\pi-\sigma_0)\geq\abs{\alpha_3}{\theta_b}\geq-(\alpha_1\pi+\sigma_0)
\end{eqnarray}
then the cross cap occupying the region 
Im$\rho\in[\,\abs{\alpha_3}{\theta_b}-\sigma_0,\,\abs{\alpha_3}{\theta_b}+\sigma_0\,]$ 
on the $\rho$ plane overlaps with the 3-closed-string 
interaction point $\rho_0=\pm i\alpha_1\pi$, and the resultant configurations become 
of the types (b-2) and (b-1) in Fig.~\ref{fig:vinfv3c-2}, respectively. 
One easily recognizes (b-1) to be the configuration in $R_-$ region,
but (b-2), at first sight, might not look like the configuration in the 
$R_+$ region. However, if 
one redraws the (b-2) diagram in Fig.~\ref{fig:vinfv3c-2} by exchanging 
the places of two handles $y$ and $z$, then the self-intersecting point 
of string 3 originally present at the bottom comes to the top in the 
diagram and can be recognized to be really the configuration in the $R_+$ 
region.

From the (b-1) (or, (a-1)) diagram in Fig.~\ref{fig:vinfv3c-2}, 
the lengths of the handles $x$ and $y$ of the (b-1) 
diagram in the region $R_-$ are found to be
\begin{equation}
\abs{x^-}=\alpha_1\pi+\sigma^-_0+\abs{\alpha_3}{\theta^-_b}, \qquad 
\abs{y^-}=-\abs{\alpha_3}{\theta^-_b}+\sigma^-_0-\alpha_1\pi\,,
\end{equation}
where we have put the superfix $-$ to $\theta_b$ and $\sigma_0$ in this 
$R^-$ case for distinction 
from the $R^+$ case below. Note that $\theta^-_b<0$ in this region $R_-$.
Similarly, taking account of the exchange of the $y$ and $z$ handles 
explained above, the lengths of the handles $x$ and $y$ of the (b-2) 
diagram in the region $R_+$ are found to be
\begin{equation}
\abs{x^+}=\alpha_1\pi+\sigma^+_0-\abs{\alpha_3}{\theta^+_b}, \qquad 
\abs{y^+}=2\alpha_2\pi-(\abs{\alpha_3}{\theta^+_b}-\alpha_1\pi+\sigma^+_0)\,.
\end{equation}
So, in order for the (b-1) and (b-2) give the same glued configuration, 
these lengths must coincide,
$\abs{x^-}=\abs{x^+}$ and $\abs{y^-}=\abs{y^+}$, from which we find the 
correspondence:  
\begin{eqnarray}
&&\rho_{0-}^{(1)}=\rho_{0+}^{(2)}-2i\abs{\alpha_3}\pi 
\qquad \rho_{0-}^{(2)}=-\rho_{0+}^{(1)}\,, \nn
&&\hbox{with}\qquad 
\rho^{(1)}_{0\pm}\equiv i(\abs{\alpha_3}{\theta^\pm_b}-\sigma^\pm_0), \qquad 
\rho^{(2)}_{0\pm}\equiv i(\abs{\alpha_3}{\theta^\pm_b}+\sigma^\pm_0) \,.
\label{eq:correspond}
\end{eqnarray}
Here $\rho_0^{(1)}=i(\abs{\alpha_3}{\theta_b}-\sigma_0)$ and 
$\rho_0^{(2)}=i(\abs{\alpha_3}{\theta_b}+\sigma_0)$ are the coordinates of the 
two end-points of the cross cap on the $\rho$ plane. 
However, we should note that the LPP vertices with these parameter sets 
$(\sigma^+_0, {\theta^+_b})$ and $(\sigma^-_0, {\theta^-_b})$ are {\it not} equal.
This is because the orientation of string $2^\c$ has been reversed in the 
above exchange process of the $x$ and $y$ handles, and so the 
precise relationship between the LPP vertices for these two 
configurations is given by
\begin{equation}
\bra{\tilde v(1^\c2^\c3^\c;\sigma^+_0,\theta^+_b)}=
\bra{\tilde v(1^\c2^\c3^\c;\sigma^-_0,\theta^-_b)}\Omega^{(2^c)}\,.
\label{eq:LPPrel}
\end{equation}
with the understanding that the parameters $(\sigma^+_0, {\theta^+_b})$ 
and $(\sigma^-_0, {\theta^-_b})$ are related with each other by 
Eq.~(\ref{eq:correspond}).

\begin{figure}[tb]
   \epsfxsize= 6cm   %or \epsfysize= HEIGHT cm
   \centerline{\epsfbox{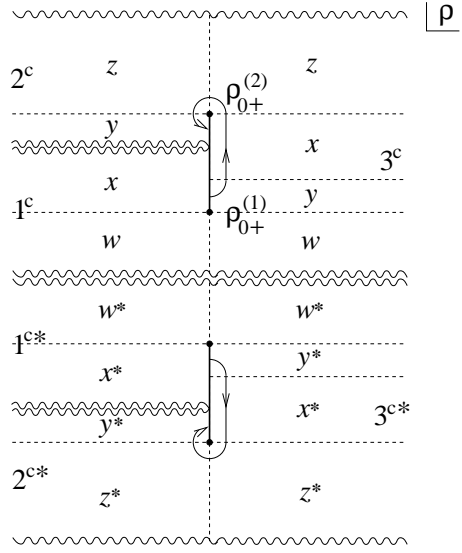}}
 \caption{Integration contour for $b_{\rho_{0+}^{(2)}}$ 
on the $\rho$ plane of $\bra{\tilde v(1^\c2^\c3^\c;\sigma^+_0,\theta^+_b)}$.}
 \label{fig:brho-1}
\end{figure}

\begin{figure}[bt]
   \epsfxsize= 13cm   %or \epsfysize= HEIGHT cm
   \centerline{\epsfbox{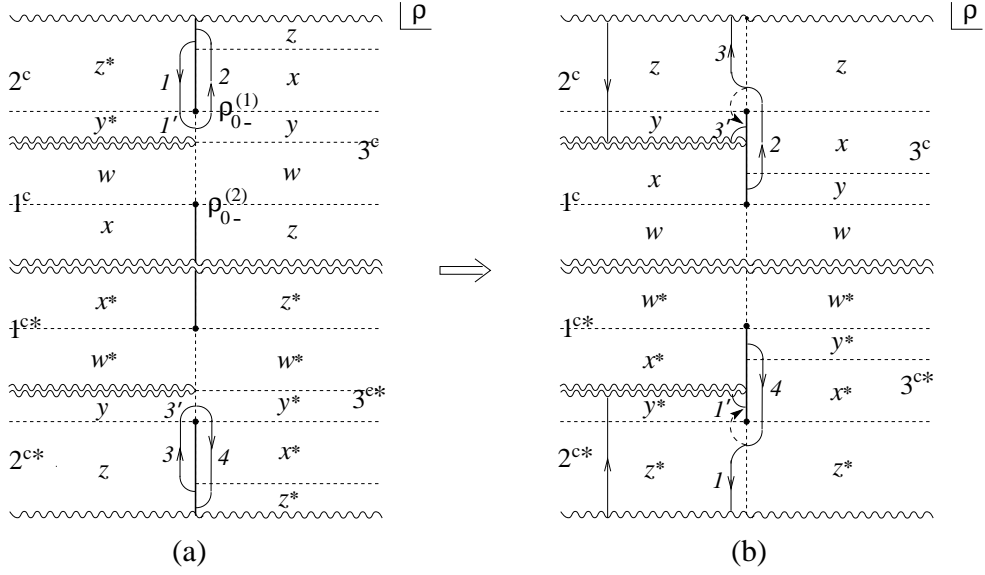}}
 \caption{(a) Integration contour for 
%\protect\raisebox{0pt}[0pt][1ex]{$b_{\rho_{0-}^{(1)}}$}
$b_{\rho_{0-}^{(1)}}$ 
on the $\rho$ plane of $\langle\tilde v(1^\c2^\c3^\c;\sigma^-_0,\theta^-_b)|$.
Going to (b), the $\rho$ plane is rearranged first by twisting the closed 
strings $1^\c$ and $3^\c$ by an amount $\abs{x}$ ($\sigma$ length of the $x$ 
region), and then by exchanging the regions $x{+}y \leftrightarrow 
y^*{+}x^*$ of string $2^\c$ (i.e., acting $\Omega^{(2^c)}$). The resultant 
plane (b) becomes the $\rho$ plane of 
$\langle\tilde v(1^\c2^\c3^\c;\sigma^-_0,\theta^-_b)|\,\Omega^{(2^c)}$. 
The integration contour for 
$b_{\rho_{0-}^{(1)}}$ on this plane 
is seen to coincide with that of $b_{\rho_{0+}^{(2)}}$ in 
Fig.~\protect\ref{fig:brho-1} by deformation after adding $\bzm{2}$. } 
\label{fig:brho-2}
\end{figure}

Since both (b-1) and (b-2) come from the same (a-1) term, we have now only 
to compare the relative sign of the  the anti-ghost factor 
$b_{\sigma_0}\bzm{b}$ in Eq.~(\ref{eq:4.49}) for the two cases of $R^{\pm}$ 
with the angle relations (\ref{eq:correspond}).
As was performed explicitly in the previous paper for the glued vertex 
$\sbra{\hat U}\bra{V_\infty}\ket{R^\c}$, the anti-ghost factor 
$b_{\sigma_0}\bzm{b}$ can be replaced by 
$b_{\sht{\rho_0^{(1)}}}b_{\sht{\rho_0^{(2)}}}$ up 
to an irrelevant proportionality factor (independent of $\theta$ and $\sigma_0$),
where $b_{\sht{\rho_0^{(1)}}}$ and $b_{\sht{\rho_0^{(2)}}}$ 
are the anti-ghost factors 
corresponding to the shifts of the two end-points 
$\rho_0^{(1)}$ and $\rho_0^{(2)}$ of the cross cap
(see the diagram (a-1) in Fig.~\ref{fig:vinfv3c-2}). 
This can be easily understood: $\bzm{b}$ is essentially the 
anti-ghost factor for the shift of $\theta_b$ and hence 
$\bzm{b}\propto(b_{\sht{\rho_0^{(1)}}}+b_{\sht{\rho_0^{(2)}}})$, 
and $b_{\sigma_0}$ is the anti-ghost factor for 
the shift of $\sigma_0$ and hence 
$b_{\sigma_0}\propto(b_{\sht{\rho_0^{(1)}}}-b_{\sht{\rho_0^{(2)}}})$. Thus the 
product gives $\propto b_{\sht{\rho_0^{(1)}}}b_{\sht{\rho_0^{(2)}}}$.
Coming back to the comparison of the anti-ghost factor 
$b_{\sigma_0}\bzm{b}$, from the angle relations (\ref{eq:correspond}), we 
are tempted to immediately write
\begin{equation}
\cases{
b_{\rho_{0-}^{(1)}}=+b_{\rho_{0+}^{(2)}}\cr
b_{\rho_{0-}^{(2)}}= -b_{\rho_{0+}^{(1)}} \cr
} 
\qquad \Rightarrow\qquad 
b_{\rho_{0-}^{(1)}}b_{\rho_{0-}^{(2)}} = +b_{\rho_{0+}^{(1)}}b_{\rho_{0+}^{(2)}}
\,. 
\label{eq:4.59}
\end{equation}
But these are not quite correct. This is because the $\rho$ planes for the 
two cases of $R^\pm$ equal with each other only under the 
twist operation $\Omega^{(2^c)}$, as noted in Eq.~(\ref{eq:LPPrel}). Therefore, 
the precise form of Eq.~(\ref{eq:4.59}) reads
\begin{equation}
\cases{
{\Omega^{(2^c)}}^\m1 b_{\rho_{0-}^{(1)}}{\Omega^{(2^c)}}=+b_{\rho_{0+}^{(2)}}\cr
{\Omega^{(2^c)}}^\m1 b_{\rho_{0-}^{(2)}}{\Omega^{(2^c)}} = -b_{\rho_{0+}^{(1)}} \cr
} 
\qquad \Rightarrow\qquad 
{\Omega^{(2^c)}}^\m1 b_{\rho_{0-}^{(1)}}b_{\rho_{0-}^{(2)}}{\Omega^{(2^c)}} 
= +b_{\rho_{0+}^{(1)}}b_{\rho_{0+}^{(2)}}
\,. 
\end{equation}
These hold in the presence of the factor $\prod_{r=1^\c,2^\c,3^\c}\bP{r}$.
Indeed, these relations can be directly confirmed by comparing the 
integration contours (of `8' shape across the cross cap cut) 
defining the anti-ghost factors $b_{\sht{\rho_{0\pm}^{(i)}}}$ 
($i=1,2$) on the two $\rho$ planes for $R^\pm$ cases. 
(See Figs.~\ref{fig:brho-1} and \ref{fig:brho-2} to confirm 
the equality ${\Omega^{(2^c)}}^\m1 b_{\sht{\rho_{0-}^{(1)}}}\Omega^{(2^c)}
= +b_{\sht{\rho_{0+}^{(2)}}}$, for instance.)
Hence, together with Eq.~(\ref{eq:LPPrel}), we obtain
%\begin{wrapfigure}{r}{\halftext}
%\end{wrapfigure}
%\begin{wrapfigure}[n]{r}{6.6cm}
%\end{wrapfigure}
\begin{equation}
\bra{\tilde v(1^\c2^\c3^\c;\sigma^+_0,\theta^+_b)}
b_{\rho_{0+}^{(1)}}b_{\rho_{0+}^{(2)}}
= +\bra{\tilde v(1^\c2^\c3^\c;\sigma^-_0,\theta^-_b)}
b_{\rho_{0-}^{(1)}}b_{\rho_{0-}^{(2)}}\Omega^{(2^c)}
\end{equation}
There appears no relative minus sign unlike the cases up to 
here. However, we should note that $b_0^{(2)}$ is odd under $\Omega^{(2^c)}$, 
i.e., ${\Omega^{(2^c)}}^\m1b_0^{(2)}\Omega^{(2^c)}=-b_0^{(2)}$, so that we have 
a relative minus sign:
\begin{eqnarray}
&&\bra{\tilde v(1^\c2^\c3^\c;\sigma^+_0,\theta^+_b)}
b_{\rho_{0+}^{(1)}}b_{\rho_{0+}^{(2)}}
\prod_{r=1,2,3}\bP{r} \nn
&& \qquad \quad = -\bra{\tilde v(1^\c2^\c3^\c;\sigma^-_0,\theta^-_b)}
b_{\rho_{0-}^{(1)}}b_{\rho_{0-}^{(2)}}\prod_{r=1,2,3}\bP{r}\Omega^{(2^c)}.
\end{eqnarray}
The $\Omega^{(2^c)}$ on the right-hand side disappears in the actual vertex 
since unoriented projection operators $\Pi^{(r)}=(1+\Omega^{(r)})/2$ 
are acting on each external string. We thus have shown the cancellation 
of (b-1) and (b-2) terms, % between $R^+$ and $R^-$ regions, 
finishing the proof for the complete cancellation of (T15) terms.

\subsubsection{T16 terms}%{Eq.~(\ref{eq:4-7})}

\begin{figure}[tb]
   \epsfxsize= 14cm   %or \epsfysize= HEIGHT cm
   \centerline{\epsfbox{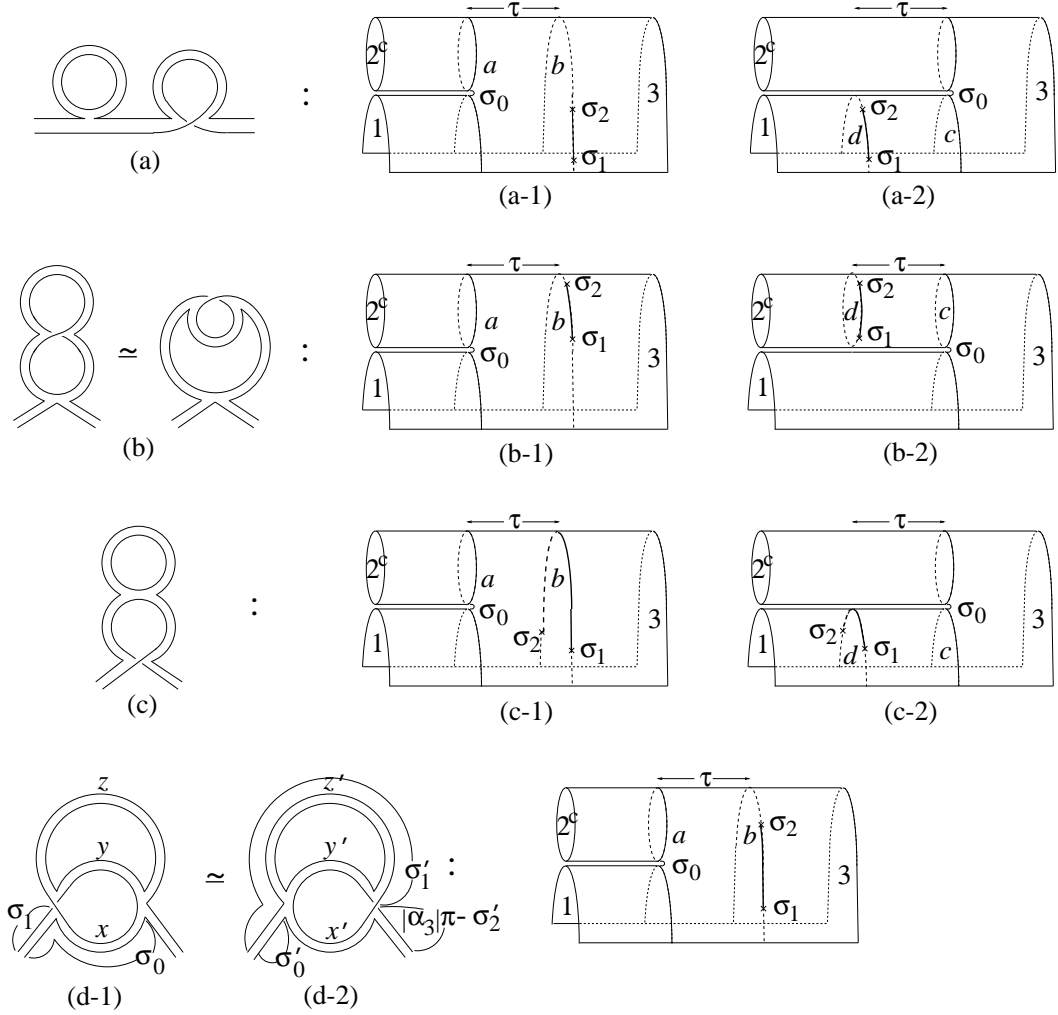}}
 \caption{Four configurations for (T16) terms. For brevity, 
$\sigma_0$, $\sigma_1$ and $\sigma_2$ on the $\rho$ planes, 
denote $\sigma_0^{(1)}$, $\sigma_1^{(b)}$ and $\sigma_2^{(b)}$ for (a-1), (b-1), 
(c-1) and (d) diagrams, 
and $\sigma_0^{(3)}$, $\sigma_1^{(d)}$ and $\sigma_2^{(d)}$ for (a-2), (b-2) 
and (c-2) diagrams, respectively.}
 \label{fig:t16}
\end{figure}

The generic configurations resultant from the contraction of 
the two vertices $\bra{U_\Omega}$ and $\bra{V_\propto}$, or 
$\bra{U_\Omega}$ and $\bra{V_\infty}$, fall into 
four types, (a), (b) (c) and (d), depicted in Fig.~\ref{fig:t16}.
%\begin{wrapfigure}[n]{r}{6.6cm}
%\end{wrapfigure}
Only (b-2) is given by gluing $\bra{U_\Omega}$ and $\bra{V_\infty}$ and all the 
others are by gluing $\bra{U_\Omega}$ and $\bra{V_\propto}$.
As always, cancellations occur between the two 
ways of gluing for a given type configuration. 

The type (a) configuration 
is realized by (a-1) diagram at $\tau=0$ in Fig.~\ref{fig:t16} 
in restricted regions with 
$0\leq\sigma_1^{(b)}\leq\sigma_2^{(b)}\leq\sigma_0^{(1)}$ or 
$2\alpha_{2^\c}\pi+\sigma_0^{(1)}\leq\sigma_1^{(b)}\leq\sigma_2^{(b)}\leq\abs{\alpha_{3^\c}}\pi$, and 
by (a-2) diagram in the region satisfying
$0\leq\sigma_1^{(d)}\leq\sigma_2^{(d)}\leq\sigma_0^{(3)}$ or
$\sigma_0^{(3)}\leq\sigma_1^{(d)}\leq\sigma_2^{(d)}\leq\alpha_1\pi$.
The terms (a-1) and (a-2) are contained in the first term in (T16) 
in the forms
\begin{eqnarray}
&& \Uomg{1,a,2^\c}\V\propto(b3xx)\ketRo(ab)\ket{\Phi}_{2^\c}\ket{\Psi}_{31} \nn
&& \qquad =\int d\sigma_0^{(1)}d\sigma_1^{(b)}d\sigma_2^{(b)}
\uomg{1,a,2^\c;\sigma_0^{(1)}}b_{\sigma_0^{(1)}}\bP{2} \nn
&&\qquad \qquad \qquad \times\v\propto(b3{\sigma_1^{(b)}}{\sigma_2^{(b)}}x)
b_{\sigma_1^{(b)}}b_{\sigma_2^{(b)}}\ketRo(ab)\ket{\Phi}_{2^\c}\ket{\Psi}_{31} 
\label{eq:t16a1}
\\
&& \Uomg{3,c,2^\c}\V\propto(d1xx)\ketRo(cd)\ket{\Phi}_{2^\c}\ket{\Psi}_{13} \nn
&& \qquad =\int d\sigma_0^{(3)}d\sigma_1^{(d)}d\sigma_2^{(d)}
\uomg{3,c,2^\c;\sigma_0^{(3)}}b_{\sigma_0^{(3)}}\bP{2} \nn
&&\qquad \qquad \qquad \times 
\v\propto(d1{\sigma_1^{(d)}}{\sigma_2^{(d)}}x)b_{\sigma_1^{(d)}}b_{\sigma_2^{(d)}}
\ketRo(cd)\ket{\Phi}_{2^\c}\ket{\Psi}_{13} 
\label{eq:t16a2}
\end{eqnarray}
From the diagrams (a-1) and (a-2) at $\tau=0$ in Fig.~\ref{fig:t16}, 
we see that 
the increasing directions of $\sigma$ are opposite for strings $b$ and 
$d$, and also for strings 1 and 3, so that we have 
$b_{\sht{\sigma_1^{(d)}}}=-b_{\sht{\sigma_2^{(b)}}}$, 
$b_{\sht{\sigma_2^{(d)}}}=-b_{\sht{\sigma_1^{(b)}}}$ and 
$b_{\sht{\sigma_0^{(3)}}}=-b_{\sht{\sigma_0^{(1)}}}$. Thus the products of 
anti-ghost factors have the same sign, 
$b_{\sht{\sigma_0^{(1)}}}b_{\sht{\sigma_1^{(b)}}}b_{\sht{\sigma_2^{(b)}}}
=b_{\sht{\sigma_0^{(3)}}}b_{\sht{\sigma_1^{(d)}}}b_{\sht{\sigma_2^{(d)}}}$, 
but the states have opposite 
signs $\ket{\Psi}_{13}=-\ket{\Psi}_{31}$, and hence the (a-1) and (a-2) terms   
cancel each other.

Next consider the type (b) configuration. The 
(b-1) diagram corresponds to $\bra{U_\Omega}\bra{V_\propto}\ket{R^\o}$ and 
(b-2) to $\bra{U_\Omega}\bra{V_\infty}\ket{R^\c}$. 
The former (b-1) is just the (a-1) diagram 
in the region with 
$\sigma_0^{(1)}\leq\sigma_1^{(b)}\leq\sigma_2^{(b)}\leq\sigma_0^{(1)}+2\alpha_{2^\c}\pi$. 
In this region, the presence of string 1 plays no 
important role. If we forget string 1, then the vertex 
$\bra{U_\Omega}$ reduces to $\bra{U}$, and, in fact the diagrams (b-1) and 
(b-2) are the same as those we encountered in the previous paper I for 
proving the cancellation between $\bra{U}\bra{V_\propto}\ket{R^\o}$ and 
$\bra{U}\bra{V_\infty}\ket{R^\c}$, that is (T10) in Eq.~(\ref{eq:3-5}). 
Therefore, the same calculation as there proves the 
cancellation of (b-1) and (b-2) terms here 
when the coupling relation 
\begin{equation}
\cc=4\pi i\bb
\label{eq:couplingrel}
\end{equation}
is satisfied. One may wonder the discrepancy of the relative weights of 
the coefficients, $-2x_ux_\propto: 2x_ux_\infty$ in Eq.~(\ref{eq:3-5}) and 
the present one $2x_\Omega x_\propto: -x_\Omega x_\infty$ in Eq.~(\ref{eq:4-7}). However, 
from the second term $\Uomg{1,3,\check{a^\c}}\V\infty(b2xx)\ketRc(ab)$ in the 
latter, the specific diagram (b-2) with a definite set of string lengths  
appear twice since the term with 3 and 1 exchanged in 
$\Uomg{1,3,\check{a^\c}}$ also give the same diagram. So the actual 
relative weight of the present case is also 
$-2x_\propto: 2x_\infty=-x_\propto: x_\infty$, thus leading 
to the same coupling relation (\ref{eq:couplingrel}) as before.

Next is the type (c) configuration, which is realized in two 
ways of gluing (c-1) and (c-2) at $\tau=0$ in Fig.~\ref{fig:t16}. 
They are nothing but (a-1) and (a-2) terms with moduli 
parameters in different regions, respectively, so can be described by 
the same equations as (\ref{eq:t16a1}) and (\ref{eq:t16a2}): 
with the LPP vertices for the glued configurations
\begin{eqnarray}
\sbra{\tilde v(132^\c;\sigma_0^{(1)},\sigma_1^{(b)},\sigma_2^{(b)})}
&\equiv&\sbra{u_\Omega(1a2^\c;\sigma_0^{(1)})}
\sbra{v_\propto(b3;\sigma_1^{(b)},\sigma_2^{(b)})}\ket{R^\o(ab)}\nn
\sbra{\tilde v(132^\c;\sigma_0^{(3)},\sigma_1^{(d)},\sigma_2^{(d)})}
&\equiv&
\sbra{u_\Omega(3c2^\c;\sigma_0^{(3)})}
\sbra{v_\propto(d1;{\sigma_1^{(d)}},{\sigma_2^{(d)}})}\ket{R^\o(cd)},
\hspace{3em}
\end{eqnarray} 
(c-1) and (c-2) appear in the forms
\begin{eqnarray}
&& \hbox{(c-1)}=\int\!d\sigma_0^{(1)}d\sigma_1^{(b)}d\sigma_2^{(b)}
\sbra{\tilde v(132^\c;\sigma_0^{(1)},\sigma_1^{(b)},\sigma_2^{(b)})}
b_{\sigma_1^{(b)}}b_{\sigma_2^{(b)}}b_{\sigma_0^{(1)}}\bP{2}
\label{eq:t16c1}
\\
&& \hbox{(c-2)}=-\!\int\!d\sigma_0^{(3)}d\sigma_1^{(d)}d\sigma_2^{(d)}
\sbra{\tilde v(132^\c;\sigma_0^{(3)},\sigma_1^{(d)},\sigma_2^{(d)})}
b_{\sigma_1^{(d)}}b_{\sigma_2^{(d)}}b_{\sigma_0^{(3)}}\bP{2}\!,
\hspace{3em}
\label{eq:t16c2}
\end{eqnarray}
where the common external states $\ket{\Phi}_{2^\c}\ket{\Psi}_{31}$ are 
omitted. 
The minus sign of (c-2) has come from $\ket{\Psi}_{13}=-\ket{\Psi}_{31}$.
So we have only to 
compare the anti-ghost factors
$b_{\sht{\sigma_1^{(b)}}}b_{\sht{\sigma_2^{(b)}}}b_{\sht{\sigma_0^{(1)}}}$ and 
$b_{\sht{\sigma_1^{(d)}}}b_{\sht{\sigma_2^{(d)}}}b_{\sht{\sigma_0^{(3)}}}$. 
Note that, since strings 3 and $d$ carry negative string lengths, 
$\sigma_1^{(d)},\ \sigma_2^{(d)}$ and $\sigma_0^{(3)}$ represent distances 
on the $\rho$ plane measured from the opposite edge of the open string.
Then the condition for these two diagrams (c-1) and (c-2) to
reduce to a common glued configuration at $\tau=0$ is that the positions of 
those interaction points coincide:
\begin{equation}
\sigma_1^{(b)}=\alpha_1\pi-\sigma_2^{(d)}, \quad \ 
\abs{\alpha_3}\pi-\sigma_2^{(b)} = \sigma_1^{(d)}, \quad\ 
\sigma_1^{(b)}+\sigma_2^{(b)}-\sigma_0^{(1)}=\abs{\alpha_3}\pi-\sigma_0^{(3)}
\label{eq:correspond2}
\end{equation}
The left-hand side of the third condition means that 
the interaction point $\sigma_0^{(1)}$ appears 
at the place $\sigma_1^{(b)}+\sigma_2^{(b)}-\sigma_0^{(1)}$ when crossing the cross 
cap cut [$\sigma_1^{(b)},\ \sigma_2^{(b)}$] on the (c-1) diagram. 
As is clear from the diagrams, however, 
the configurations of (c-1) and (c-2) at $\tau=0$ are not quite equal
to each other as they stand, 
since the whole region of closed string $2^\c$ in (c-1) case 
is wrapped by the cross cap cut. Therefore the glued vertices 
coincide with each other only when the twist operator $\Omega^{(2^\c)}$ 
acts on the (c-2) glued vertex:
\begin{eqnarray}
\bra{\tilde v(1,3,2^\c;\sigma_0^{(1)},\sigma_1^{(b)},\sigma_2^{(b)})}
=\bra{\tilde v(1,3,2^\c;\sigma_0^{(3)},\sigma_1^{(d)},\sigma_2^{(d)})}\Omega^{(2^\c)}
\end{eqnarray} 
Similarly to the previous (T15) [type (b)] case,
we can see from Eq.~(\ref{eq:correspond2}) the relation
\begin{equation}
b_{\sigma_1^{(b)}}b_{\sigma_2^{(b)}}b_{\sigma_0^{(1)}}
={\Omega^{(2^c)}}^\m1 (-b_{\sigma_2^{(d)}})(-b_{\sigma_1^{(d)}})b_{\sigma_0^{(3)}}{\Omega^{(2^c)}}
=-{\Omega^{(2^c)}}^\m1 b_{\sigma_1^{(d)}}b_{\sigma_2^{(d)}}b_{\sigma_0^{(3)}}{\Omega^{(2^c)}}
\end{equation}
in the presence of $\bP{2}$, and hence obtain, noting also that $\bzm{2}$ 
is odd under ${\Omega^{(2^c)}}$,
\begin{eqnarray}
&&\bra{\tilde v(1,3,2^\c;\sigma_0^{(1)},\sigma_1^{(b)},\sigma_2^{(b)})}
b_{\sigma_1^{(b)}}b_{\sigma_2^{(b)}}b_{\sigma_0^{(1)}}\bzm{2} \nn
&&\qquad =-\bra{\tilde v(1,3,2^\c;\sigma_0^{(3)},\sigma_1^{(d)},\sigma_2^{(d)})}
b_{\sigma_1^{(d)}}b_{\sigma_2^{(d)}}b_{\sigma_0^{(3)}}{\Omega^{(2^c)}}\bzm{2} \nn
&&\qquad =+\bra{\tilde v(1,3,2^\c;\sigma_0^{(3)},\sigma_1^{(d)},\sigma_2^{(d)})}
b_{\sigma_1^{(d)}}b_{\sigma_2^{(d)}}b_{\sigma_0^{(3)}}\bzm{2}{\Omega^{(2^c)}}.
\end{eqnarray} 
This implies that the (c-1) and (c-2) terms cancel each other.

Finally consider the type (d) configuration, which
corresponds to (a-1) term with the moduli parameters in the region
\begin{eqnarray}
R&:& \qquad \sigma_1^{(b)}\leq\sigma_0^{(1)}\leq\sigma_2^{(b)} \nn
R'&:& \qquad \abs{\alpha_3}\pi-\sigma_2^{(b)}\leq\sigma_0^{(1)}\leq\abs{\alpha_3}\pi-\sigma_1^{(b)}
\end{eqnarray}
This case is more similar to the (T15) type (b) case. The cancellation 
occurs between the configurations in these two regions $R$ and $R'$.
These two configurations can be seen to give the same pattern 
of gluing; indeed, if we redraw the (d-2) diagram in the region $R'$ by
exchanging the two handles $y'$ and $z'$, then it can be recognized as
possessing the same pattern as the (d-1) diagram in the region $R$. It is
also the same as before that the orientation of the closed string $2^\c$ is 
reversed in this exchanging process of $y'$ and $z'$. Therefore, we find
the conditions for these two to give a common configuration:
\begin{eqnarray}
\hbox{the left leg}&:&\qquad 
\sigma_1 = \sigma'_0\,,\nn
\hbox{the right leg}&:&\qquad 
{\alpha_1}\pi-\sigma_0 = \abs{\alpha_3}\pi-\sigma'_2\,,\nn
\hbox{length of $y$ and $z'$}&:&\qquad 
\sigma_2-\sigma_0 = \sigma'_1-\sigma'_0 \,,
\end{eqnarray}
where we have omitted the superscripts $(b)$ and $(1)$ and 
put prime to denote the parameters in the $R'$ region to make distinction 
from those in $R$. 
The glued vertices with these corresponding parameter sets coincide 
with each other when the twist operator ${\Omega^{(2^c)}}$ is acting:
\begin{eqnarray}
\bra{\tilde v(1,3,2^\c;\sigma_0,\sigma_1,\sigma_2)}
=\bra{\tilde v(1,3,2^\c;\sigma'_0,\sigma'_1,\sigma'_2)}\Omega^{(2^\c)}.
\end{eqnarray} 
Noting the presence of the twist operation, we have the correspondence 
of the anti-ghost factors:
\begin{equation}
b_{\sigma_1}b_{\sigma_2}b_{\sigma_0}
={\Omega^{(2^c)}}^\m1 (b_{\sigma'_0})(b_{\sigma'_1})(b_{\sigma'_2}){\Omega^{(2^c)}}
=+{\Omega^{(2^c)}}^\m1 b_{\sigma'_1}b_{\sigma'_2}b_{\sigma'_0}{\Omega^{(2^c)}}
\end{equation}
which is again valid in the presence of $\bP{2}$ factor. 
We thus find
\begin{eqnarray}
&&\bra{\tilde v(1,3,2^\c;\sigma_0,\sigma_1,\sigma_2)}
b_{\sigma_1}b_{\sigma_2}b_{\sigma_0}\bzm{2} \nn
&&\qquad =\bra{\tilde v(1,3,2^\c;\sigma'_0,\sigma'_1,\sigma'_2)}
b_{\sigma'_1}b_{\sigma'_2}b_{\sigma'_0}{\Omega^{(2^c)}}\bzm{2} \nn
&&\qquad =-\bra{\tilde v(1,3,2^\c;\sigma'_0,\sigma'_1,\sigma'_2)}
b_{\sigma'_1}b_{\sigma'_2}b_{\sigma'_0}\bzm{2}{\Omega^{(2^c)}} \ ,
\end{eqnarray} 
implying that (d) type configurations also cancel between $R$ and $R'$ 
parameter regions. This finishes the proof for (T16) case.

\section{Summary}

We have presented the full SFT action (\ref{eq:action}) 
for the unoriented open-closed mixed system and determined the BRS/gauge 
transformation laws for the open and closed string fields.
We have shown that the action (\ref{eq:action}) indeed satisfies the BRS 
invariance at the `tree' level; namely, all the terms (T1) --- (T16) 
vanish provided that the coupling constants satisfy the relations 
(\ref{eq:rel-1}) -- (\ref{eq:rel-5}).
Also for the other remaining terms (L1) --- (L5) we have identified 
which one loop diagrams they are expected to cancel.

The task to show that those loop diagrams are indeed anomalous and 
the terms (L1) --- (L5) really cancel them, are left to the forthcoming 
paper. Because of this, two of the coupling constants, say 
$x_u$ and $x_\infty=-nx_u^2$, are still left as free parameters at this stage. 
We will show that these are indeed determined by the requirements of 
anomaly cancellations in the next paper. In particular, this 
determines that the gauge group SO($n$) must be SO($2^{13}$) in this 
bosonic unoriented theory case.\cite{rf:DougGrin,rf:Weinberg}

\section*{Acknowledgements}
The authors would like to express their sincere thanks to 
H.\ Hata, H.\ Itoyama, M.\ Kato, Y.\ Kazama, K.\ Kikkawa, N.\ Ohta, 
M.\ Maeno, S.\ Sawada, K.\ Suehiro, Y.\ Watabiki and T.\ Yoneya for 
valuable and helpful discussions. They also acknowledge hospitality at 
the Summer Institute Kyoto '97. T.~K.\ and T.~T.\ are supported in part 
by the Grant-in-Aid for Scientific Research (\#10640261) and the 
Grant-in-Aid (\#6844), respectively, from the Ministry of Education, 
Science, Sports and Culture.

\appendix
\section{BRS and gauge transformations}

We here summarize the general rule for obtaining the BRS and gauge 
transformation laws from the action with a precise treatment of the 
statistics of the fields.
\cite{rf:Hata1,rf:Hata2,rf:Zwie1,rf:HataZwie}
The BRS invariance of the action implies 
what is called BV master equation\cite{rf:BV}
and automatically means the 
gauge invariance of the action. 

\subsection{Notations and differentiation rules}
We introduce the notation $\Phi_I$ and $\Phi^I$, denoting the open and 
closed field unifiedly:
\begin{equation}
\left.\matrix{\ket\Phi\cr \ket\Psi\cr}\right\} \LRarrow \Phi_I \,,
\qquad \left.\matrix{\bra\Phi\cr \bra\Psi\cr}\right\} \LRarrow \Phi^I 
\end{equation}
As a convention, we take 
SL(2;C) ket vacuum $\ket{0}$ Grassmann {\it even} and so 
ket Fock vacuum $\ket{1}$ Grassmann {\it odd}. Then, SL(2;C) bra vacuum 
$\bra{0}$ must be Grassmann {\it odd} and 
the bra Fock vacuum be Grassmann {\it even} as is enforced by
\begin{equation}
\bra{0}c_{-1}c_0c_1\ket{0}=\bra{1}c_0\ket{1}=1.
\end{equation}
Taking this into account we have the following 
Grassmann even-odd property: we cite here the statistics indices 
also for the quantities appearing below.
\begin{eqnarray}
\Phi_I, \ {\delta\over\delta\Phi_I} \quad  &:& \qquad  
1\ \hbox{(odd) always odd independently of open or closed}\nn
\Phi^I, \ {\delta\over\delta\Phi^I} \quad  &:& \qquad  
I \equiv\cases { 0\ \hbox{(even)} & if $\Phi_I$ is open \cr
            1\ \hbox{(odd)} & if $\Phi_I$ is closed \cr} \nn
R_{IJ},\ R^{IJ} \quad &:& \qquad  I+1 = J+1 \quad \hbox{since no open-closed transition} \nn
\delta^I_{\ \,J},\ \delta_I^{\ \,J}\quad  &:& \qquad  0\ \hbox{(even)}
\end{eqnarray}

Introduce a metric $R_{IJ}$ and $R^{IJ}$ for lowering and raising the 
indices, which are the same as the reflector:
\begin{eqnarray}
\left.\matrix{
\bidx{1}\bra{\Phi} &=& \bra{R^\c(1,2)}\ket{\Phi}_2 \cr
\bidx{1}\bra{\Psi} &=& \bra{R^\o(1,2)}\ket{\Psi}_2 \cr}
\right\}&\LRarrow&  
\Phi^I =  R^{IJ}\Phi_J
  \nn
\noalign{\vskip 1ex}
\left.\matrix{
\ket{\Phi}_1 &=& \bidx{2}\bra{\Phi}\ket{R^\c(2,1)} \cr
\ket{\Psi}_1 &=& \bidx{2}\bra{\Psi}\ket{R^\o(2,1)} \cr}
\right\}&\LRarrow&  
\Phi_I =  \Phi^JR_{JI},
\label{eq:Metric}
\end{eqnarray}
where the (pair of upper and lower) repeated indices imply the contractions 
(or summations). Note that $R_{IJ}$ and $R^{IJ}$ have only open-open and 
closed-closed diagonal components.
We have the property of the reflector (or metric):
\begin{eqnarray}
R_{IJ} = (-)^{I+1}R_{JI} \quad &:& \quad 
\left\{\matrix{\hbox{anti-symmetric for open}\cr
       \hbox{symmetric for closed} \cr}\right. \nn
R^{IJ} = R^{JI} \quad &:& \qquad \hbox{symmetric}
\end{eqnarray}
and satisfies
\begin{eqnarray}
 R^{IJ}R_{JK} = \delta^I_{\ \,K} = (-)^{I+1}\delta_K^{\ \ I}.
\end{eqnarray}
A care should be taken of the order of the indices of the 
Kronecker deltas $\delta_K^{\ \ I}$ and $\delta^I_{\ K}$, in particular, 
for the {\it open string} case, for which $(-)^{I+1}$ is negative. 
The Kronecker deltas $\delta_K^{\ \ I}$ and $\delta^I_{\ K}$ are defined by the 
following property:
\begin{eqnarray}
&&\delta^{\ \,J}_{I}\, \Phi_J = \Phi_I , \quad {\rm but} \quad 
\delta^{J}_{\ \,I}\,\Phi_J = (-)^{I+1}\Phi_I, \nn
&&\Phi^{J} \delta^{\ \,I}_{J} = \Phi^I , \quad {\rm but} \quad 
\Phi^J\delta^{I}_{\ \,J} = (-)^{I+1}\Phi^I. 
\end{eqnarray}

For making it easy to translate into the bra-ket notation, 
we take always the convention:
\begin{eqnarray}
{\delta\over\delta\Phi_I}\  \sim\ {\delta\over\delta\ket{\Phi}_1}
\quad  &:& \quad \hbox{differentiation from {\it Right} } \nn
{\delta\over\delta\Phi^I}\ \sim\ {\delta\over\delta\bidx{1}\bra{\Phi}}
\quad  &:& \quad \hbox{differentiation from {\it Left}}
\end{eqnarray}
We have the following rule:
\begin{eqnarray}
&&%{\delta\ket{\Phi}_1\over\delta\ket{\Phi}_2}= \delta_1^{\ 2} &\LRarrow & 
{\delta\Phi_I\over\delta\Phi_J}=\delta_I^{\ J},
\qquad \quad {\delta\Phi^I\over\delta\Phi_J}=R^{IJ} \nn 
&&{\delta\Phi^I\over\delta\Phi^J}=\delta_J^{\ \,I},
\qquad \quad {\delta\Phi_I\over\delta\Phi^J}= R_{JI} 
\end{eqnarray}
It should be noted that, as seen from ${\delta\Phi_I/\delta\Phi_J}\propto\delta_I^{\ J}$ and 
${(\delta/\delta\Phi^J)\Phi^I}\propto\delta_J^{\ \,I}$,  
the derivatives ${\delta/\delta\Phi}$ have the same Grassmann even-odd properties as
$\Phi$ in the denominator: that is, ${\delta/\delta\Phi_I}$ is always odd and 
${\delta/\delta\Phi^I}$ is $I$ (even for open and odd for closed).

Note: if we use the notation
\begin{equation}
{\delta F\over\delta\Phi_I} \equiv F^I, \qquad  
{\delta F\over\delta\Phi^I} \equiv F_I,  
\end{equation}
then, for any (Grassmann even) derivation $\delta$, we have
\begin{eqnarray}
\delta F = F^I\delta\Phi_I , \qquad \delta F = \delta\Phi^J F_J.
\label{eq:DiffIdentity}
\end{eqnarray}
For (Grassmann odd) anti-derivation $\mib\delta_A$, 
we can convert $\mib\delta_A$ into the usual derivation $\lambda\mib\delta_A$ by 
multiplying a Grassmann odd constant $\lambda$ and then we have 
\begin{equation}
\lambda\mib\delta_A F =   F^I\lambda\mib\delta_A\Phi_I = \lambda(-)^{(F+1)} F^I\mib\delta_A\Phi_I, 
\qquad \lambda\mib\delta_AF = \lambda\mib\delta_A\Phi^J F_J 
\end{equation}
from which follows
\begin{equation}
\mib\delta_A F = (-)^{(F+1)} F^I\mib\delta_A\Phi_I, \qquad 
\mib\delta_AF = \mib\delta_A\Phi^J F_J .
\label{eq:AntiDiffIdentity}
\end{equation}

From this Eq.~(\ref{eq:DiffIdentity}), 
we can also derive an identity which gives a relation 
between $F_I$ and $F^I$: from Eq.~(\ref{eq:Metric})
\begin{eqnarray}
\delta\Phi^J = R^{JI} \delta\Phi_I, \qquad 
\delta\Phi_I =  \delta\Phi^J R_{JI}.
\end{eqnarray}
Substituting this into
\begin{equation}
F^I\delta\Phi_I = \delta\Phi^J F_J,
\end{equation}
we have, noting $\abs{F^I}=F+1$ and $\abs{F_J}=F+J$,
\begin{eqnarray}
F^I\delta\Phi_I &=& \delta\Phi^J F_J
=  R^{JI}\delta\Phi_I F_J \nn
&=& (-)^{F+J}  R^{JI} F_J \delta\Phi_I= 
(-)^{F+I}  R^{IJ} F_J \delta\Phi_I\,\nn
\Rightarrow&& F^I = (-)^{F+I}  R^{IJ} F_J \cr
\delta\Phi^J F_J &=& F^I\delta\Phi_I =  F^I \delta\Phi^J R_{JI}\nn
&=&  (-)^{J(F+1)}\delta\Phi^J F^I R_{JI}
=  (-)^{I(F+1)}(-)^{I+1}\delta\Phi^J F^I R_{IJ}\nn
\Rightarrow&& F_J =  (-)^{IF+1} F^I R_{IJ}.
\label{eq:RaisingLowering}
\end{eqnarray}

Let us introduce a generic notation:
\begin{equation}
F_{I_1\cdots I_n}^{J_1\cdots J_m}
\equiv{\delta^{n+m}F\over\delta\Phi^{I_1}\cdots\delta\Phi^{I_n}\delta\Phi_{J_1}\cdots\cdots \delta\Phi_{J_m}}
\equiv{\overrightarrow\delta^n\over\delta\Phi^{I_1}\cdots\delta\Phi^{I_n}}
F{\overleftarrow\delta^m\over\delta\Phi_{J_1}\cdots\cdots \delta\Phi_{J_m}}
\end{equation} 
Then, since the left and right derivatives commute, we clearly have
\begin{equation}
F^J_I
={\overrightarrow\delta\over\delta\Phi^{I}}F{\overleftarrow\delta\over\delta\Phi_{J}}
={\delta\over\delta\Phi^{I}}F^J={\delta\over\delta\Phi_J}F_I
\end{equation} 

\subsection{BRS transformation}

Define the \BRS transformation from the action $S$ by
\begin{equation}
\delta_\B \Phi_I \equiv{\delta S\over\delta\Phi^I} \equiv S_I,
\label{eq:BRStrf}
\end{equation}
which actually stands for 
\begin{equation}
\delta_\B\Phi_I =
\cases{ \mib\delta_\B \Phi_I &for open ($I=0$) \cr
        \mib\delta_\B b_0^-\Phi_I &for closed ($I=1$) \cr}
\end{equation}
where $\mib\delta_\B$ is the true BRS transformation.
This BRS transformation $\mib\delta_\B$ is an {\it anti-derivation} 
so that it obeys the rule (\ref{eq:AntiDiffIdentity}). 
Using that rule, we have
\begin{equation}
\mib\delta_\B S = - S^I\mib\delta_\B\Phi_I. 
\end{equation} 
We note that $S^I$ for $I=1$ (closed case) always has a $b_0^-$ factor 
on the most right and so that we can multiply $c_0^-b_0^-$ to it from the 
right since $b_0^-c_0^-b_0^-=b_0^-$. So, 
inserting $(c_0^-)^I(b_0^-)^I$ which is $c_0^-b_0^-$ for $I=1$ and 1 for
$I=0$, we obtain
\begin{eqnarray}
\label{eq:dBS-BV}
\mib\delta_\B S &=& - S^I (c_0^-)^I(b_0^-)^I  \mib\delta_\B \Phi_I 
= - S^I (-c_0^-)^I  (\mib\delta_\B (b_0^-)^I\Phi_I) \nn
&=& -S^I (-c_0^-)^I  \delta_\B \Phi_I 
= -S^I (-c_0^-)^I  S_I. 
\end{eqnarray} 
So, if the action is BRS invariant, 
we have an identity, usually called BV master equation:
\begin{equation}
  S^I(-c_0^-)^IS_I =0.
\end{equation}

If the BV master equation is satisfied, the nilpotency of the BRS
transformation automatically follows as follows:
\begin{eqnarray}
(\mib\delta_\B)^2(b_0^-)^I \Phi_I &=& \mib\delta_\B \delta_\B\Phi_I = \mib\delta_\B S_I = 
(-)^{S_I+1} {\delta S_I\over\delta\Phi_J}\mib\delta_\B\Phi_J \nn
&=&(-)^{I+1} {\delta S_I\over\delta\Phi_J}(c_0^-)^J(b_0^-)^J\mib\delta_\B\Phi_J
=(-)^{I+1} {\delta S_I\over\delta\Phi_J}(-c_0^-)^J(\mib\delta_\B(b_0^-)^J\Phi_J) \nn
&=&(-)^{I+1} {\delta S_I\over\delta\Phi_J}(-c_0^-)^J\delta_\B\Phi_J
= (-)^{I+1} S_I^J (-c_0^-)^JS_J \nn
&=&(-)^{I+1}\half{\delta\over\delta\Phi^I}\left(S^J (-c_0^-)^JS_J\right) =0.
\end{eqnarray}

\subsection{Gauge invariance}

The gauge transformation is defined by
\begin{eqnarray}
\label{eq:dS-BV}
\delta(\Lambda)(b_0^-)^I\Phi_I \equiv(-)^{I+J}{\delta(\delta_\B \Phi_I)\over\delta\Phi_J}\Lambda_J = 
(-)^{I+J}S_I^J\Lambda_J.
\end{eqnarray}
Then, if the BV master equation is satisfied, the gauge invariance of
the action also follows automatically:
\begin{eqnarray}
\delta(\Lambda)S &=& S^I \delta(\Lambda)\Phi_I
= S^I (c_0^-)^I(b_0^-)^I \delta(\Lambda)\Phi_I
= S^I \bigl((c_0^-)^I \delta(\Lambda)(b_0^-)^I\Phi_I\bigr) \nn
&=& (-)^{J}S^I (-c_0^-)^I S_I^J\Lambda_J
= (-)^{J}\half {\delta\over\delta\Phi^J}\left(S^I (-c_0^-)^I S_I\right) \Lambda_J =0. 
\end{eqnarray}

It should hold that
\begin{equation}
 S^I (-c_0^-)^IS_I\ = S_I \cdot S^I (-c_0^-)^I
\end{equation}
since $\abs{S_I}=I$ and $\abs{S^I (-c_0^-)^I}=I+1=0$.
We can confirm this directly by the lowering and raising index identities 
(\ref{eq:RaisingLowering}) which now read
\begin{equation}
S^I = (-)^I R^{IJ}S_J, \qquad 
S_I = (-)^1 S^JR_{JI}.
\end{equation}
Using these, we see
\begin{eqnarray}
&&\hspace{-2em}S^I (-c_0^-)^IS_I \nn
&=& (-)^{I+1}R^{IJ}S_J (-c_0^-)^I S^KR_{KI}
= (-)^{I+1}R^{IJ}S_J \cdot S^K(+c_0^-)^IR_{KI} \nn
&=& (-)^{I+1}R^{IJ}S_J \cdot (-)^{IK}S^K(-c_0^-)^KR_{KI} \qquad \leftarrow 
(+c_0^-)^IR_{KI} = (-c_0^-)^KR_{KI} \nn
&=& R^{IJ}R_{KI}S_J  \cdot S^K(-c_0^-)^K 
\qquad \leftarrow\abs{S_J S^K(-c_0^-)^K}=1+J+K=1+2I=1 \nn
&=& (-)^{I+1}R^{JI}R_{IK}S_J  \cdot S^K(-c_0^-)^K \qquad 
\leftarrow R^{IJ}=R^{JI}, \quad R_{KI}=(-)^{I+1}R_{IK}, \nn
&=& \delta_K^{\ \ J}S_J  \cdot S^K(-c_0^-)^K 
= S_K \cdot S^K(-c_0^-)^K
\end{eqnarray}
Essentially the same procedures prove the above used identities:
\begin{eqnarray}
S_J^I (-c_0^-)^IS_I
&=&\half{\delta\over\delta\Phi^J}\left(S^I (-c_0^-)^IS_I\right), \nn
S^I (-c_0^-)^I S_I^J
&=& \half {\delta\over\delta\Phi_J}\left(S^I(-c_0^-)^IS_I\right)
\end{eqnarray}

\subsection{Anomaly}

If the integration measure is not BRS invariant, then the BV master 
equation gets a contribution from it and modified into
\begin{equation}
S^I(-c_0^-)^IS_I 
= \hbar\bigl({\delta\over\delta\Phi^I}(-c_0^-)^I{\delta\over\delta\Phi_I}\bigr)S 
= \hbar S^{IJ} (-c_0^-)^J R_{JI} .
\end{equation}
But this expression is too formal and it needs a suitable regularization
to properly define the RHS. We shall discuss this point in the forthcoming 
paper.

\section{GGRT}

The GGRT formulas have been proved by LPP\cite{rf:LPP} and the 
present authors (AKT).\cite{rf:AKT}\ \  However, they 
restricted to the simplest situation in which only open strings exist.
To apply the formulas in this mixed system of open and closed strings, 
we need some generalizations of the original GGRT, which we shall give 
in this appendix.

In our SFT, there appear seven LPP vertices, which we can classify into 
the following three different classes, depending on the type of CFT 
which is referred to in the definition of the vertex:
\begin{enumerate}
\item[I.] tree level open-type vertices $\bra{v_{\rm I}}$ (Grassmann odd)
\begin{eqnarray}
&&\hbox{3-pt:}\qquad \v3o(123xx), \quad \v2x(12xxx), \nn
&&\hbox{4-pt:}\qquad \v4o(1234{\sigma_0}), \quad 
\v\infty x(12{\sigma_0}xx), \quad 
\v\Omega x(123{\sigma_0}x).
\hspace{5em} 
\end{eqnarray}

\item[II.] tree level closed-type vertex $\bra{v_{\rm II}^\c}$ 
(Grassmann even)
\begin{equation}
\v3c(123xx)
\end{equation}

\item[III.] 1-loop level open-type vertex $\bra{v_{\rm L}}$ (Grassmann even)
\begin{equation}
\v\propto x(12{\sigma_1}{\sigma_2}x)
\end{equation}
\end{enumerate}

Here the tree and 1-loop level vertices refer to the CFT on sphere and 
torus, respectively. The closed-type vertex (which is now uniquely 
$\v3c(123xx)$) refers to a pair of CFT's corresponding to the 
holomorphic and ant-holomorphic degrees of freedoms, separately, while 
the open-type one refers to a single CFT on a complex $z$ plane. (This 
explains why $\v\infty x(12{\sigma_0}xx)$, for instance, is classified 
into the open-type vertex although it is the vertex of purely closed 
strings.)

Because of this, the closed-type vertex is always given by a tensor 
product of a pair of `open-type' vertices representing the
holomorphic and anti-holomorphic parts: in the present case, the 
closed 3-string vertex takes the form
\begin{equation}
\v3c(123xx) = \bra{\bar v_3(\bar1,\bar2,\bar3)}\otimes \bra{v_3(1,2,3)}
\end{equation}
The reflectors $\braRo(12)$ and $\braRc(12)$ are, of course, 2-point 
vertices, and similarly can be classified to the open-type and closed-type 
vertices, respectively. Indeed the latter closed reflector, 
and its ket counterpart also, are given in the following tensor product 
forms:
\begin{eqnarray}
\braRc(12) &=& \bra{\bar R(\bar1,\bar2)}\otimes \bra{R(1,2)} \nn
\ketRc(12) &=& \ket{R(1,2)}\otimes \ket{\bar R(\bar1,\bar2)}
\label{eq:factorize}
\end{eqnarray}
This can be confirmed by inspecting the explicit expressions
(2$\cdot$11) and (2$\cdot$12) in the previous paper I:
precisely speaking, Eq.~(\ref{eq:factorize}) holds if 
the exponent $E^\c_{12}$ in (2$\cdot$12) of I is replaced by
\begin{equation}
E^\c_{12} = \sum_{n\geq1}(-)^{n+1}\left(\frac{1}{n}
{\alpha_n^\mu}^{(1)}{\alpha_n}_\mu^{(2)}
+{c_n}^{(1)}{b_n}^{(2)}
-{b_n}^{(1)}{c_n}^{(2)}\right)+{\rm a.h.},
\end{equation}
where the alternating sign factor $(-)^{n}$ has been put additionally. 
Although the reflectors with and without this sign factor are equivalent
to each other in the presence of the closed string projection operator 
${\cal P}$, it is necessary for the equality (\ref{eq:factorize}) itself
to hold in the absence of ${\cal P}$.

The loop-level vertex is defined by the CFT on the torus:
\begin{eqnarray}
\bra{v_L(\{\Phi_i\};\tau)}\prod_i\ket{\calO_{\Phi_i}}_{\Phi_i} 
&=& 
\bigl\langle\, \prod_i\hat h_{\Phi_i}[\calO_{\Phi_i}]\, \big\rangle_{{\rm torus}
\ \tau} \nn
&\equiv& 
-{\rm Tr}\left[\,(\m1)^{N_{\rm FP}}\,q^{2L_0}\,
\prod_i\hat h_{\Phi_i}[\calO_{\Phi_i}]\, \right]\,,
\end{eqnarray}
where $q=e^{i\pi\tau}$ and 
$(\m1)^{N_{\rm FP}}$ is the FP ghost number defined by
\begin{equation}
N_{\rm FP} = c_0b_0 + \sum_{n\geq1}(c_{-n}b_n - b_{-n}c_n) 
\end{equation}
which counts the ghost number from the Fock vacuum.

As was shown in LPP and AKT, we have the following tree level GGRT which 
holds for any two tree level open-type vertices 
$\Gv I(\{B_j\},D)$ and $\Gv I(C,\{A_i\})$
\begin{equation}
\Gv I(\{B_j\},D) \Gv I(C,\{A_i\}) \ketRo(DC) = 
\Gv I(\{B_j\},\{A_i\})\,.
\label{eq:GGRT}
\end{equation}
The resultant LPP vertex $\Gv I(\{B_j\},\{A_i\})$ for the glued 
configuration also becomes a tree level open-type vertex. 

We need another type of gluing already at the tree level; 
the gluing of a tree level open-type vertex 
$\bra{v_{\rm I}(\{B_j\}, D^\c)}$ 
containing at least one closed string $D^\c$ 
and the tree level closed-type vertex 
$\bra{v_{\rm II}^\c(C^\c,\{A^\c_i\})}$, 
by contraction using $\ketRc(DC)$.
However, applying the above GGRT twice, we can show that 
the same form of GGRT holds also for this case:
\begin{equation}
\bra{v_{\rm I}(\{B_j\}, D^\c)}
\bra{v_{\rm II}^\c(C^\c,\{A^\c_i\})}\ketRc(DC) 
=\bra{v_{\rm I}(\{B_j\}, \{A^\c_i\})}.
\end{equation}
Indeed, separating the various closed string quantities into the 
holomorphic and anti-holomorphic parts and writing $D^\c = (\bar D, D)$, 
etc, we have
\begin{eqnarray}
&&\bra{v_{\rm I}(\{B_j\}, D^\c)}
\bra{v_{\rm II}^\c(C^\c,\{A^\c_i\})}\ketRc(DC) \nn
&&\qquad =\bra{v_{\rm I}(\{B_j\}, \bar D, D)} 
\bra{\bar v(\bar C,\{\bar A_i\})}\bra{v(C,\{A_i\})}
\ket{R(D,C)}\ket{\bar R(\bar D,\bar C)} \nn
&&\qquad 
=
\bra{v_{\rm I}(\{B_j\}, \bar D, D)} \bra{v(C,\{A_i\})}
\ket{R(D,C)} 
\cdot \bra{\bar v(\bar C,\{\bar A_i\})}\ket{\bar R(\bar D,\bar C)} \nn
&&\qquad 
=
\bra{\tilde v_{\rm I}(\{B_j\}, \bar D, \{A_i\})}
\bra{\bar v(\bar C,\{\bar A_i\})}\ket{\bar R(\bar D,\bar C)} \nn
&&\qquad 
=\bra{v_{\rm I}(\{B_j\}, \{\bar A_i\}, \{A_i\})}
=\bra{v_{\rm I}(\{B_j\}, \{A^\c_i\})}.
\end{eqnarray}

Next consider the gluing of two tree level open-type vertices 
each containing closed string 
by contraction using $\ket{R^\c}$:
\begin{eqnarray}
&&\bra{v_{\rm I}(D^\c,\{B_j\})}\bra{v_{\rm I}(C^\c,\{A_i\})}\ketRc(DC) \nn
&&=\bra{v_{\rm I}(\bar D, D, \{B_j\})}\bra{v_{\rm I}(\bar C, C, \{A_i\})}
\ket{R(D,C)}\ket{\bar R(\bar D,\bar C)}
\end{eqnarray}
But this has exactly the same form as the 1-loop level GGRT proved in 
AKT:
\begin{equation}
\Gv I(D,F,\{ B_j\}) \Gv I(C,\{ A_k\},E) \ketRo(DC)\ketRo(EF) = 
\bra{v_{\rm L}(\{ B_j\},\{ A_k\};\tau)}\,.
\label{eq:GGRTloop}
\end{equation}
Therefore, we immediately obtain
\begin{equation}
\bra{v_{\rm I}(D^\c,\{B_j\})}\bra{v_{\rm I}(C^\c,\{A_i\})}\ketRc(DC) 
=-\bra{v_{\rm L}(\{B_j\}, \{A_i\})}
\end{equation}
with $\bra{v_{\rm L}(\{B_j\}, \{A_i\})}$ being the 1-loop level 
LPP vertex resultant from this gluing. (Note, however, that there is 
actually a sign ambiguity here in the right-hand side 
%defining $\bra{v_{\rm L}(\{B_j\}, \{A_i\})}$ 
since the exchange of the two vertices $\bra{v_{\rm I}}$ 
on the left-hand side gives rise to a sign change contrary to the case of 
tree level GGRT formula (\ref{eq:GGRT}).)

Finally consider the gluing of 1-loop level vertex and tree level 
open-type vertex by contraction using $\ket{R^\o}$. 
\begin{equation}
\bra{v_{\rm L}(D,\{B_j\})}\Gv I(C,\{A_i\})\ketRo(DC)
=\bra{v_{\rm L}(\{B_j\}, \{A_i\})}
\label{eq:looptree}
\end{equation}
This can also be proved by using the tree level GGRT. To do this, we
first note that the loop level vertex $\bra{v_{\rm L}(\{\Phi_i\})}$ 
can generally be reduced to a tree level vertex in the 
following form:
\begin{equation}
\bra{v_{\rm L}(\{\Phi_i\})} = \Gv I(F, \{\Phi_i\}, E)\ketRo(EF) \,.
\end{equation}
This is clear since if we cut the loop of the 1-loop 
diagram corresponding to the vertex $\bra{v_{\rm L}(\{\Phi_i\})}$ 
then the both sides of the cutting line correspond to the 
intermediate (open) strings $E$ and $F$ and the diagram becomes a tree 
level vertex $\Gv I(F, \{\Phi_i\}, E)$ before contraction by $\ketRo(EF)$.
Then Eq.~(\ref{eq:looptree}) is proved by the tree level GGRT as follows:
\begin{eqnarray}
&&\bra{v_{\rm L}(\{B_j\}),D}\Gv I(C,\{A_i\})\ketRo(DC) \nn
&&\qquad =\Gv I(F,\{B_j\},D,E)\ketRo(EF)\Gv I(C,\{A_i\})\ketRo(DC) \nn
&&\qquad =\Gv I(F,\{B_j\},D,E)\Gv I(C,\{A_i\})\ketRo(DC)\ketRo(EF) \nn
&&\qquad =\Gv I(F,\{B_j\},\{A_i\},E)\ketRo(EF)
=\bra{v_{\rm L}(\{B_j\}, \{A_i\})}.
\end{eqnarray}

In summary, we have shown that we can apply the naive GGRT formula to 
all the cases we are discussing in the text.


\newpage
\begin{thebibliography}{99}
%%%%%%%%%%%%%%%%%%%%%%%%%%%%%%%%%%%%%%%%%%%%%%%%%%%%%%%%%%%%%
% Some macros are available for the bibliography:
%   o for general use
%      \JL : general journals          \andvol : Vol (Year) Page
%   o for individual journal 
%      \PR  : Phys. Rev.               \PRL : Phys. Rev. Lett.
%      \NP  : Nucl. Phys.              \PL  : Phys. Lett.
%      \JMP : J. Math. Phys.           \CMP : Commun. Math. Phys.
%      \PTP : Prog. Theor. Phys.       \JPSJ: J. Phys. Soc. Jpn.
%      \JP  : J. of Phys.              \NC  : Nouvo Cim.
%      \IJMP: Int. J. Mod. Phys.       \ANN : Ann. of Phys.
% Usage:
%   \PR{D45,1990,345}            ==> Phys.~Rev.\ {\bf D45} (1990), 345
%   \JL{Phys.~Lett.,A30,1981,56} ==> Phys.~Lett.\ {\bf A30} (1981), 56
%   \andvol{B123,1995,1020}      ==> {\bf B123} (1995), 1020
%%%%%%%%%%%%%%%%%%%%%%%%%%%%%%%%%%%%%%%%%%%%%%%%%%%%%%%%%%%%%
\bibitem{rf:KugoTaka}
T.~Kugo and T.~Takahashi,
  \PTP{99,1998,649}.
\bibitem{rf:GreenSchwarz}
M.~B.~Green and J.~H.~Schwarz,
  \PL{151B,1985,21}.
\bibitem{rf:DougGrin}
M.~R.~Douglas and B.~Grinstein,
  \PL{183B,1987,52}.
\bibitem{rf:Weinberg}
S.~Weinberg,
  \PL{187B,1987,278}.
\bibitem{rf:ItoyamaMoxhay}
H.~Itoyama and P.~Moxhay,
  \NP{B293,1987,685}.
\bibitem{rf:Ohta}
N.~Ohta, 
  \PRL{59,1987,176}.
\bibitem{rf:DasRey}
S.R.~Das and S-J.~Rey,
  \PL{186B,1987,328}.
\bibitem{rf:Tseytlin1}
A.A.~Tseytlin,
  \PL{208B,1988,228}.
\bibitem{rf:FischSusskind1}
W.~Fischler and L.~Susskind,
  \PL{171B,1986,383}.
\bibitem{rf:FischSusskind2}
W.~Fischler and L.~Susskind,
  \PL{173B,1986,262}.
\bibitem{rf:CLNY}
C.~G.~Callan, C.~Lovelace, C.~R.~Nappi and S.~A.~Yost,
  \NP{B288,1987,525}.
\bibitem{rf:PolCai}
J.~Polchinski and Y.~Cai,
  \NP{B296,1987,91}.
\bibitem{rf:FischSussKlebanov}
W.~Fischler, I.~Klebanov and L.~Susskind,
  \NP{B306,1988,271}. 
\bibitem{rf:Tseytlin2}
A.A.~Tseytlin,
  \IJMP{A3,1988,365}.
\bibitem{rf:Pol}
J.~Polchinski,
  \NP{B307,1988,61}.
\bibitem{rf:kugozwie}
T.~Kugo and B.~Zwiebach,
  \PTP{87,1992,801}.
\bibitem{rf:KK} 
M.~Kaku and K.~Kikkawa, 
  \PR{D10,1974,1823}.
\bibitem{rf:ST1}
Y.~Saitoh and Y.~Tanii,
  \NP{B325,1989,161}.
\bibitem{rf:ST2}
Y.~Saitoh and Y.~Tanii,
  \NP{B331,1990,744}.
\bibitem{rf:KikkawaSawada}
K.~Kikkawa and S.~Sawada,
  \NP{B335,1990,677}.
\bibitem{rf:GreenSchwarz2}
M.~B.~Green and J.~H.~Schwarz,
  \NP{B243,1984,475}.
\bibitem{rf:Zwie2}
B.~Zwiebach, ``Oriented Open-Closed String Theory Revisited'',
 {hep-th/9705241}.
\bibitem{rf:LPP}
A.~LeClair, M.E.~Peskin and C.R.~Preitschopf,
  \NP{B317,1989,411}.
\bibitem{rf:AKT}
T.~Asakawa, T.~Kugo and T.~Takahashi, 
  to appear in Prog.\ Theor.\ Phys.,  {\tt hep-th/9805119}.
\bibitem{rf:GM}
S.B.~Giddings and E.~Martinec,
  \NP{B278,1986,91}.
\bibitem{rf:Martinec}
E.~Martinec,
  \NP{B281,1987,157}.
\bibitem{rf:DHokerPhong}
E. D'Hoker and D.H.~Phong,
  \NP{B296,1986,205}.
\bibitem{rf:AGMV}
L.~Alvarez-Gaum\'e, C.~Gomez,
G.~Moore and C.~Vafa,
  \NP{B303,1988,411}.
\bibitem{rf:KugoSuehiro}
T.~Kugo and K.~Suehiro,
  \NP{B337,1990,434}.
\bibitem{rf:HIKKO1}
H.~Hata, K.~Itoh, T.~Kugo, H.~Kunitomo and K.~Ogawa,
  \PR{D34,1986,2360}.
\bibitem{rf:HIKKO2}
H.~Hata, K.~Itoh, T.~Kugo, H.~Kunitomo and K.~Ogawa,
  \PR{D35,1987,1318}.
\bibitem{rf:ShapThorn}
J.~A.~Shapiro and C.~B.~Thorn,
  \PR{D36,1987,432}.
\bibitem{rf:Hata-Nojiri}
H.~Hata and M.~M.~Nojori,
  \PR{D36,1987,1193}.
\bibitem{rf:Hata1}
H.~Hata,
   \NP{B329,1990,698}.
\bibitem{rf:Hata2}
H.~Hata,
   \NP{B339,1990,663}.
\bibitem{rf:Zwie1}
B.~Zwiebach,
  \NP{B390,1993,33}.
\bibitem{rf:HataZwie}
B.~Zwiebach and H.~Hata,
  \ANN{229,1994,177}.
\bibitem{rf:BV}
I.A.~Batalin and G.A.~Vilkovisky,
  \PR{D28,1983,2567}.
\end{thebibliography}
\end{document}